\def\imgdir{images}

\documentclass[fleqn,final,5p,times,twocolumn]{elsarticle}

\usepackage{amsmath,amsfonts,txfonts}

\usepackage{amssymb}
\usepackage{amsfonts,amssymb}
\usepackage{textcomp}
\usepackage{graphicx}
\usepackage[english]{babel}
\usepackage{epsfig}
\usepackage{graphicx}
\usepackage{amsmath}
\usepackage{amsthm}
\usepackage{amssymb}
\usepackage{subcaption}
\usepackage{xcolor}
\usepackage{todonotes}
\usepackage{booktabs}
\usepackage{pifont}
\usepackage[linesnumbered,ruled]{algorithm2e}
\usepackage{algorithmic}
\usepackage{multirow}
\usepackage{diagbox}
\usepackage{lineno}
\usepackage{mathtools}
\usepackage{lipsum}
\usepackage{wrapfig}

\newtheorem{proposition}{Proposition}[section]
\newtheorem{lemma}{Lemma}[proposition]

\SetKwComment{Comment}{/* }{ */}

\journal{Additive Manufacturing}

\begin{document}
\setlength{\mathindent}{0pt}

\begin{frontmatter}
\title{Inference of highly time-resolved melt pool visual characteristics and spatially-dependent lack-of-fusion defects in laser powder bed fusion using acoustic and thermal emission data}

\author[label1,label2,label3]{Haolin Liu}
\author[label1]{Christian Gobert}
\author[label1]{Kevin Ferguson}
\author[label1]{Brandon Abranovic}
\author[label1]{Hongrui Chen}
\author[label1,label2,label3]{Jack L. Beuth\corref{cor1}}
\author[label2,label3]{Anthony D. Rollett\corref{cor1}}
\author[label1,label3]{Levent Burak Kara\corref{cor1}}

\address[label1]{Department of Mechanical Engineering, Carnegie Mellon University, Pittsburgh, Pennsylvania, USA}

\address[label2]{Department of Materials Science and Engineering, Carnegie Mellon University, Pittsburgh, Pennsylvania, USA}

\address[label3]{NextManufacturing Center, Carnegie Mellon University, Pittsburgh, Pennsylvania, USA}

\cortext[cor1]{Corresponding authors. \\\hspace*{1.6em} E-mail address: \{beuth, rollett, lkara\}@andrew.cmu.edu}

\begin{abstract}
With a growing demand for high-quality fabrication, the interest in real-time process and defect monitoring of laser powder bed fusion (LPBF) has increased, leading manufacturers to incorporate a variety of online sensing methods including acoustic sensing, photodiode sensing, and high-speed imaging. However, real-time acquisition of high-resolution melt pool images in particular remains computationally demanding in practice due to the high variability of melt pool morphologies and the limitation of data caching and transfer, making it challenging to detect the local lack-of-fusion (LOF) defect occurrences. In this work, we propose a new acoustic and thermal information-based monitoring method that can robustly infer critical LPBF melt pool morphological features in image forms and detect spatially-dependent LOF defects within a short time period. We utilize wavelet scalogram matrices of acoustic and photodiode clip data to identify and predict highly time-resolved (within a 1.0~$\mathrm{ms}$ window) visual melt pool characteristics via a well-trained data-driven pipeline. With merely the acoustic and photodiode-collected thermal emission data as the input, the proposed pipeline enables data-driven inference and tracking of highly variable melt pool visual characteristics with $R^2\geq 0.8$. We subsequently validate our proposed approach to infer local LOF defects between two adjacent scanlines, showing that our proposed approach can outperform our selected baseline theoretical model based on previous literature. Revealing the physical correlation between airborne acoustic emission, thermal emission, and melt pool morphology, our work demonstrates the feasibility of creating an efficient and cost-effective acoustic- and thermal-based approach to facilitate online visual melt pool characterization and LOF defect detection. We believe that our work can further contribute to the advances in quality control for LPBF. 
\end{abstract}

\begin{graphicalabstract}
\label{sec:graphical}
\begin{figure*}[!ht]
\center{\includegraphics[width=\linewidth]
{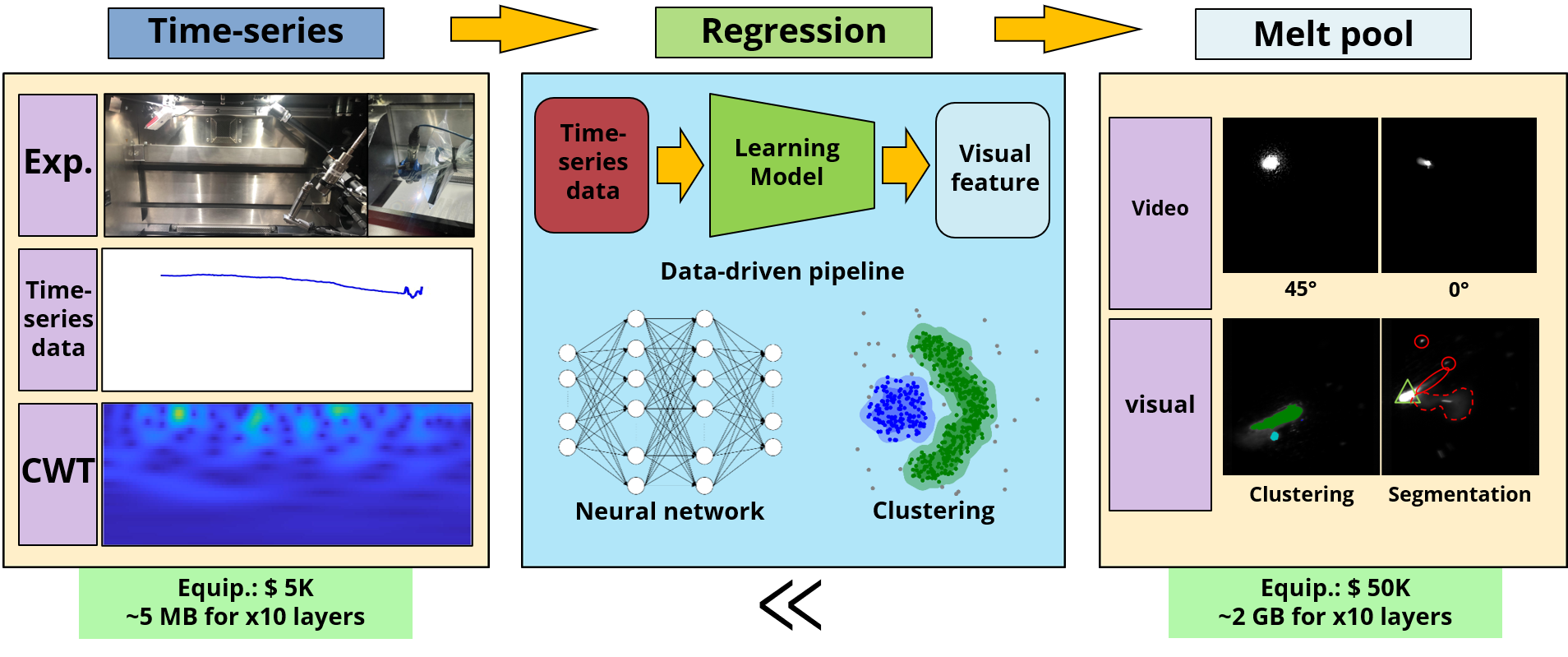}
\caption{Graphical Abstract}
\label{fig:graphical_abstract}
}
\end{figure*}
\end{graphicalabstract}

\begin{highlights}
\item In this work, we investigate the capability and efficacy of highly time-resolved LPBF visual tracking and process monitoring by employing a deep learning approach and using acoustic and photodiode-collected data. 

\item Accurate melt pool image results can be predicted and reconstructed within a time window as short as 1.0~$\mathrm{ms}$. 

\item Our work also performs knowledge discoveries and implies a potential underlying relationship between acoustic signatures, thermal emission, and corresponding fusion dynamics.  

\item We validate our proposed approach to local LOF defect detection between two adjacent scanlines, showing that our proposed approach can outperform our selected baseline theoretical model. 
\end{highlights}

\end{frontmatter}


\section{Introduction}
\label{sec:intro}
Metallic laser powder bed fusion (LPBF) is one of the most widely applied and well-developed additive manufacturing (AM) technologies of the last decade. Almost every build with LPBF has printing defects caused by multiple flaw formation mechanisms, including keyholing (KH), lack-of-fusion (LOF), and bead-up (BU) ({\it aka} balling). These three types of defects occur under various printing conditions and can eventually lead to high porosity, shortened fatigue life, and poor mechanical properties of the as-built part~\cite{Narra:2023aa}. As a result, a demand from both industry and academia naturally emerges for the quantification and optimization of LPBF manufacturing qualities. To that end, manufacturers must attentively perform process monitoring, a critical step in the workflow in which data is collected, analyzed, and fed back to the manufacturing process. 

Process monitoring and quantification have been developed in many manufacturing processes for quality and efficiency improvement~\cite{forien2020detecting, repossini2017use, lough2020situ, zhang2021data, guo2019situ, hojjatzadeh2021situ, javadi2020continuous, grasso2018situ, tenbrock2021effect}. For some well-known manufacturing processes like computer numerical control (CNC) machining, monitoring techniques such as force/vibration sensing are utilized to detect machining defects and tool wear~\cite{downey2016real, plaza2019efficiency}. Other heat-conduction-driven processes, such as injection molding and casting, also perform ex-situ post-processing quality quantification with sectioning, computed tomography (CT) scanning, or non-destructive material testing to optimize process parameters such as material inflow rate, cooling time, or mold dimension. For the LPBF process, the melt pool morphology -- the size and shape the melt pool adopts at any given moment and how it varies along a scanline or in a layer -- is one of the most important characteristics that concerns process engineers~\cite{grasso2021situ, everton2016review, mccann2021situ}. Dynamic melt pool oscillation might cause undesired changes in the melt pool morphology over time and may cause a variety of printing defects, such as spatter ejection, reduction of laser absorption, unstable remelt ratio, plume-induced laser blockage, end-of-track anomalies, and even LOF~\cite{grasso2021situ, everton2016review, khairallah2021onset}. However, despite wide applications of some well-known theoretical models and simulation tools such as the Rosenthal or Eagar-Tsai equations ~\cite{bachmann2022elucidation, shahabi2022statistical, yang2020scan, jones2021hybrid, dong2022part, meier2021physics, li2021deformations}, it is still challenging to precisely capture the transient melt pool status as only a function of initial printing conditions and the time stamp during the print. Moreover, since the LPBF process is highly mechanized and automated, manufacturers are also interested in combining melt pool monitoring with process design to achieve online printing characterization and optimization, such that a build with minimum defects can be accomplished with robust and delicate closed-loop process control~\cite{mccann2021situ, yeung2020meltpool, yeung2020residual, liao2022simulation}. Owing to the highly complex and dynamic physical nature, as well as the value of online process awareness and quality management, it is crucial to introduce real-time in-situ visual monitoring of the melt pool in LPBF~\cite{yeung2020residual}. 

\begin{figure}[!htb]
\center{\includegraphics[width=\linewidth]
{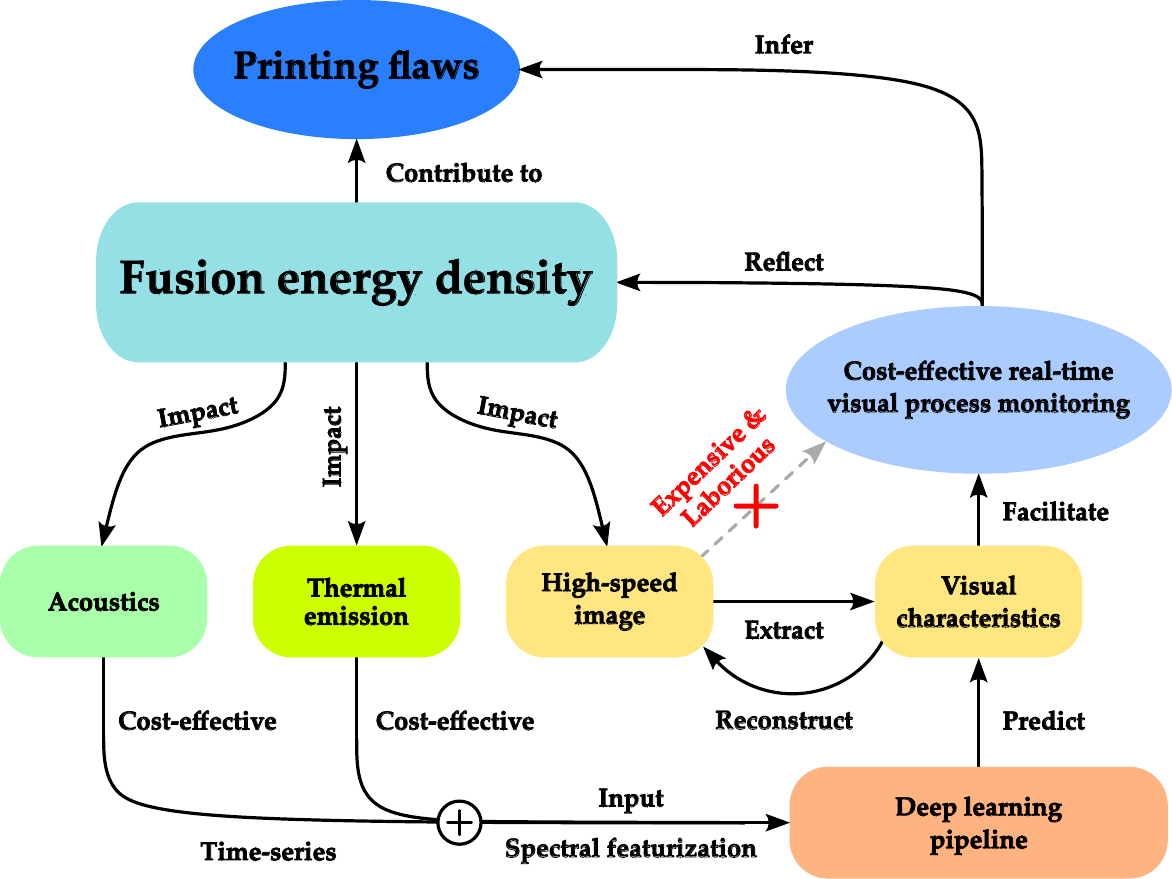}
\caption{The general pipeline and big picture of the work. We propose the use of acoustic and photodiode-collected thermal information to identify and track the melt pool visual characteristics in the LPBF process efficiently and nearly in real-time. We also utilize the proposed acoustic-and-thermal-driven monitoring method to infer printing flaws ({\it e.g.}, the occurrence of LOF defects) at the scanline-level. }
\label{fig:flowchart}
}
\end{figure}

Given LPBF's fast laser scanning speed and small melt pool dimensions, high-resolution high-speed imaging (HSI) with either a co-axial or an off-the-axis setup is a very natural and popular approach to observing and capturing the melt pool morphology data above the printing bed during the build. By selecting optimal exposure times, photo bandpass, observation angle, viewing frame dimension, and imaging frequency parameters, HSI is able to collect a set of images that include a melt pool, associated plume, scanned melting track, ejected spatters, and surrounding powder with very high resolution. Nevertheless, despite the excellent capability of the HSI monitoring method, significant limitations in HSI data collection -- such as limited caching memory and slow data transfer -- prevent its real-time use for most practical print jobs. This motivates the use of more easily accessible process monitoring~\cite{Ren:2023aa} methods that are complementary to HSI. For instance, the ability of acoustic emission (AE) and fusion-induced photodiode signals to detect printing defects, first studied in the laser welding community~\cite{drouet1982acoustic, prezelj2008use, wang2009analysis, cayo2009non, asif2022machine, zhang2019real}, has been investigated in recent years for LPBF and also other AM processes~\cite{tempelman2022detection, drissi2022differentiation, pandiyan2020analysis, pandiyan2021semi, taheri2019situ, wasmer2018situ, wu2016situ}. AE- and photodiode-based methods are two of the most frequently studied process monitoring approaches in LPBF and they have been successfully validated in many promising applications~\cite{Ren:2023aa, tempelman2022detection, drissi2022differentiation, wasmer2018situ, hossain2020situ, gillespie2021situ}. However, despite the efficacy of AE and photodiode-collected thermal emission data for process monitoring, high-speed images are still necessary for capturing the rapidly varying melt pool morphology. This motivates the present study, which is aimed at obtaining high-speed images at lower cost. 

Owing to the advanced development of machine learning (ML) methodologies in recent years, it is possible to select and train a data-driven deep learning model given the collected process data and then use it to construct pipelines that facilitate process monitoring as well as post-analysis tasks~\cite{mccann2021situ, jones2021hybrid, tan2020neural, fang2022data, demir2021laser, qin2022research, doh2022bayesian, rohe2021detecting, acevedo2020residual, smoqi2022monitoring, liu2021physics, fang2021situ, pandiyan2022deep}.  Wasmer {\it et al.}~\cite{wasmer2018situ} and Shevchik {\it et al.}~\cite{shevchik2018acoustic} show how the AE signals can be correlated with different levels of an as-built part's porosity using a convolutional neural network (CNN)-based ML model. Tempelman {\it et al.}~\cite{tempelman2022detection} also fits a support vector machine (SVM) to demonstrate that AE has the potential to be used for online keyhole occurrence detection during the LPBF process. Laser reflection and thermal emission data have also been exploited to characterize KH modes and laser absorptivity in a data-driven manner~\cite{Ren:2023aa}. However, many of these previous works mainly use ML models to perform classification tasks, and fewer efforts have been invested in highly time-resolved ({\it i.e.}, within a time window of a few milliseconds) melt pool visual feature regression using any of the aforementioned methods. Thus, this gap of cost-effective visual monitoring for LPBF remains to be filled, and its application deserves to be explored further. 

Hence, we propose an ML-based pipeline that infers the melt pool visual characteristics given the AE and photodiode-collected thermal emission time-series data in a highly time-resolved manner, as shown in Fig.~\ref{fig:flowchart}. Our results not only demonstrate the feasibility of utilizing acoustic and photodiode-based signals to quantify critical dynamic melt pool features, but also further explore the physical correlations between the acoustic, thermal, and transient melt pool image data. Our results indicate that by inputting the combination of acoustic and thermal emission data, our model achieves a performance with an $R^2$ score of 0.85 for melt pool geometric feature prediction. The parametric study result of the time window shows that our pipeline is capable of achieving accurate visual tracking of the melt pool in a time window as short as 1.0~$\mathrm{ms}$. Moreover, we also developed a new melt pool tracking evaluation metric $Q$ to assess the prediction quality that includes critical melt pool physical features such as width and variability. As a validation case, we demonstrate that our proposed data-driven approach can detect LOF defect occurrences at the scanline level under different local heat accumulation statuses, showcasing the approach's potential capability of real-time defect characterization. We believe this work is a valuable step toward multi-modal and cost-effective LPBF process monitoring. 
\section{Experiment and Methods}
\label{sec:method}
\begin{figure*}[!ht]
\center{\includegraphics[width=\linewidth]
{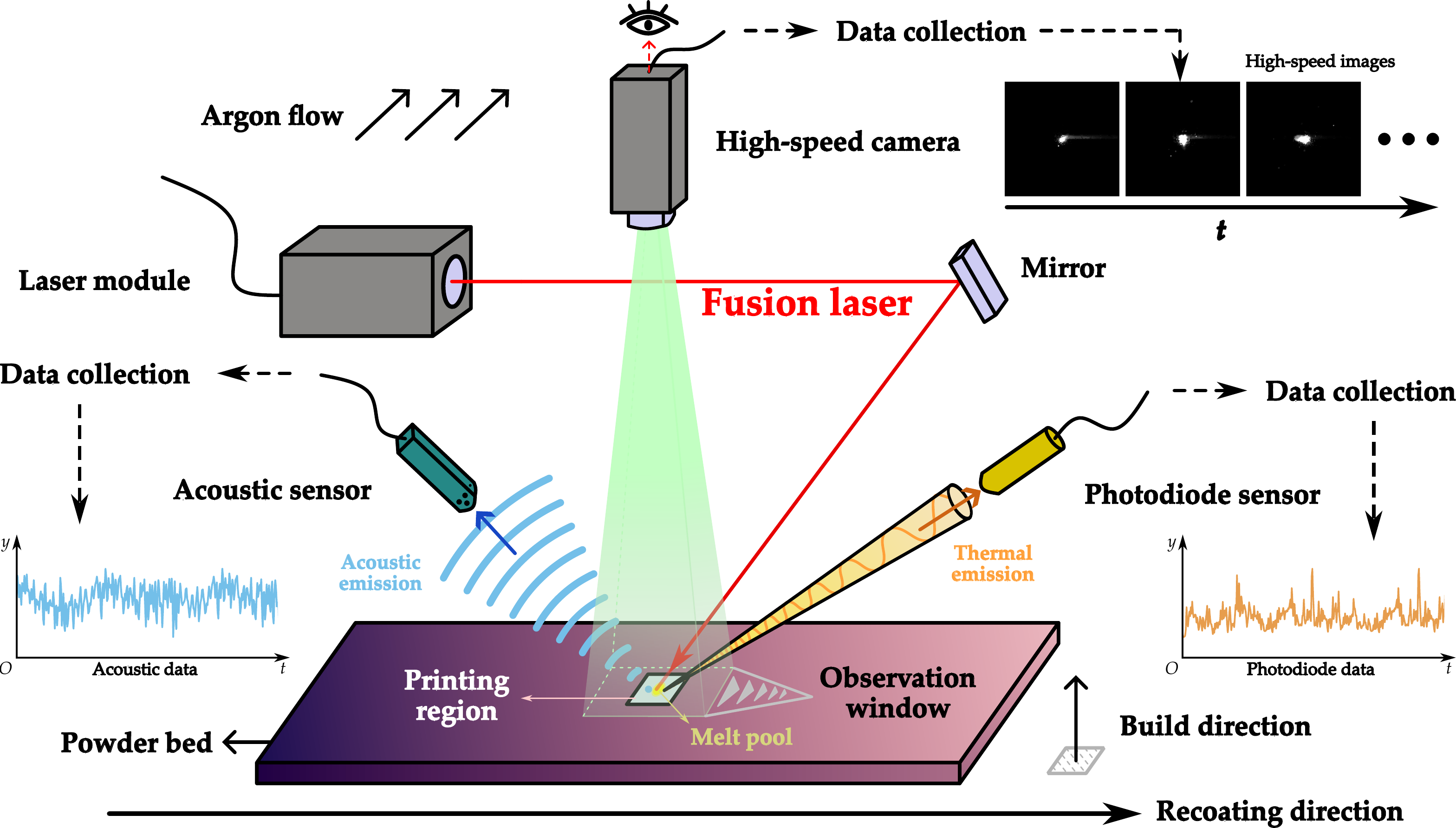}
\caption{Depiction of experiment and sensor setup inside the LPBF printing chamber. }
\label{fig:experimental_setup}
}
\end{figure*}

In this section, we first elaborate on the experimental details of acoustic data, photodiode sensor-captured thermal emission data, and high-speed data collection. Next, we introduce our data synchronization and processing workflow. Finally, we discuss our ML model and how we use acoustic and photodiode data to predict the transient visual features of melt pools. 

\subsection{Experiment setup and data collection}
\label{subsec:exp}
The LPBF printing experiments were conducted on Carnegie Mellon University's EOS M290  equipped with a 1064~$\mathrm{nm}$ Yb-fiber laser with a Gaussian intensity distribution, a maximum output power of 400~$W$, and a spot diameter of 100~$\mu\mathrm{m}$. Unless otherwise stated, all machine settings were set to the EOS recommended values for Ti-6Al-4V (Ti-64). The inert gas used in the process was Argon (Ar). Ti-64 powder was gas atomized with a D10 of 28~$\mu\mathrm{m}$, D50 of 39~$\mu\mathrm{m}$, and D90 of 55~$\mu\mathrm{m}$, obtained from Allegheny Technologies Incorporated (ATI Inc.). The chemistry of the powder provided by ATI Inc. is detailed in Tab.~\ref{tab:powder_chemistry}.

\begin{table}[!ht]
    \centering{
    \caption{Composition in weight percentage ($wt\%$) of Ti-64 powder. }
    \vspace{-2mm}
    \resizebox{\linewidth}{!}{
    \begin{tabular}{c|c|c|c|c|c|c|c|c|c}
        \toprule
        \textbf{Element} & Al & C & Fe & H & N & O & V & Y & Ti \\
        \midrule
        \textbf{wt\%} & $6.0300$ & $0.0080$ & $0.2000$ & $0.0007$ & $0.0100$ & $0.1520$ & $4.0400$ & $<0.0009$ & $\mathrm{Balance}$ \\
        \bottomrule
    \end{tabular}}
    \label{tab:powder_chemistry}}
\end{table}

\begin{figure}[!htb]
\center{\includegraphics[width=\linewidth]
{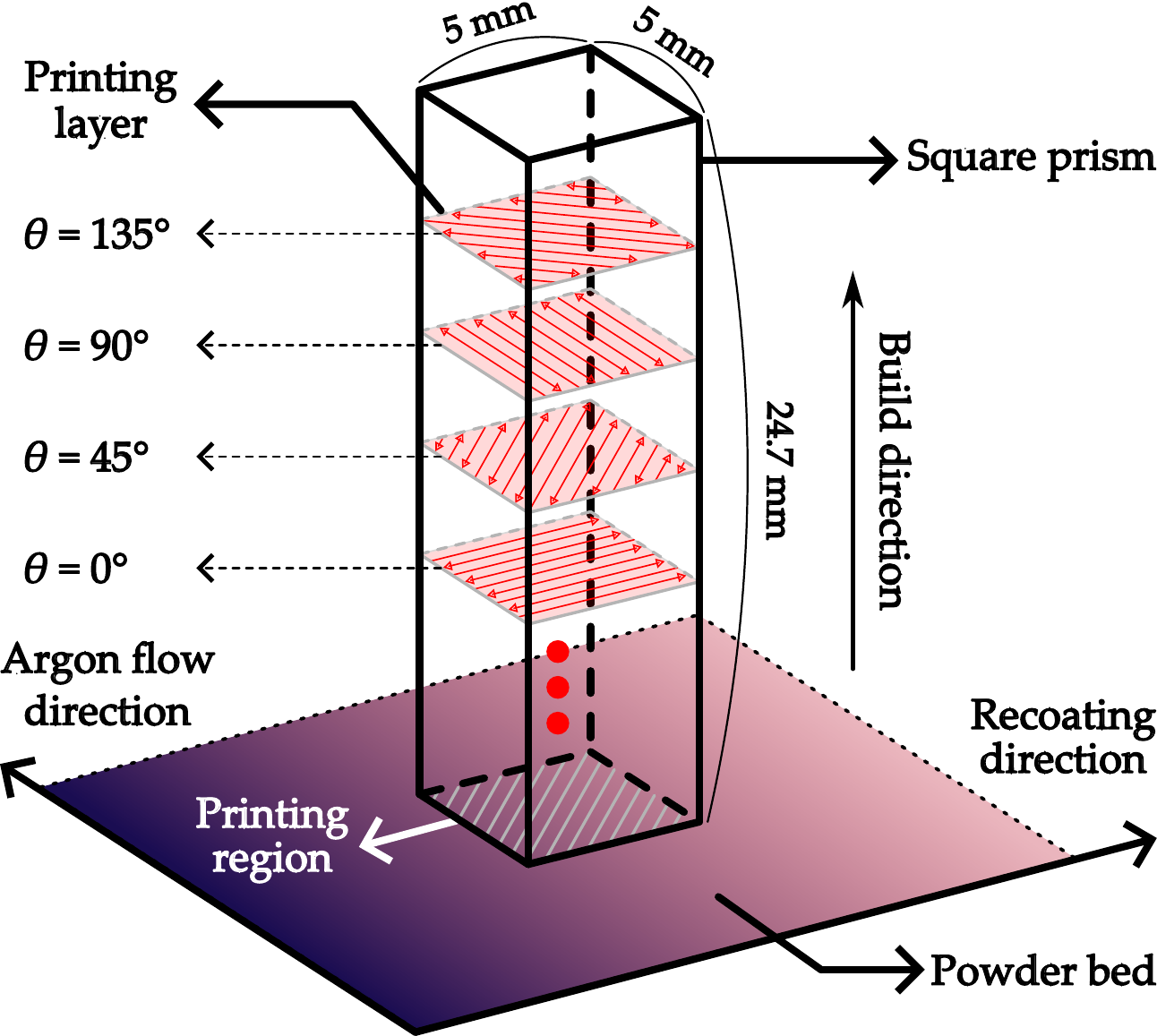}
\caption{Printed square prism and different laser scanning orientations. }
\label{fig:printed_sample}
}
\end{figure}

The printing chamber setup, sensor installation, and collected data forms are shown in Fig.~\ref{fig:experimental_setup}. We conducted a set of experiments with different printing conditions to collect process data with three sensors. In our experiments, we designed and printed a $5\times 5\times 24.7$~$\mathrm{mm}^{3}$ square prism using Ti-64 alloy, as shown in Fig. \ref{fig:printed_sample}. The layer thickness and hatch spacing of the Ti-64 powder were 30~$\mu\mathrm{m}$ and 140~$\mu\mathrm{m}$, respectively. 

During the printing time, three types of sensors -- an acoustic sensor, photodiode sensors, and a high-speed camera -- were set up to collect the acoustic data, thermal radiant emissions, and top-view melt pool image data, respectively. The acoustic and photodiode sensors were pointed directly at the print location. The high-speed camera was aimed downward at the build plate from the top of the build chamber through an observation window. We tuned the focus of the HSI framing camera to ensure that it covered the entire $5\times 5$~$\mathrm{mm}^{2}$ printing region and had a clear view of the sample at the build location. We expound the data collection process in detail below: 

\textbf{Acoustic data collection. }A PCB Piezotronics HT378A06 microphone system was placed inside the EOS build chamber. The acoustic sensor was connected to a National Instruments CDAQ NI9232 data collection module with a sampling frequency set to 100~$\mathrm{kHz}$. We ensured that the frequency range of our acoustic sensor could reach at least 40~$\mathrm{kHz}$, such that it satisfied the Nyquist-Shannon sampling theorem and was able to capture the high-frequency acoustic signatures reported in previous literature~\cite{shevchik2018acoustic}. The acoustic sensor was located $50$~$\mathrm{mm}$ above the build plate, and the distance between the sensor and the build plate was fixed. For the convenience of later synchronization between different data streams and data processing, we attached a mechanical switch on the recoater to automatically trigger the partitioning of the collected data into individual layers. 

\textbf{Photodiode data collection. } As shown in Fig. \ref{fig:experimental_setup}, two photodiodes were mounted inside the EOS M290 build chamber, a Thorlabs PDA10CS2 (A10) and PDA20CS2 (A20). The Indium Gallium Arsenide (InGaAs) photodiodes were sensitive to wavelengths between 800 and 1700~$\mathrm{nm}$. The A10 photodiode had a 1~$\mathrm{mm}$ diameter sensor, and the A20 had a 2~$\mathrm{mm}$ diameter sensor. The A10 photodiode had a 15~$\mathrm{ns}$ rise time, and the A20 had a 25~$\mathrm{ns}$ rise time, where rise time was defined as the time required to rise from 10\% to 90\% of a step input. The A10 photodiode was fitted with a 1064~$\mathrm{nm}$ band-stop filter to prevent laser light from damaging the sensor, whereas the A20 was fitted with a band-pass $1300\pm 30~\mathrm{nm}$ filter. The smaller A10 sensor recorded thermal emissions across the full spectrum 800-1700~nm (except near and at 1064~nm) whereas the larger A20 sensor only recorded thermal emissions in a small infrared band. The photodiode sampling rate was 100~$\mathrm{kHz}$, and data was recorded with a National Instruments CDAQ NI922. Since the photodiode sensor shared the same sampling frequency as the acoustic sensor, we later used it as a critical time stamp to synchronize the collected acoustic data and the high-speed melt pool image data. 

\textbf{HSI data collection. } As shown in Fig. \ref{fig:experimental_setup}, a monochrome Photron Mini AX200-900K-M-16GB (AX200) high-speed camera was used to observe the printing process. The AX200 high-speed camera had a $1024\times 1024$ silicon pixel array sensitive to wavelengths from 400~$\mathrm{nm}$ to 1000~$\mathrm{nm}$. A high magnification optical train was placed on the high-speed camera that enabled up to $5\times 5~\mu\mathrm{m}^2$ resolution of the build surface when mounted above the printing chamber. A Thorlabs FESH950 short-pass filter was placed in the optical train to limit incident light wavelengths from 500~$\mathrm{nm}$ to 950~$\mathrm{nm}$, which prevented high-energy laser reflectance from damaging the sensor. The camera recording rate was set to 22,500 frames per second with a frame integration time equivalent to 50,000 frames per second. Light recorded by the high-speed camera was from the thermal incandescence of the hot material. The field of view was approximated at $7\times 7~\text{mm}^2$ with $512\times 512$ pixels at a resolution of nearly 14~$\mu\mathrm{m}$ per pixel. The high-speed camera remained fixed with a slight shooting angle.

Given the ratio between sampling frequencies of three types of sensors (acoustic, photodiode, and high-speed), we can roughly get: $\frac{f_A}{f_I} = \frac{f_P}{f_I} \approx 4.545$, which means 100 acoustic or photodiode signal sample points can be synchronized and registered to 22 melt pool image frames. We explain the details of data synchronization in Sec. \ref{subsubsec:data_sync}.

\begin{figure}[!htb]
\center{\includegraphics[width=\linewidth]
{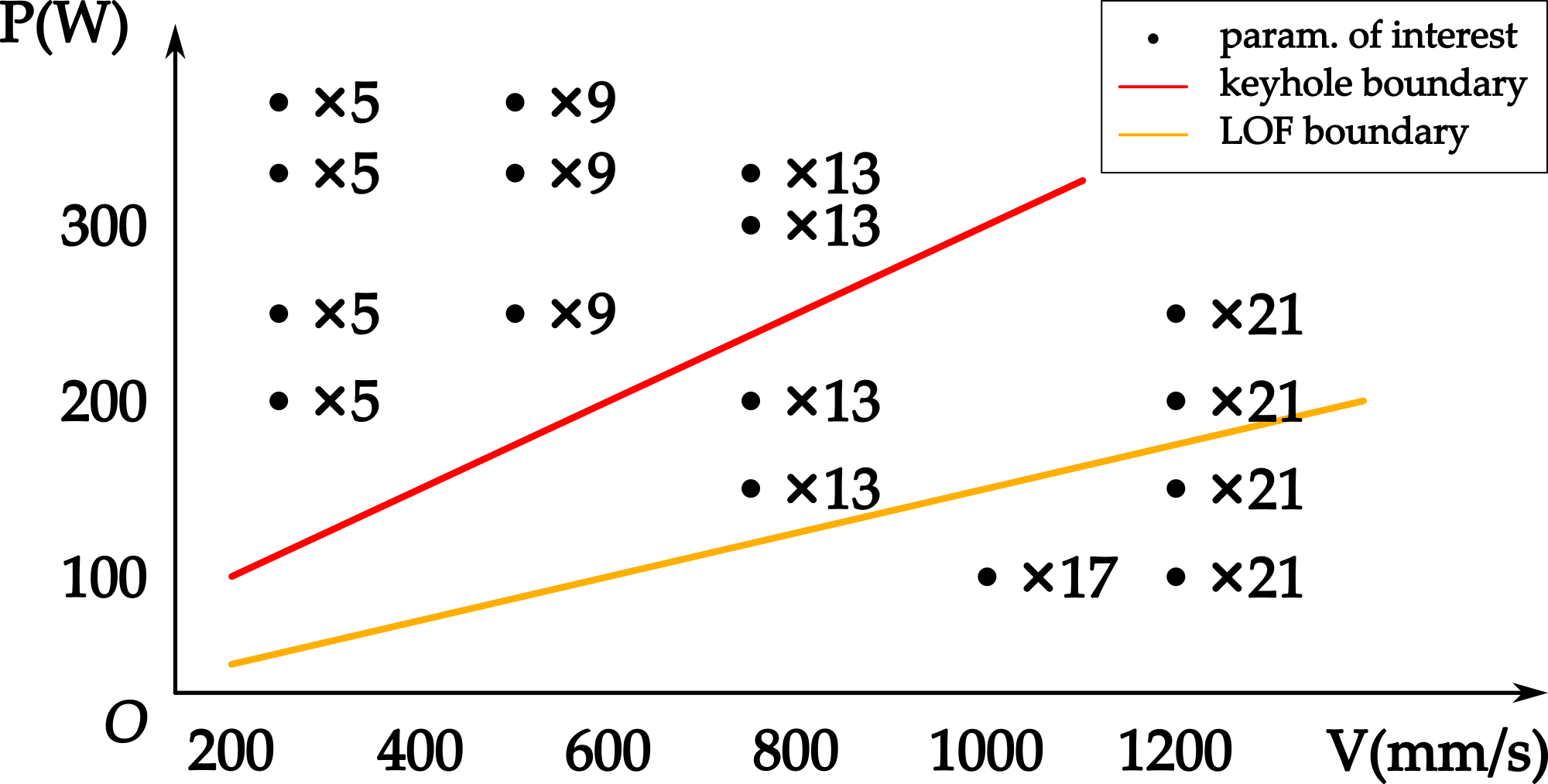}
\caption{Ti-64 LPBF $P$-$V$ process map~\cite{GORDON2020101552} and process parameter selection of the build. Each selected $(P, V)$ couple has a certain number of repetitions. }
\label{fig:process_design}
}
\end{figure}

Our process parameter selection is shown in Fig. \ref{fig:process_design}. We chose a set of process parameter couples, laser power $(P)$ and laser scanning velocity $(V)$, that spanned across the $P$-$V$ process map of Ti-64 and covered all sub-regions of flaw formation mechanisms -- KH, LOF, and BU~\cite{shevchik2018acoustic}. A detailed process parameter couple selection map is shown in  Fig.~\ref{fig:process_design}. To increase the variance of melt pool morphology, we intentionally varied the fusion energy densities by controlling the $\frac{P}{V}$ ratio. Since our designed square prism can be partitioned into roughly 800 layers along with the build direction, we shuffled and assigned our selected $P$-$V$ couples to different layers of the build. As shown in Fig.~\ref{fig:process_design}, each $P$-$V$ couple had been prescribed a certain number of repetitions -- in general, higher $V$ tends to have more repetitions -- so that we can get a roughly similar amount of monitoring data for each printing condition. As shown in Fig.~\ref{fig:printed_sample}, we also explored different scanning orientations ($\theta$), namely $\theta=0^{\circ}$, $\theta=45^{\circ}$, $\theta=90^{\circ}$ and $\theta=135^{\circ}$, for each P-V couple, such that various heat accumulation situations can be included when the laser scans different parts of the square region. Each layer was rasterized following the aforementioned $\left(P, V\right)$ and specified $\theta$ without contour and skins.

\subsection{Data synchronization and processing}
\label{subsec:data_process}
In this section, we separately introduce our data synchronization workflow, high-speed image segmentation pipeline, and acoustic/thermal data spectral analysis. 

\subsubsection{Data synchronization}
\label{subsubsec:data_sync}
\begin{figure*}[!htb]
\center{\includegraphics[width=\linewidth]
{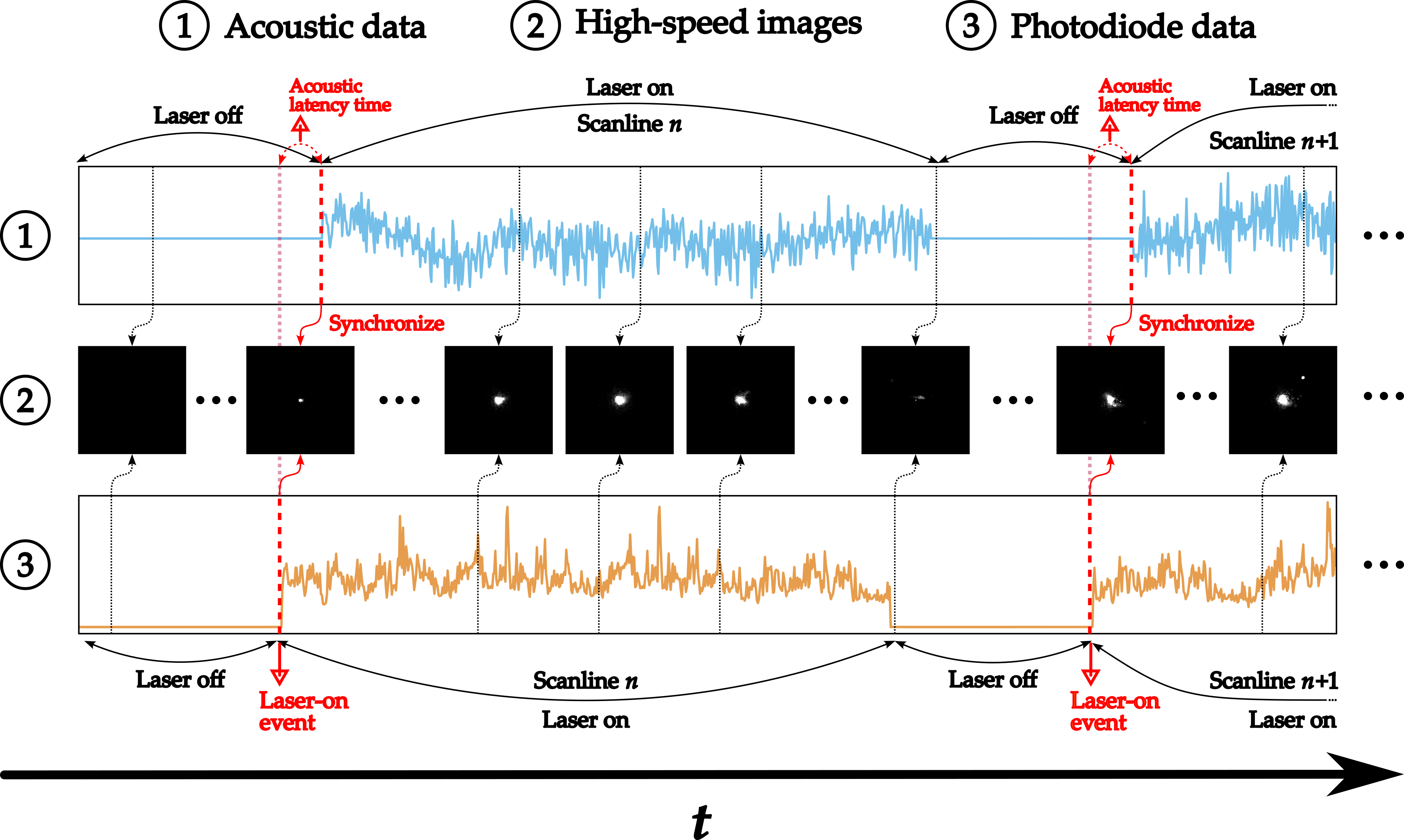}
\caption{Data synchronization. The laser-on events were utilized as the synchronization anchor points for three separate data channels. Acoustic latency time was characterized manually according to the distance between the acoustic receiver and the build spot. }
\label{fig:data_sync}
}
\end{figure*}

Because three sensors were operated independently and there was limited communication between the sensors and the control computer, data synchronization was considered a crucial post-processing step. Figure~\ref{fig:data_sync} depicts our data synchronization workflow. First, we located the layer numbers for the three data types and extracted them to obtain three sections. Next, we aligned the beginnings of these sections to quantify the latency of the acoustic signal, a short time delay caused by the difference between the speed of sound in the Argon flow and the speed of light for thermal emissions. In this work, the acoustic signal latency was measured to be 0.8~$\mathrm{ms}$. Since the distance between the acoustic sensor and the build plate was fixed throughout the printing, we applied this same latency to the data synchronization of all other layers. 

After determining the latency time of the acoustic signal, we paid attention to the occurrence of laser-on events. As shown in Fig.~\ref{fig:data_sync}, the laser-on events were the moments when a layer or a scanline started to print and the melt pool immediately appeared. We treated these laser-on events as valid data synchronizing anchor points and then registered the three data streams to them. This laser-on-event-based synchronization was used to partition data sections into smaller data groups consisting of only one or a few consecutive scanlines, achieving accurate scanline-to-scanline registration. By re-synchronizing the data streams at each laser-on-event, we guaranteed that the effect of desynchronization over time was minimal, which also facilitated the data labeling process. 

\subsubsection{Continuous wavelet transform of time-series data}
\label{subsubsec:wavelet}
We applied spectral analysis methods suitable for 1D time-series data to the acoustic and photodiode signals. Extracting and tracking spectral features for the duration of each signal, these methods yield 2D matrix representations of the data known as scalograms. For our acoustic and thermal data, we chose continuous wavelet transformation (CWT) from among a large variety of spectral analysis methods since CWT is advantageous in both temporal and spectral feature extraction~\cite{arts2022fast}. In our acoustic and photodiode data processing pipeline, we used the Morlet function as the mother wavelet and a set of discrete scales from 0.9~$\mathrm{kHz}$ to 36.8~$\mathrm{kHz}$ to cover the acoustic sensor's frequency range. Wavelet transformation data processing was implemented in Matlab using the Wavelet Toolbox. 

\begin{figure}[!htb]
\center{\includegraphics[width=\linewidth]
{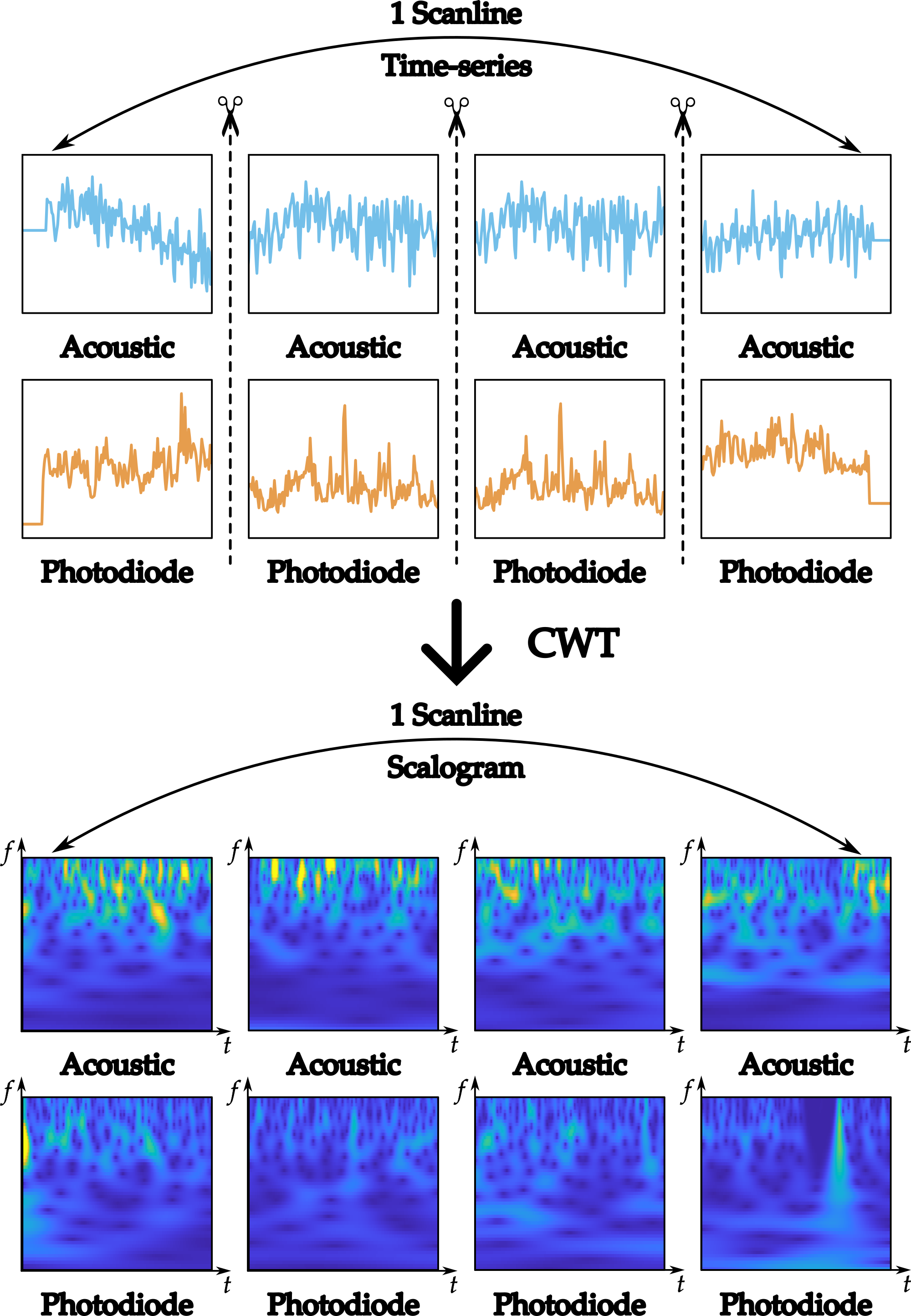}
\caption{CWT scalogram transformation of 4~ms acoustic and photodiode data clips. }
\label{fig:wavelet}
}
\end{figure}

In order to achieve nearly real-time process monitoring with acoustic and photodiode data, we performed feature extraction on short, informative segments of the time-series data, which were obtained by partitioning the original raw time-series data using a sliding time window with a window length ($t_w$). As shown in Fig.~\ref{fig:wavelet}, we partitioned the acoustic and photodiode data into short clips. Next, we applied the CWT method to the segmented acoustic and thermal data clips using the aforementioned mother wavelet function and corresponding scaling factors. Scalograms of short acoustic and photodiode data clips were obtained in matrix forms, with the horizontal ($x$-) axis being the time axis $T$ and the vertical ($y$-) axis being the frequency axis $F$. Intensities of scalogram pixels represent the signal energy density at that specific $(t, f)$ coordinate. In other words, scalograms of acoustic/photodiode data can reveal how the signal energy was distributed across different frequency bands and how this energy distribution evolved with time. Therefore, we used the scalogram matrices as the representation of the acoustic and photodiode data signature for our ML model training. 

\begin{figure}[!htb]
\center{\includegraphics[width=\linewidth]
{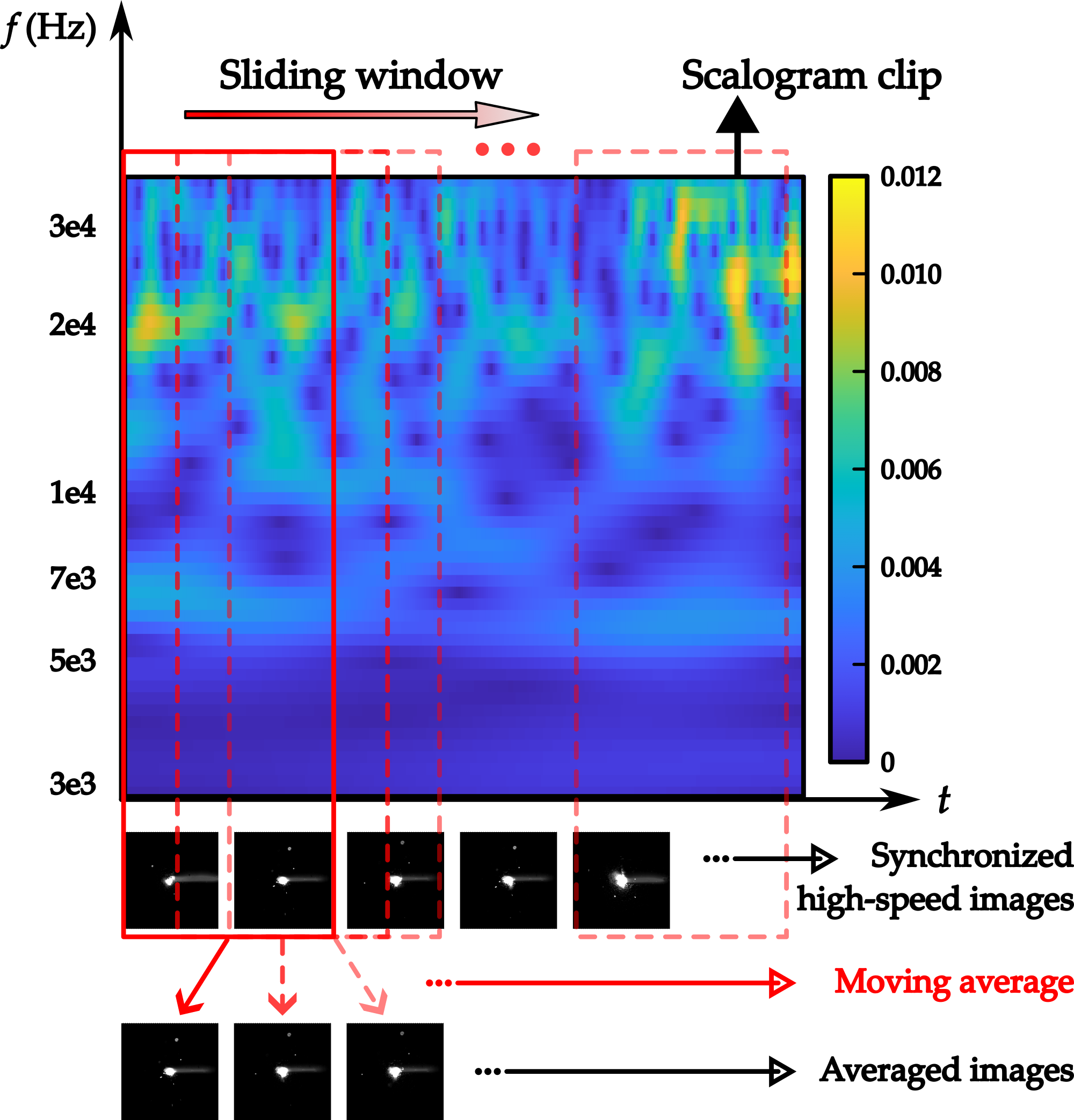}
\caption{Implementation of moving average data labeling over an example of an acoustic scalogram clip. The frequency axis of the scalogram is plotted on a logarithmic scale. Observations of the correlation between frequency bands on the corresponding acoustic wavelet scalogram and melt pool features are detailed in Sec.~\ref{sec:discussion}. }
\label{fig:moving_avg}
}
\end{figure}

For each pair of the acoustic and thermal scalograms, we matched the consecutive sequence of high-speed images to its corresponding set of extracted feature vectors along with the $x$-axis of the scalogram. Since the melt pool morphology can have a large variance within the time duration of a scalogram, we applied the moving average technique with a certain window size and sliding stride to generate averaged versions of a few high-speed image subsets at different temporal coordinates; each subset yields only one averaged image, and there was no overlap between adjacent subsets. As depicted in Fig.~\ref{fig:moving_avg}, this implementation significantly increases the approach's tolerance of melt pool geometric variance and reduces the dimension of visual features. In our training pipeline, we set both the window size and the sliding stride of the moving average to 11 image frames, and we labeled each pair of acoustic and photodiode scalograms with the averaged image or visual features of the middle consecutive 11 high-speed melt pool images. 

\subsubsection{High-speed image segmentation and feature extraction}
\label{subsubsec:segmentation}
\begin{figure}[!htb]
\center{\includegraphics[width=\linewidth]
{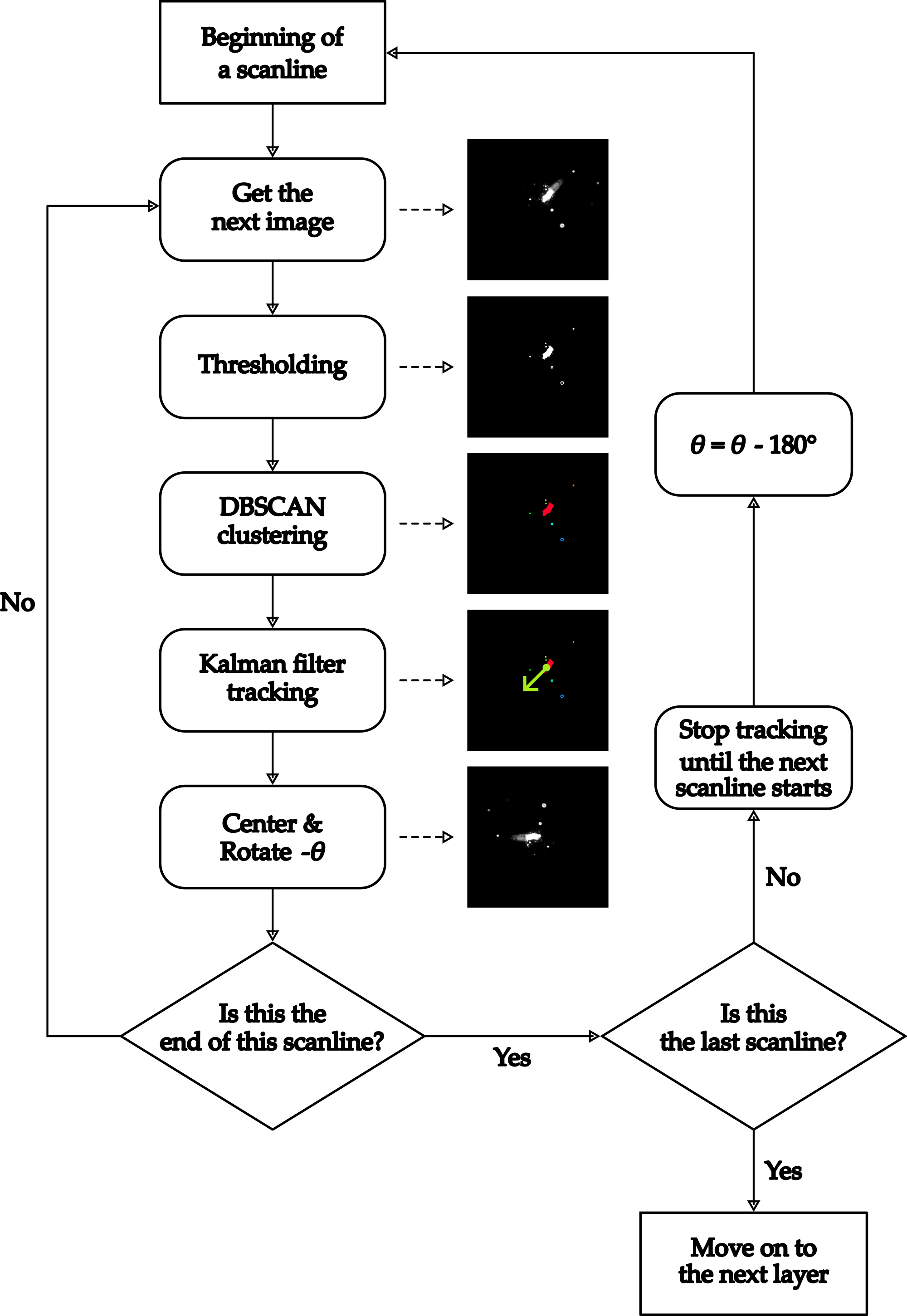}
\caption{A flowchart of iterative high-speed melt pool image data processing within a layer. Each melt pool image was first clustered via DBSCAN, then tracked by Kalman filtering, and finally straightened. The algorithm was set to make sure all the scanlines were processed consecutively until the next layer of the high-speed image dataset came in. }
\label{fig:straighten}
}
\end{figure}

We obtained melt pool images via the high-speed camera while the laser rasterized as prescribed within the HSI viewing window during the printing process. To keep only the melt pool morphological characteristics, we straightened our high-speed images, coinciding the image center with the melt pool center and orienting the scanning direction of the melt pool to the right, and eliminated any scanning orientation- and position-related image variations. Figure~\ref{fig:straighten} shows a flowchart of how we extracted, tracked, and straightened the melt pool in our high-speed images. Moreover, for the convenience of the data synchronization in the next step, we also grouped the high-speed images and processed them scanline by scanline. 

To briefly explain Fig.~\ref{fig:straighten}, we first applied a constant intensity threshold to the grayscale high-speed image to extract the bright pixels containing the melt pool, affiliated melting track, plumes, and spatters. The threshold was set to $0.8$ so that all the pixels with intensity values in the $[0, 0.8]$ range were masked out. After thresholding, we performed image segmentation by implementing the density-based spatial clustering of applications with noise (DBSCAN) method~\cite{schubert2017dbscan} to further group the bright pixels into multiple clusters. The DBSCAN algorithm is detailed in~\ref{appendix:dbscan}. DBSCAN assigned the melt pool, spatters, plume, and melting track separate cluster labels; thus, we can directly pick the melt pool pixels by selecting the correct blob-like cluster out of each image. To collect deterministic melt pool features, we specifically focused on properties like cluster size, length, and width, each of which was quantified in units of number of pixels. 

Next, we tracked the melt pool movement ({\it i.e.}, location and scanning orientation information) within each scanline. Since the scanning of each scanline or layer commonly started with a clean melt pool (very few spatters and plumes), we can easily locate the first melt pool appearance by selecting the largest pixel cluster from the DBSCAN results (shown as Fig.~\ref{fig:straighten}). Then, we located the melt pool's center point ($c_M(x,y)$) at the spatial centroid of pixels of the melt pool cluster. For the rest of the high-speed images, it was hypothesized that we could track the melt pool motion by employing a linear model with no acceleration (rare melt pool acceleration at the beginning and the end of each scanline was ignored). The model explicitly knew the melt pool's scanning orientation $\theta$ and scanning velocity $V$, which were prescribed by the printer within each scanline before the experiments. Based on the above hypotheses, we modeled the melt pool motion using the Kalman filter algorithm~\cite{maybeck1990kalman}. We accurately tracked the melt pool and updated the estimate of $c_M$ and an uncertainty covariance matrix $P$ by taking measurements throughout a sequence of consecutive high-speed images. 

Assuming the observation of $c_M$ through our high-speed camera followed a Gaussian distribution, the Kalman filter model of our melt pool motion can be formulated as: 

\begin{equation}
\label{eqn:kalman}
\begin{aligned}
\begin{cases}
    &\hat{c_M}\left(x_k,y_k\right) = F_k c_M\left(x_{k-1},y_{k-1}\right) \\
    &P_k \left(\hat{c_M}\right) = F_kP_{k-1}(c_M)F_k^T+J \\
    &\hat{c_M}^{\prime}\left(x_k,y_k\right) = \hat{c_M}\left(x_k,y_k\right)+K\left(\overrightarrow{z_k}-H\hat{c_M}\left(x_k,y_k\right)\right) \\
    &P_k^{\prime}\left(\hat{c_M}\right) = P_k \left(\hat{c_M}\right) - KH P_k \left(\hat{c_M}\right)
\end{cases}
\end{aligned}
\end{equation}

\noindent where $\hat{c_M}$ and $P_k$ denote the simulated melt pool center position and its covariance at time step $k$, and $\hat{c_M}^{\prime}$ and $P_k^{\prime}$ denote the updated melt pool center position estimate and the updated covariance given the Kalman gain $K$, which was inferred by the real measurement $\overrightarrow{z_k}$ and the sensor model $H$ at time step $k$. $K$ and the rest of the model parameters were defined and calculated as follows (see~\ref{appendix:kalman} for more details of our Kalman filter implementation):

\begin{align}
\label{eqn:kalman_2}
    & K = P_K\left(\hat{c_M}\right)H^{\top}\left(H P_k H^{\top}+R\right)^{-1} && F = \begin{bmatrix}
          1 & 0 & \frac{1}{f_I} & 0 \\
          0 & 1 & 0 & \frac{1}{f_I} \\
          0 & 0 & 1 & 0 \\
          0 & 0 & 0 & 1 
          \end{bmatrix} 
\end{align}

\begin{align}
\label{eqn:coeff}
    & P_0 = 0.01\cdot I_{4\times 4} && J = 0.01\cdot I_{4\times 4} && H = I_{4\times 4} && R = 0.01\cdot I_{4\times 4}
\end{align}

\noindent Moreover, it's worth noting that $P$ was associated with $c_M$ since the shape and size of the melt pool kept oscillating with varying heat accumulations when it moved along with scanlines. 

After obtaining $\hat{c_M}$ of the melt pool using Eqn.~\ref{eqn:kalman} and Eqn.~\ref{eqn:kalman_2} given the initial condition and model parameters from Eqn.~\ref{eqn:coeff}, we straightened the high-speed image by centering the image at $\hat{c_M}$ and rotating the image by $-\theta$. We call the final processed high-speed image the ``straightened" image, from which we can extract either deterministic features including melt pool size (i.e., the total number of pixels that belong to the melt pool cluster), length (i.e., the maximum number of melt pool pixels along with the horizontal axis), and width (i.e., the maximum number of melt pool pixels along with the vertical axis), or data-driven features such as the latent embedding from a trained autoencoder. It is worth noting that the entire workflow demonstrated in Fig.~\ref{fig:straighten} was not time-consuming when processing only one high-speed image at a time, which enabled real-time process monitoring on an experimental setup. We used the straightened high-speed images for the feature extraction and ML tasks in the next step.

\subsection{Deep learning pipeline}
\label{subsec:ML_model}
\begin{figure*}[!htb]
\center{\includegraphics[width=\linewidth]
{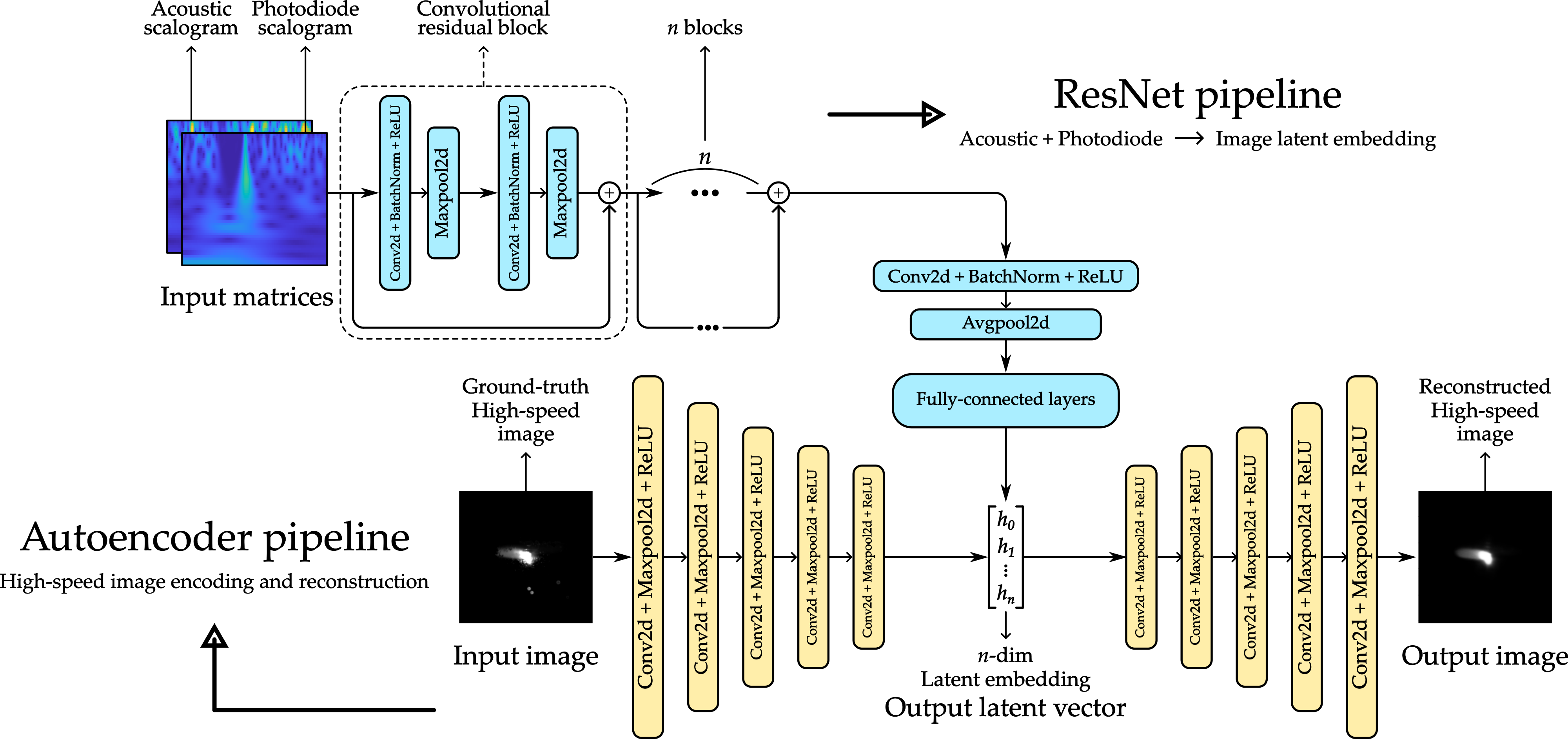} 
\caption{Machine learning model architecture, consisting of an autoencoder and a deep CNN model (ResNet). }
\label{fig:ML_archi}
}
\end{figure*}

We trained a deep learning pipeline to predict and reconstruct high-speed images given the input of the acoustic and thermal scalograms. Through a parametric study on machine learning models, we found that the best deep learning architecture for this task was a combination of ResNet34 and an autoencoder, which converted high-speed melt pool images to latent embeddings. 

Figure~\ref{fig:ML_archi} depicts the overall architecture of our proposed deep learning pipeline. We first trained the autoencoder part of the architecture, for which straightened high-speed images served as both the input and the output for model training. We exploited our autoencoder training process to construct a latent representation of each high-speed image. In this way, the trained autoencoder can be regarded as a data-driven image feature extractor, and the latent space constructed by it mainly reflected the morphological features of the melt pool. 

After the autoencoder was trained, we trained a deep learning model that takes in the acoustic and thermal scalograms as the input and outputs the pre-extracted latent vectors of high-speed images. The dimension of the final layer of our deep-learning model depends on the latent dimension prescribed by the autoencoder. Each input scalogram was given a latent vector representation as a label, as described in Sec.~\ref{subsubsec:wavelet}. For each acoustic and thermal scalogram clip input, we ensured that there would be only a single latent vector that corresponded to it. Using the moving average technique, we can assign each pair of input scalograms an averaged latent vector that was derived from the middle scalogram slice. 
\section{Results of melt pool visual detection}
\label{sec:results}
In this section, we demonstrate the critical results of our work. In essence, our entire data-driven pipeline takes in the processed acoustic and thermal scalogram data and eventually reconstructs melt pool images that contain critical melt pool visual features. After pushing the collected raw data through the data processing workflow described in Sec.~\ref{subsec:data_process}, we obtain a dataset with a size of 20,000 that consists of pairs of input scalograms and their corresponding averaged latent vectors. 

Our ML pipeline was implemented and executed with PyTorch 1.13.1 and was trained using a consumer-grade Intel(R) Core(TM) i7-10700K CPU @ 3.80 GHz and an NVIDIA GeForce RTX 3070 with 8.0 GB VRAM and CUDA 12.0. The learning rate for the training of both the autoencoder and the deep learning model was $10^{-4}$. The batch sizes for the autoencoder training and the deep learning model training were 128 and 64, respectively. All the models discussed in this paper were trained until the loss function values converged to a stable level. 

\begin{figure*}[!htb]
\center{\includegraphics[width=\linewidth]
{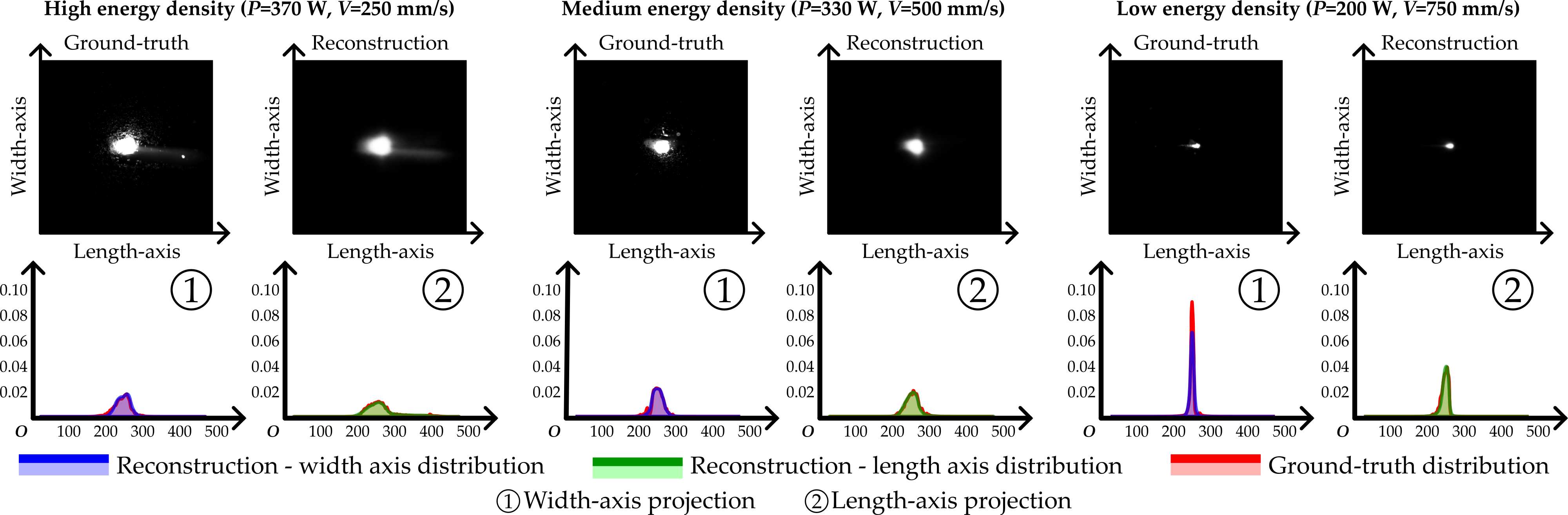}
\caption{Autoencoder reconstruction result of three different levels of fusion energy density ({\it i.e.}, high, medium, and low). \ding{172} and \ding{173} label subfigures of width-axis and length-axis image intensity distributions, respectively. The results show that the reconstructed high-speed images match well with their corresponding ground-truth images. }
\label{fig:ae_reconstruction}
}
\end{figure*}

\subsection{Image similarity evaluation metric}
\label{subsec:evaluation_metric}
In the AM community, researchers and process engineers care about many melt pool visual characteristics, such as melt pool area $A$, melt pool length $L$, and melt pool width $W$. In this work, we propose a new metric $Q(u,v)$ that serves in particular as a similarity metric for a pair of melt-pool images $u$ and $v$ and covers the above critical visual features of the melt-pool as well as thermal-related information. We formulate $Q(u,v)$ as follows: 

\begin{equation}
\label{eqn:q}
\begin{aligned}[b]
    \resizebox{1.0\linewidth}{!}{$Q(u,v) = \frac{1}{4}\times \left(\sqrt{\|w_0\left(u,v\right)\|_2^2 + \|w_1\left(u,v\right)\|_2^2} + \sqrt{\frac{|A_u-A_v|}{\pi}} + |L_u-L_v| + |W_u-W_v| \right)$}
\end{aligned}
\end{equation}

\noindent where $w_j\left(u,v\right) = \inf_{\gamma\in\Gamma\left(u,v\right)}\mathbf{E}_{(x,y)\sim\gamma}d\left(x,y\right) \left(j\in\{0,1\}\right)$ denotes the sliced Wasserstein 1-distance (SWD), {\it aka} earth-mover distance~\cite{kolouri2018sliced}, which can assess how similar the two intensity probability distributions are. The term $\sqrt{\frac{|A_u-A_v|}{\pi}}$ is inspired by the analytic formulation of the ellipse area. The efficacy of our proposed $Q(u,v)$ is demonstrated in~\ref{appendix:wasserstein}. In other words, Eqn.~\ref{eqn:q} not only evaluates traditional deterministic melt pool geometric features but also takes into account the spatial heat distribution information in a statistical fashion. 

By including thermal information when comparing two melt pool images, we can evaluate the similarities in a way that correlates with more melt pool-related physical quantities. In general, the more similar the two melt pool images $u$ and $v$ are, the lower $Q(u,v)$ would be. Moreover, it can be proven that $Q(u,v)$ is a valid metric in image space, indicating that our proposed formulation is mathematically unambiguous and capable of evaluating the similarities of more than two image objects without any inter-contradictory issues. The proof of Eqn.~\ref{eqn:q} being a valid metric can be found in~\ref{appendix:metric}. 

In this paper, we not only evaluate melt pool visual characteristics prediction performance using geometric features but also utilize $Q(u,v)$ (hereafter referred to as ``$Q$ score") to assess the similarity between a ground-truth image and a reconstructed image, especially when the prediction accuracy of thermal accumulation and melting track needs to be assessed.

\subsection{Image reconstruction and latent space distribution}
\label{subsec:ae_results}
We specify the latent dimension as 4 for our high-speed image embeddings through a parametric study to ensure the best learning performance. Three of the autoencoder-generated reconstruction results -- namely the examples that correspond to high, medium, and low energy densities, respectively -- are shown in Fig.~\ref{fig:ae_reconstruction}. By comparing the ground-truth image with the corresponding reconstructed image, we show that the reconstruction accuracy of our autoencoder is high enough to capture critical melt pool and thermal-related visual features across different printing conditions. This gives us confidence that the embedding generated by the encoder is physically sensical and can also distinguish between high-speed melt pool images in the constructed latent space. 

\begin{figure}[!htb]
\center{\includegraphics[width=\linewidth]
{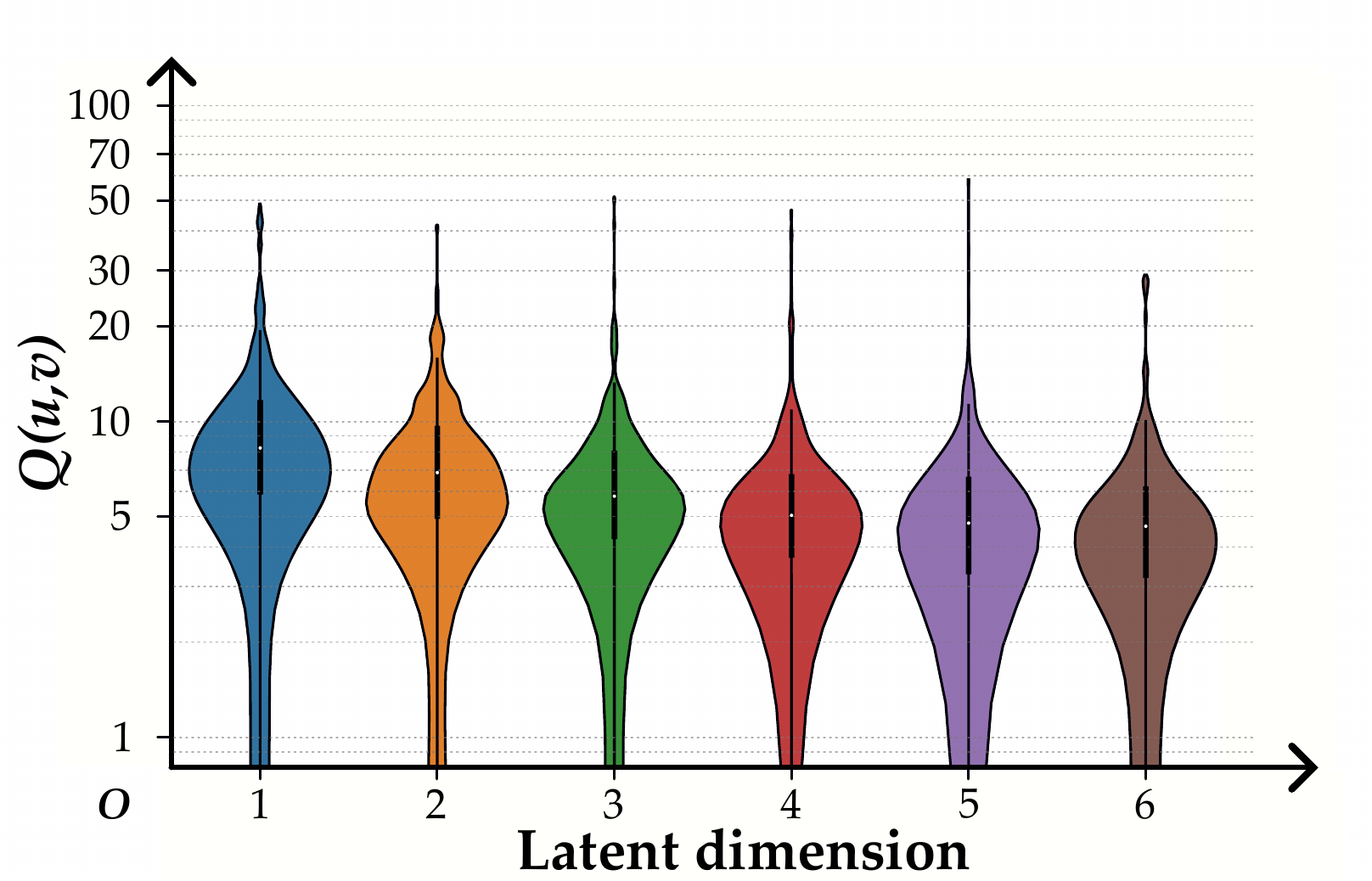} 
\caption{$Q$ score violin plot with different latent dimensions. }
\label{fig:ae_parametric}
}
\end{figure}

Figure~\ref{fig:ae_parametric} demonstrates the results of the parametric study we perform on the latent dimensions of the autoencoder. It is worth noting that there is a trade-off between the reconstruction accuracy and the deep-learning model prediction accuracy. That is, a smaller latent dimension might lead to a lower autoencoder reconstruction accuracy but result in higher deep-learning model prediction accuracy, whereas a larger latent dimension could yield better reconstruction quality while the prediction accuracy of the deep-learning model may converge to a plateau or even start to reduce. As such, as we increase the latent dimension in the parametric study, we can specify the ``optimal" latent dimension by finding a latent vector length at which the reconstruction error starts to converge ({\it i.e.}, $Q(u,v)\leq 5.0$). To this end, we eventually pick the latent dimension of 4 as the optimum according to the parametric study results from Fig.~\ref{fig:ae_parametric}. We also report the autoencoder construction results in~\ref{appendix:additional_latent}.

\subsection{Deep learning prediction of melt pool visual features}

\begin{figure*}[!htb]
\center{\includegraphics[width=0.95\linewidth]
{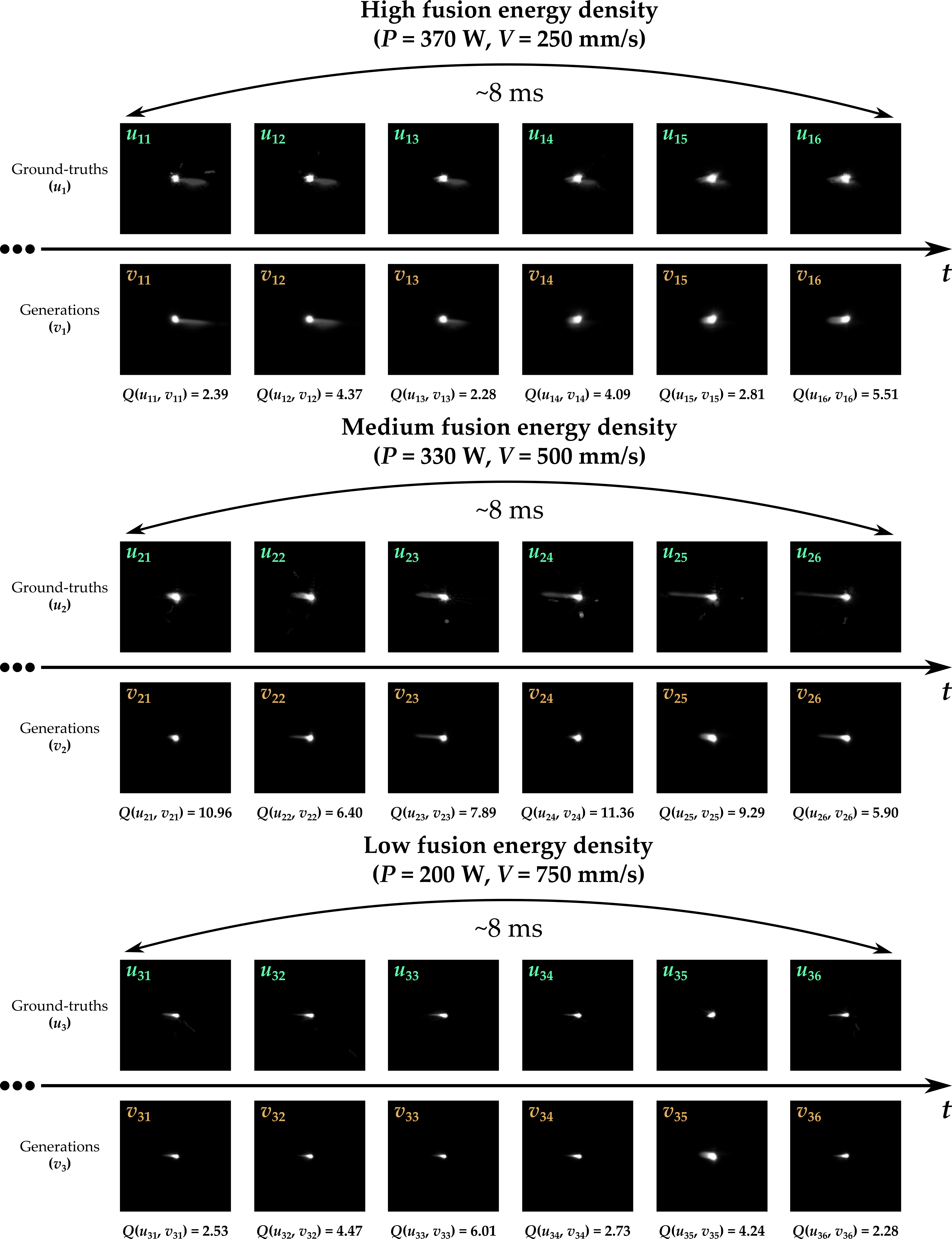}
\caption{Melt pool image prediction results via the proposed deep CNN model in an 8~$\mathrm{ms}$ consecutive frame sequence. Examples of high, medium and low fusion energy densities are showcased separately. }
\label{fig:dl_result}
}
\end{figure*}

\begin{figure*}[!htb]
\center{\includegraphics[height=0.93\textheight]
{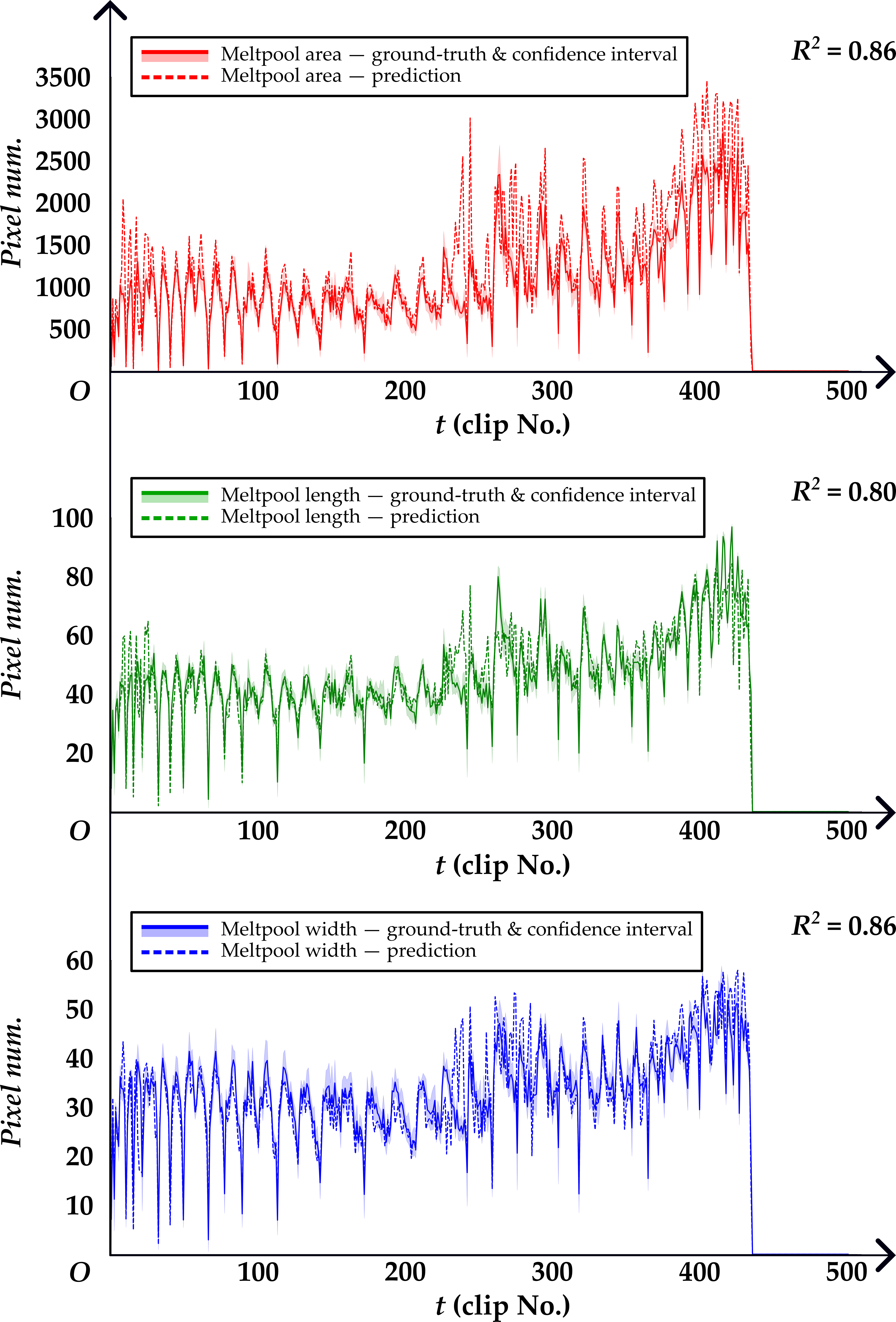}
\caption{Plots of consecutive melt pool length, width, and area predictions within a test printing layer ($P=370~\texttt{W}$, $V=250~\texttt{mm/s}$). The coefficient of determination ($R^2$ score) is also reported for each quantity. Results show that critical melt pool features can be inferred merely based on the acoustic and photodiode data information by our proposed well-trained pipeline. }
\label{fig:dl_r2}
}
\end{figure*}

\label{subsec:dnn_results}
After we construct the latent space via the autoencoder training, we train our deep-learning model to predict the high-speed melt pool image latent vector given the corresponding acoustic and thermal scalogram. Similar to the moving-average technique explained in Sec.~\ref{subsubsec:wavelet}, we generate the latent output vector for the corresponding acoustic and thermal scalogram pair by averaging vectors of the middle 11 high-speed melt pool images. We employ ResNet34 as the deep CNN model to achieve the aforementioned mapping. Figure~\ref{fig:dl_result} showcases the comparison between a consecutive sequence of ground-truth averaged high-speed images and the corresponding acoustic/photodiode data-based predicted images from the proposed method. It demonstrates that given the short clips of acoustic and thermal scalogram data, our trained deep-learning model can generate smoothed high-speed melt pool images that contain critical melt pool visual features such as melt pool geometric features and thermal-related distribution patterns. The $Q$ score for each pair of ground-truth-to-prediction comparisons shows that our reconstruction can yield high accuracy when informing the critical transient fusion status. 

Figure~\ref{fig:dl_r2} reports pipeline prediction performance utilizing consecutive plots of melt pool geometrical features, {\it i.e.}, melt pool length, width, and area, for an entire layer of the build. The result shows that our proposed method can predict these critical melt pool visual features accurately with high correlation scores. Also, the result demonstrates the capability of our proposed approach on melt pool visual feature tracking. 

\begin{figure}[!htb]
\center{\includegraphics[width=\linewidth]
{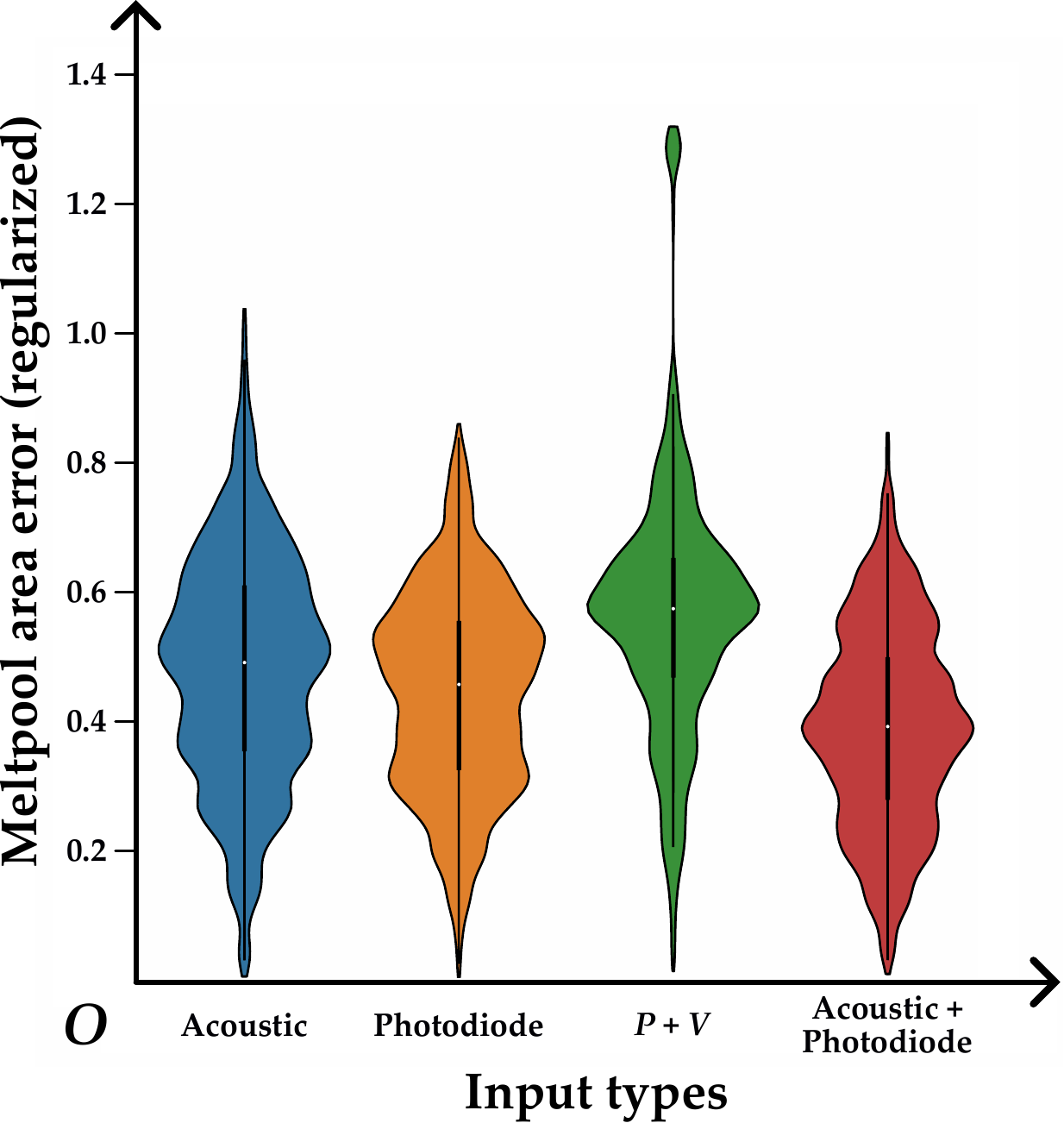}
\caption{Violin plots of regularized melt pool area error with four different types of input. We can conclude that both acoustic and photodiode data can contribute to melt pool feature identification. The input combination of acoustic and photodiode data can achieve the best performance among the four. }
\label{fig:dl_param_input}
}
\end{figure}

We are also interested in how the prediction performance changes with variations in input formats and printing conditions. Figure~\ref{fig:dl_param_input} shows a set of violin plots on the regularized error of melt-pool area with different combinations of input data. From the results, we can see that except for merely feeding in $P$ and $V$ information, each type of input data can generate good prediction accuracy, demonstrating the correlation between the melt pool morphology and the affiliated acoustic and photodiode data. Among all input types, the combination of acoustic and photodiode scalograms gives the best prediction performance and therefore stands out as the best candidate to train our proposed pipeline. 

\begin{figure}[!htb]
\center{\includegraphics[width=\linewidth]
{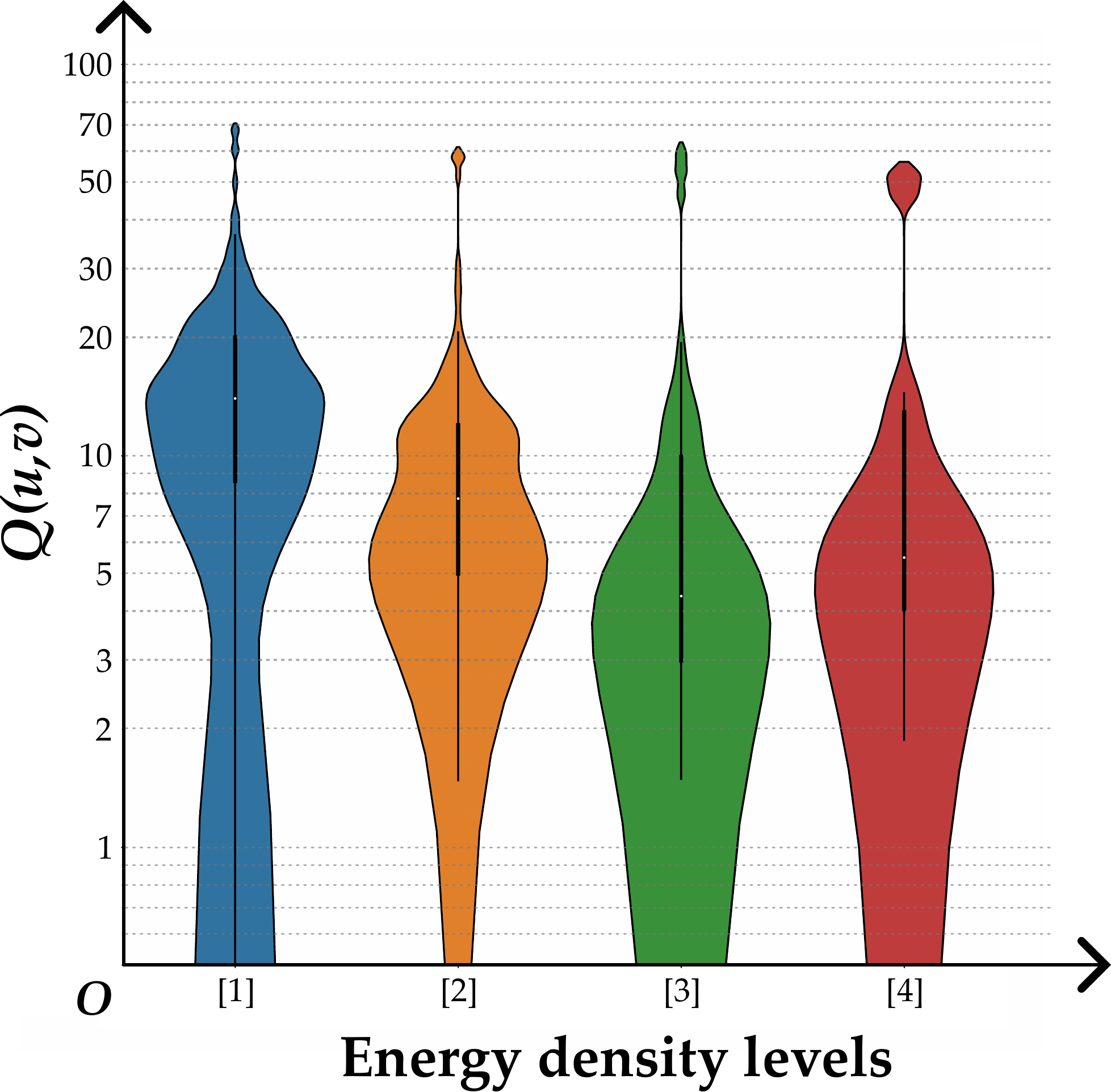}
\caption{$Q$ score violin plots with four different fusion energy densities (printing conditions). Case [1]: $P=370~W$, $V=250~\mathrm{mm/s}$. Case [2]: $P=330~W$, $V=500~\mathrm{mm/s}$. Case [3]: $P=200~W$, $V=750~\mathrm{mm/s}$. Case [4]: $P=150~W$, $V=1200~\mathrm{mm/s}$. }
\label{fig:dl_param_PV}
}
\end{figure}

Figure~\ref{fig:dl_param_PV} shows the $Q$ score violin plots of our proposed method over four different $P/V$ ratios. We can observe from the plot that our proposed pipeline generally has a good prediction quality across different printing conditions; high energy density cases such as $[1]$ in Fig.~\ref{fig:dl_param_PV} might lead to lower prediction accuracy in terms of $Q$ score.

\subsection{Parametric studies on $t_w$ and various ML models}
\label{subsec:time_window_ML_models}

\begin{figure}[!htb]
\center{\includegraphics[width=\linewidth]
{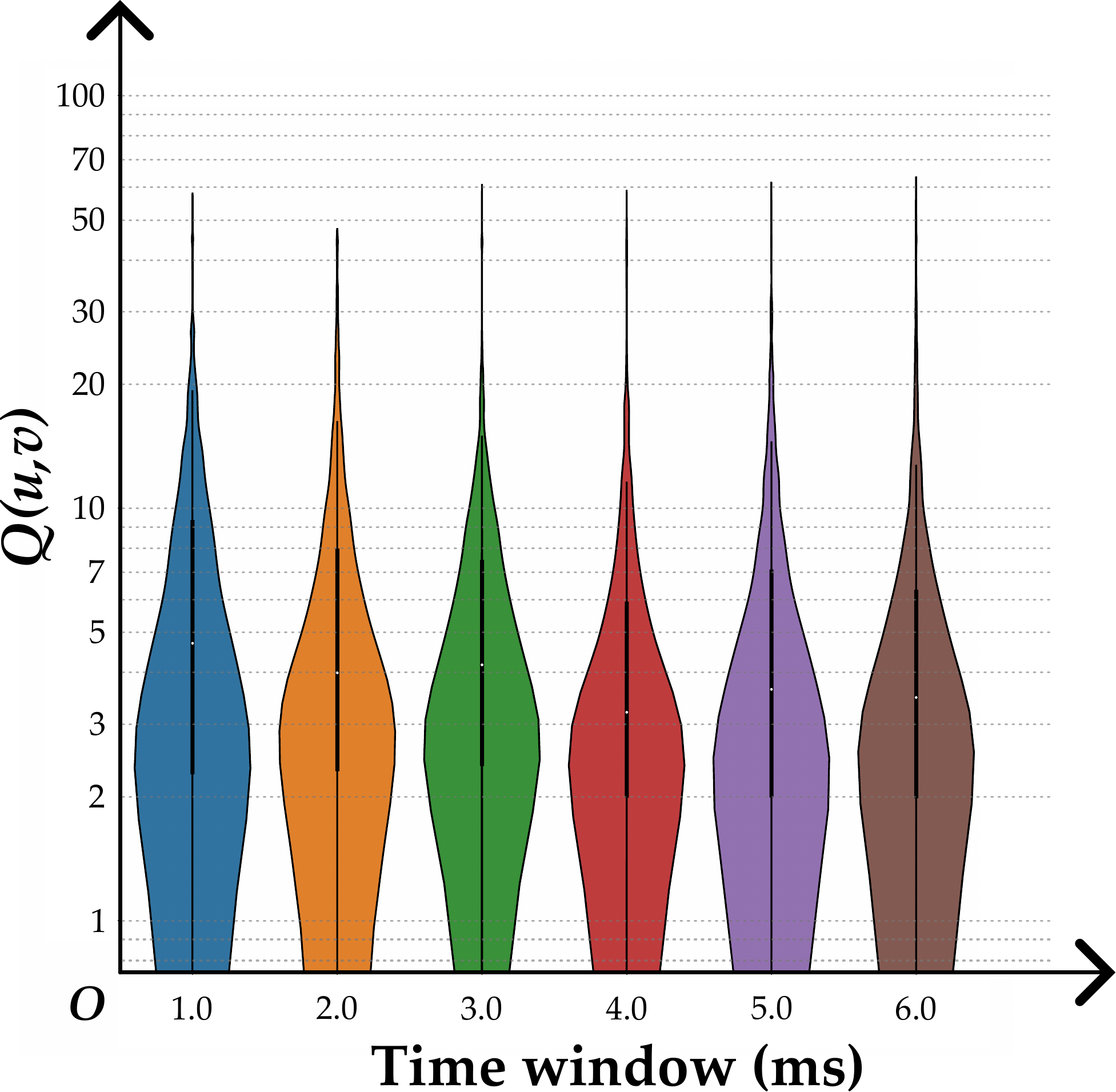}
\caption{$Q$ score violin plots with different time window lengths. }
\label{fig:dl_param_window}
}
\end{figure}

We explore the prediction accuracy sensitivity over different time window sizes ({\it i.e.}, $t_w$) of the acoustic and thermal scalograms. As shown in Fig.~\ref{fig:dl_param_window}, we study six different time window lengths and report the respective $Q$ score distributions. 

From the results, we can see that there is not much variance in prediction performance when changing the time window length. We conclude that we can achieve accurate melt pool identification within a time period as short as 1.0~$\mathrm{ms}$. 

\begin{figure}[!htb]
\center{\includegraphics[width=\linewidth]
{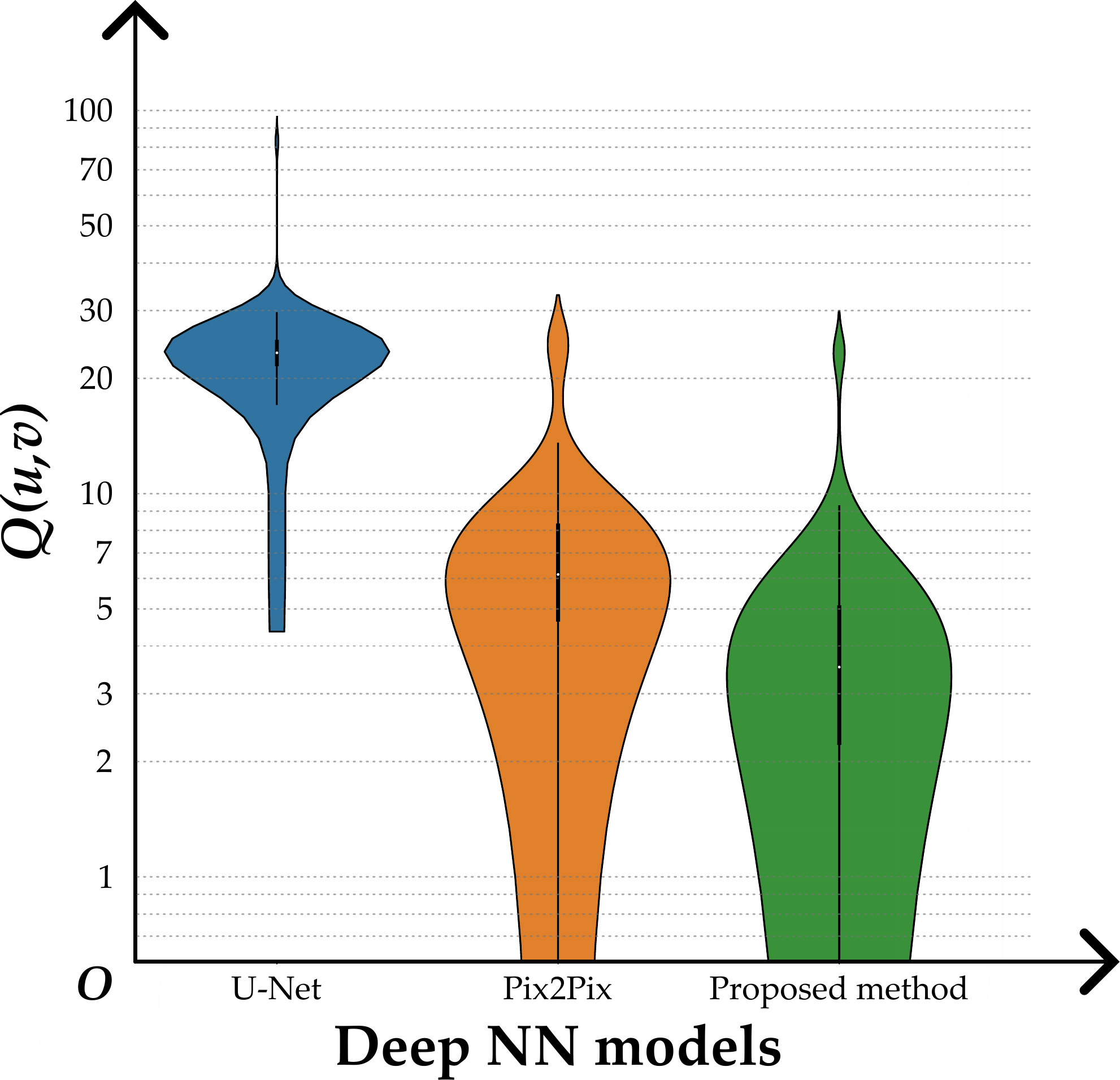}
\caption{$Q$ score violin plots with different learning models. }
\label{fig:dl_param_models}
}
\end{figure}

We compare our proposed deep-learning pipeline with some other baseline learning models to further exploit ML capability on our dataset. For the baseline models, we select U-Net~\cite{ronneberger2015u} and Pix2Pix~\cite{isola2017image} as two alternative deep-learning models that directly perform image-to-image translation from acoustic and thermal scalograms to melt pool high-speed images. Details of the two baseline models can be found in~\ref{appendix:baseline}. The comparison of the above two models and the proposed approach is demonstrated in violin plots shown in Fig.~\ref{fig:dl_param_models}. 

Figure~\ref{fig:dl_param_models} shows that U-Net cannot achieve accurate enough predictions, potentially due to end-to-end skip connections in its architecture, which introduce superimposition of the input acoustic and photodiode data on top of the melt pool output image and lead to noisy image reconstruction. The prediction of the Pix2Pix model can be inaccurate in terms of the spatial thermal distribution of the melt pool, which will increase the Wasserstein distance in our $Q$ formulation and lead to a lower $Q$ score. Therefore, we can see that our proposed pipeline has the best prediction accuracy and can capture the most critical features of the melt pool. 
\section{Prediction of spatially-dependent LOF defect formation}
\label{sec:LOF_valid}
In this section, we set up a validation case for LOF defect detection between two adjacent scanlines based on the criteria determined by the process parameters and the morphological characteristics of the melt pool. We report the comparison between the LOF occurrence from direct observations, simulated LOF occurrence from previous literature, and predicted LOF occurrence based on our proposed approach, aiming to showcase the potential application value of this work.

\subsection{Extended LOF defect determination criteria for unequal adjacent melt pools}
\label{subsec:LOF_criteria}
According to Tang \textit{et al.}~\cite{tang2017prediction}, we can determine whether there are LOF defects between two adjacent scanlines based on a simple geometric criterion computed from the depth ($D$) and width ($W$) of the two melt pools, as well as the hatch spacing ($H$) between the scanlines and thickness of the newly melted and resolidified layer ($L$). Specifically, \cite{tang2017prediction} states that for LOF-defect-free cases, we have:

\begin{equation}
\label{eqn:LOF_Tang_criteria}
    \left(\frac{H}{W}\right)^2+\left(\frac{L}{D}\right)^2\leq 1
\end{equation}

Equation~\ref{eqn:LOF_Tang_criteria} indicates that we can utilize melt pool morphological features embedded in the high-speed melt pool images to predict whether LOF defects will occur at certain points along with the scanlines from the variability of, say, melt pool width and melt pool depth. In our work, since we have accurately synchronized our dataset (see Sec.~\ref{subsubsec:data_sync}), we can deduce whether there are LOF occurrences from the extracted melt pool features of the ground-truth high-speed images (we call it the ``direct observations" of LOF defect occurrences). Meanwhile, we can also predict LOF occurrences using the melt pool features reconstructed by our proposed data-driven approach given the corresponding acoustic and thermal scalograms (referred to as the ``proposed method predictions" of LOF defect occurrences). 

Owing to the highly stochastic nature of the melt pool oscillation, the geometry of the melt pool along a scanline or within a layer varies continually despite the macroscopically constant $P$ and $V$. Moreover, relying on the thermal history prior to the scanlines and the geometric dimensions of the printing region, local heat accumulation also varies from time to time during the print and could severely impact the melt pool geometry transition as the build goes on. As a result, defect occurrences along scanlines are highly transient events that depend on spatially varying melt-pool morphologies~\cite{scime2019melt}. However, Tang's original LOF criterion~\cite{tang2017prediction} was based on an assumption of constant melt-pool dimensions and neglected the variability of melt pools, which is rarely the case in most realistic LPBF printing. Therefore, it is crucial to first generalize Tang's LOF criteria to the case where two unequal adjacent melt pools are present. We then used our newly extended LOF criteria to determine the direct observations of LOF defects by taking in locally measured $W$ and $D$ of the two adjacent melt pools from the collected ground truth high-speed images. 

\begin{figure}[!htb]
\center{\includegraphics[width=\linewidth]
{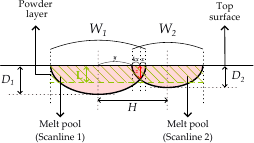}
\caption{Depiction of extended criteria on LOF defect occurrence between two adjacent melt pools (left: melt pool 1; right: melt pool 2). Red semi-ellipsoids depict melt pools beneath the current printing surface (horizontal solid black line). $W_1$ and $D_1$ are the width and depth of the relatively larger melt pool, respectively, and $W_2$ and $D_2$ are the width and depth of the relatively smaller melt pool, respectively. $L$ is the thickness of the powder layer (depicted as green diagonal stripes above the horizontal dash green line). $z$ is the co-melting depth. $x$, $y$, and $k$ denote specific segment lengths and related coefficients as illustrated in the figure. As suggested by Tang \textit{et al.}~\cite{tang2017prediction}, the fundamental criterion of full melting between two adjacent scanlines is $L\leq z$. }
\label{fig:melt_pool_overlap}
}
\end{figure}

As shown in Fig.~\ref{fig:melt_pool_overlap}, two semi-elliptical melt pools, namely melt pool 1 (the right one in Fig.~\ref{fig:melt_pool_overlap}) and melt pool 2 (the left one in Fig.~\ref{fig:melt_pool_overlap}), overlap with each other by a small portion, defined by $W$ ($W_1$ and $W_2$) and $D$ ($D_1$ and $D_2$) of two melt pools and $H$ between them. For the convenience of calculation, $W_1$ and $D_1$ denote the width and depth of the relatively larger melt pool, respectively, and $W_2$ and $D_2$ denote the width and depth of the relatively smaller melt pool, respectively. Therefore, we have $W_1 \geq W_2$ and $D_1 \geq D_2$. Then, we can solve for $x$, $y$, $z$, and $k$ (shown in Fig.~\ref{fig:melt_pool_overlap}) using the following system of equations: 

\begin{equation}
\label{eqn:melt_pool_overlap}
\begin{aligned}
\begin{cases}
    &y+\frac{W_2}{2} = H \\
    &(k+1)x+y = \frac{W_1}{2} \\
    &\left(1-\frac{2kx}{W_2}\right)^2 + \frac{z^2}{D_2^2} = 1 \\
    &\frac{4\left(kx+y\right)^2}{W_1^2} + \frac{z^2}{D_1^2} = 1
\end{cases}
\end{aligned}
\end{equation}

The solutions of $x$, $y$, $z$, and $k$ can be obtained numerically using the Python SciPy function $\texttt{scipy.optimize.fsolve}$~\cite{2020SciPy-NMeth}. As a result, the extended LOF defect occurrence criteria for $L\leq z$ (full melting) should be: 

\begin{equation}
\label{eqn:LOF_criteria}
    \left(\frac{h_1}{W_1}\right)^2+\left(\frac{L}{D_1}\right)^2\leq 1, h_1=y+kx=\frac{2H+kW_1-W_2}{k+1}
\end{equation}

\noindent or alternatively, 

\begin{equation}
\label{eqn:LOF_criteria_alt}
    \left(\frac{h_2}{W_2}\right)^2+\left(\frac{L}{D_2}\right)^2\leq 1, h_2=\frac{W_2}{2}-kx=\frac{2kH-kW_1+W_2}{k+1}
\end{equation}

\noindent where $H$ and $L$ are 140~$\mu\mathrm{m}$ and 30~$\mu\mathrm{m}$ in our experiments, respectively, and $k$ is the zero of the following function:

\begin{equation}
\label{eqn:k}
    f(k)=D_1^2\left[1-\left(\frac{h_1(k)}{W_1}\right)^2\right]-D_2^2\left[1-\left(\frac{h_2(k)}{W_2}\right)^2\right]
\end{equation}

We name $h_1$ and $h_2$ the effective hatch spacing (EHS) of melt pool 1 and melt pool 2, respectively. In Fig.~\ref{fig:melt_pool_overlap}, since $x$, $y$, and $k$ are defined for the melt pool 1, we call $h_1$ the subjective EHS and $h_2$ the objective EHS between the two adjacent melt pools. In essence, EHS represents the distance from the center of the melt pool to the intersection of the two adjacent melt pools, and $k$ reflects how the overlapping region is divided by the two overlapped melt pools. Obviously, we have:

\begin{equation}
\label{eqn:h}
    h_1(k)+h_2(k)=\frac{2H+kW_1-W_2}{k+1}+\frac{2kH-kW_1+W_2}{k+1}=2H
\end{equation}

Therefore, in a more general case where the two adjacent melt pools are unequal either due to the discrepancy of the respective local fusion energy densities or stemming from the stochastic nature of the melt pool oscillation, we should apply the more general criteria proposed as Eqn.~\ref{eqn:LOF_criteria} and Eqn.~\ref{eqn:LOF_criteria_alt} and evaluate the derived EHS (i.e., $h_1$ or $h_2$, rather than $H$) to determine whether there are potential local LOF defect occurrences between the two adjacent scanlines. 

From the above derivations, we see that the (subjective) EHS of a melt pool reflects how much the melt pool size influences its overlap with the other one, and in turn how the melt pool contributes to the local LOF defects. When $W_1=W_2$ and $D_1=D_2$, Eqn.~\ref{eqn:LOF_criteria} and Eqn.~\ref{eqn:LOF_criteria_alt} will both revert to Tang's original LOF criterion, as it is a special case with $k=1$ being the zero of $f(k)$ in Eqn.~\ref{eqn:k}. In this case, both subjective and objective EHSs between the two adjacent scanlines will revert to the prescribed hatch spacing, namely: 

\begin{equation}
\label{eqn:h_revert_1}
    h_1^{\left(W_1=W_2, D_1=D_2\right)}\bigg\vert_{k=1}=h_1^{(0)}=H
\end{equation}

\begin{equation}
\label{eqn:h_revert_2}
    h_2^{\left(W_1=W_2, D_1=D_2\right)}\bigg\vert_{k=1}=h_2^{(0)}=H
\end{equation}

Since LOF-affiliated $(P, V)$ is far from the boundary of the KH regime on the process map and its melt pool morphologies are mainly driven by the heat conduction mode, we reasonably hypothesize $D_1\approx\frac{W_1}{2}$ and $D_2\approx\frac{W_2}{2}$. We also assume that we can accurately capture the planar shape of the melt pool by applying an intensity thresholding method to the collected high-speed melt pool images (see Sec.~\ref{subsec:data_process} for implementation details). 

\subsection{Results on the prediction of highly time-resolved spatially-dependent local LOF occurrences}
\label{subsec:LOF_result}
In this work, we ideally assume that we can determine the direct observations of the LOF defects occurrences based on the melt pool dimension measurements on the ground truth high-speed images. To demonstrate the effectiveness of our proposed approach, we chose Rosenthal Eqn., which was used in~\cite{tang2017prediction} to simulate the width and depth of the melt pool (see Sec.~\ref{appendix:LOF_baseline} for the formulation), as the theoretical baseline model. We believe that it is suitable for LOF-affiliated printing cases since it is capable of simulating melt pool geometries for pure heat conduction mode. Together with the extended criteria proposed in Sec.~\ref{subsec:LOF_criteria}, we use the selected baseline model to determine theoretical predictions on LOF defect occurrences (we call it ``Tang's theoretical predictions" of LOF defect occurrences) given merely $P$, $V$, and material-related parameters. 

\begin{figure}[!htb]
\center{\includegraphics[width=\linewidth]
{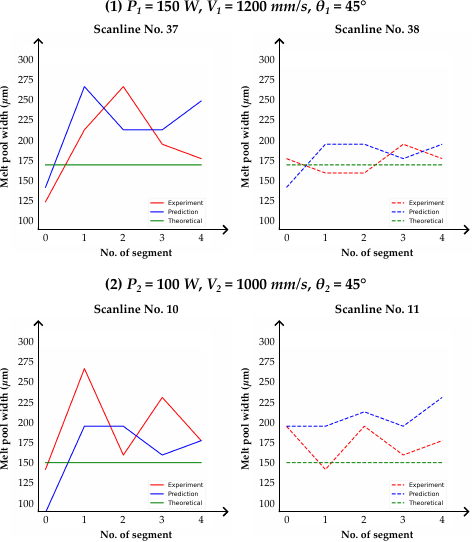}
\caption{Melt pool width from ground truth, theoretical baseline model, and our proposed method at two pairs of adjacent scanlines. Case (1) represents low local heat accumulation status where the print nears the beginning of the layer printing. Case (2) represents high local heat accumulation status where the print nears the end of the layer printing. }
\label{fig:LOF_MP_width}
}
\end{figure}

We aim to show that our proposed method can predict the occurrence of LOF defects with higher accuracy compared to the selected baseline model. Under certain local heat accumulation cases (either low or high), we expect that melt pool size prediction of the baseline theoretical model becomes unreliable, which might lead to false predictions of LOF defect occurrences. To this end, we intentionally select two pairs of adjacent scanlines with specific process parameters: (1) scanline No.~$37$ and scanline No.~$38$ with $P=100~\texttt{W}$, $V=1000~\texttt{mm/s}$, where it is supposed to be LOF along with the entire two scanlines; (2) scanline No.~$10$ and scanline No.~$11$ with $P=150~\texttt{W}$, $V=1200~\texttt{mm/s}$, where it is supposed to be LOF-free along the entire two scanlines. Scanline pair~(1) has lower local heat accumulation since it is near the beginning of the printing of this layer and there is not much prior thermal history before it. Likewise, scanline pair~(2) has higher local heat accumulation since it is near the end of the printing of this layer and there has been a lot of potential heat accumulation before it as well as close to its location. 

We used the proposed pre-trained approach with $t_w=1.0~\texttt{ms}$ to yield predictions of the melt pool width and subsequent predictions of LOF occurrences by the proposed method. Since the proposed approach can only predict the features of the average melt pool within $t_w$, we should apply an equivalent treatment of the moving average for the ground truth melt pool width data. In other words, we need to partition the scanline pair into shorter segments, each of which has a length equal to the laser traveling distance in $t_w$ and will be accordingly labeled the LOF occurrence status. Figure~\ref{fig:LOF_MP_width} shows the melt pool width results of the consecutive scanline segments from the experiment ground truth, the baseline theoretical model, and our proposed method, respectively. The results demonstrate that our proposed model can track the melt pool width variation with better accuracy, which implies that it can successfully extract local heat accumulation information to some extent. 

\begin{figure}[!htb]
\center{\includegraphics[width=\linewidth]
{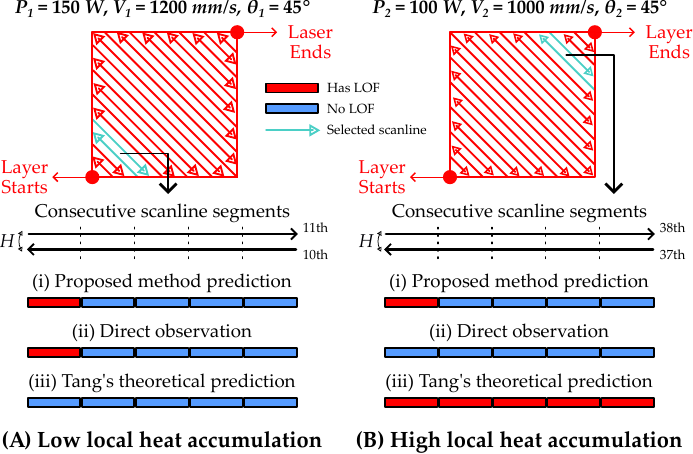}
\caption{Comparison of scanline-scale LOF occurrence detection between direct observations (obtained from the ground truth high-speed melt pool images and Eqn.~\ref{eqn:LOF_criteria}), Tang's theoretical predictions (obtained from the theoretically simulated melt pool geometries~\cite{tang2017prediction} and Eqn.~\ref{eqn:LOF_criteria}), and proposed method predictions (obtained from the reconstructed high-speed melt pool images and Eqn.~\ref{eqn:LOF_criteria}). (A) Low local heat accumulation case, where $P_1=150~\texttt{W}$, $V_1=1200~\texttt{mm/s}$, and $\theta_1=45^{\circ}$. (B) High local heat accumulation case, where $P_2=100~\texttt{W}$, $V_2=1000~\texttt{mm/s}$, and $\theta_2=45^{\circ}$. }
\label{fig:LOF_results}
}
\end{figure}

For each scanline segment, we utilized the extended LOF criteria ({\it i.e.}, Eqn.~\ref{eqn:LOF_criteria}) to label proposed method predictions, direct observations, and Tang's theoretical predictions of LOF defect occurrences based on the local melt pool width result from proposed method-reconstructed images, ground truth high-speed images, and baseline-model-simulated images, respectively. As shown in Fig.~\ref{fig:LOF_results}, Tang's theoretical predictions fail to capture the LOF defect occurrences accurately since the selected theoretical baseline model yields consistent prediction results given merely $P$ and $V$, while our proposed method can predict the LOF defect occurrences that match the direct observations with better accuracy since it is capable of capturing the local heat accumulation information. Figure~\ref{fig:LOF_results}(A) demonstrates that lower heat accumulation and relatively short scanlines close to the beginning of the build will lead to shrinkage of the size of the melt pool, causing the occurrence of LOF defects despite the presumed complete melting given $P$ and $V$; Figure.~\ref{fig:LOF_results}(B) suggests that higher heat accumulation will reasonably increase the size of the melt pool, which might lead to the elimination of LOF defects at certain points despite the presumed occurrences of LOF defects along with these scanline segments according to the theoretical baseline model. By reporting the LOF prediction results of both low and high local heat accumulation cases, we reveal the potential capability of our proposed approach on tracking LOF defect status under different heat accumulation conditions. 
\section{Discussion}
\label{sec:discussion}
In this section, we provide a detailed discussion of the methodology, underlying physics, cost-effectiveness for monitoring, and potential application scenarios of our work. 

\subsection{Methodology selection and interpretations}
\label{subsec:disc_method}
Our results demonstrate that our proposed approach is capable of tracking visual characteristics of the melt pool in real-time given solely the acoustic and photodiode data, despite the high variability of the melt pool features. The definition of real-time can be ambiguous here since no time window with a finite width can truly be instantaneous. Nonetheless, the proposed approach can achieve swift data processing and scalogram-to-melt-pool mapping in approximately 12 milliseconds, which is fast enough to capture melt pool dynamic morphology in a highly time-resolved fashion. 

In terms of image-based melt pool predictions, we prefer using an autoencoder as the feature extractor of the straightened high-speed melt pool images rather than predetermined melt pool features, since an autoencoder can perform a mathematical latent space construction in an unsupervised manner and will not be limited by prior knowledge of melt pools. As such, it can not only include deterministic features that people are familiar with ({\it e.g.}, melt pool length, width, and area) but also embed some other features ({\it e.g.}, spatial image moment, melt pool, and track thermal distribution) that could be difficult for us to observe and identify. Moreover, by directly generating image output, our proposed approach allows subsequent image-centered feature quantification and analysis during or after the print. Therefore, we choose to use the image format as the final output form of the prediction, and we develop a prior-knowledge-based $Q$ score to assess the prediction quality of the reconstructed melt pool images. 

It is worth emphasizing that our generated low-dimensional latent space for the high-speed melt-pool images does not necessarily bear any physical interpretability. In this work, our autoencoder pipeline serves mainly as a reduced-order modeling approach that yields concise latent vectors representing data points in the corresponding image space. While the non-uniqueness of the latent space might lead to functional variations of the trained decoder and the deep convolutional network, the function that maps from the acoustic and thermal scalograms to the melt pool geometric features shall be invariant given the unique existence of their essential physical correlations. With that being noted, we have discovered from a latent space disentanglement analysis (see Sec.~\ref{appendix:additional_latent}) that our constructed latent space does successfully capture some physical quantities of the melt pool (e.g., lengths, widths, areas, and thermal distributions), showing that our generated mapping from acoustic and thermal emission scalograms to the melt pool 2D images is potentially physically interpretable. For future work, we will extensively explore the possibility of generating a physics-aware data-driven mapping between our inputs and outputs that can reveal meaningful physical correlations based on the collected data.

\subsection{Potential application scenarios}
\label{subsec:disc_app}
According to previous work~\cite{wasmer2018situ, wu2016situ}, acoustic and thermal energies can be directly correlated with fusion energy density. The real volumetric energy density of the melt pool, namely $\frac{P}{V\cdot t\cdot H}$ (where $P$, $V$, $H$, and $t$ are the local fusion power, scanning speed, hatch spacing, and molten layer thickness, respectively), changes dynamically during the printing process due to the melt pool oscillation and local heat accumulation as layers are stacked up. As shown in Fig.~\ref{fig:LOF_results}, by building a square prism geometry with various $\theta$, we intentionally introduce different levels of heat accumulation at the corners and along the sides of the layer. In severe heat accumulation cases, the actual fusion energy density increases and is higher than the expected prescribed value, which allows acoustic and photodiode data to serve as real-time fusion information that can infer local melt pool status (see Sec.~\ref{sec:LOF_valid}). 

For the detection of spatially dependent LOF defects, we directly observed the LOF occurrences by measuring the melt pool width from the ground truth high-speed images and applying the extended LOF determination criteria. As our next step, experiments of true LOF defect detection will be done to further validate the defect detection capability of our proposed approach. With that said, we believe the validation case elaborated in Sec.~\ref{sec:LOF_valid} can still demonstrate the effectiveness of our proposed method when the defects or any other AM-process-related properties are related to the melt pool morphologies. In addition, considering that the melt pool oscillation can significantly impact the cooling rate and microstructure development during the build, it is beneficial to be aware of the melt pool variations along with the printing process, especially when the change of melt pool geometry due to local heat accumulation leads to severe process deviations. Our proposed method provides a cost-effective way to provide critical melt pool morphology information by leveraging the physical correlation between airborne acoustic emission, thermal emission, and melt pool morphologies. Moreover, we believe the extended LOF determination criteria for unequal adjacent melt pools can not only be useful when taking care of varying local fusion energy densities but also be suitable when future AM machines are capable of printing scanlines with different process parameters. 

In this work, we conducted our series of experiments on the same EOS M290 machine. Therefore, it is reasonable to believe that our trained model might not be able to be directly transplanted to another machine, where the printing environment and some other physical setup-related parameters could vary, without any adaptation. As a consequence, the pipeline might require retraining or fine-tuning when it is applied to a different printing platform. More importantly, we argue that our proposed pipeline model -- consisting of an autoencoder and a deep convolutional network -- can be adapted to many different printing platforms given a certain amount of their specific experiment data. The concept and efficacy of our proposed approach should be generalizable among different platforms considering that the mapping that our approach achieves relies on the physical correlation between acoustic emission, thermal emission, and melt pool geometric characteristics. 

\subsection{Underlying physics}
\label{subsec:disc_physics}
Regarding the underlying physics of our work, photodiode-collected melt pool thermal information can infer melt pool morphology to a certain extent according to Ren {\it et al.}~\cite{Ren:2023aa}. Other previous works also suggest that AE information can be correlated with printing flaws~\cite{tempelman2022detection, drissi2022differentiation, pandiyan2020analysis}. From the acoustic scalogram matrices (shown in Fig.~\ref{fig:moving_avg}), we observe that the acoustic frequency band between 20~kHz and 30~kHz is particularly correlated with melt pool dynamics such as melt pool size variation and shape oscillation. Interestingly, other research also suggests that the critical frequency of keyhole oscillation is within the range of 20~kHz to 40~kHz, which coincides well with the highlighted acoustic and thermal frequency band in our scalograms according to our data analysis~\cite{Ren:2023aa, pandiyan2021semi, taheri2019situ}. Therefore, we reasonably hypothesize that both the airborne acoustic signature and the photodiode-collected thermal information are directly and deeply related to the melt pool dynamics ({\it e.g.}, how the melt pool and vapor channel vibrate and oscillate~\cite{khairallah2021onset, huang2022keyhole, caprio2020observing, pyeon2021time}, how the liquid metal evaporates, and how the spatters eject~\cite{bitharas2022interplay}) and can infer related melt pool characteristics with proper processing and data interpretation. With the aid of machine learning technologies, such an interpretation can be accomplished given high-quality data. However, the exact correlation between acoustic signals, thermal emissions, and melt pool morphology as well as the critical physical understanding of it is yet to be characterized and extracted. This will be part of our future work on the knowledge discovery of LPBF process monitoring. 

In this work, the orientation and location of the melt pool are omitted through data processing, and acoustic and photodiode data are used to infer the spatial shape, size, and distribution of the melt pool. Since acoustic and photodiode data are related to fusion energy density, we can use them to characterize melt pool spatial features. According to the Rosenthal equation and Eagar-Tsai model, melt pool width, length, and area are inter-correlated and are functions of both the preheating temperature and the energy density. This explains why time-series data like acoustic signals and thermal emissions can still accurately inform melt pool spatial information, which further reflects the potential occurrence of certain printing defects. 

\subsection{Cost-effectiveness for process monitoring}
\label{subsec:disc_cost}
For the cost-effectiveness of acoustic and thermal emission-based melt pool tracking, we calculated the data flow rates of the acoustic data, photodiode data, and high-speed image data. It turns out that each melt pool image frame from the high-speed camera will take nearly 88~kB; however, by using acoustic signals and photodiode-collected thermal emissions, our approach just uses 2~kB data to reconstruct a satisfactory version of the melt pool image. This demonstrates that our proposed approach can not only reduce the cost of sensor equipment but also reduce the usage of computational resources necessary for the entire monitoring process. Moreover, with a potentially successful application to local LOF defect detection, our proposed approach shows its capability of cost-effectively inferring the LOF distribution at the scanline scale and saving substantial costs from laborious sample sectioning or CT scanning. 
\section{Conclusion and Future Work}
\label{sec:conclusion}
In this work, we investigate the capability and efficacy of highly time-resolved LPBF visual tracking and LOF defect detection by employing a deep learning approach and using acoustic signals and photodiode-collected thermal emissions data. Our work first shows that our proposed approach possesses a potential capability of achieving cost-effective real-time melt pool visual characteristics given only the acoustic and photodiode data. Accurate melt pool image results can be predicted and reconstructed within a time window as short as 1.0~$\mathrm{ms}$. We then demonstrate that our proposed method can be employed to infer LOF defect occurrence at the scanline level, outperforming Tang's theoretical models that employed the Rosenthal equation. This work moves one step toward efficient and cost-effective LPBF process monitoring. Our results also demonstrate knowledge discoveries and imply the potential underlying relationship between acoustic signatures, thermal emissions, and corresponding fusion dynamics.  

For future extensions of this research, there are a few next steps worth exploring in order to build on recent progress and address current limitations. For example, other parameters related to printing conditions, such as hatch spacing, layer thickness, and powder-affiliated or powder-less setup, can be included in the study to investigate how they affect the corresponding acoustic and photodiode data signature. Moreover, due to the small scale of our print and the usage of the straightening technique for melt pool images, we reasonably neglected the melt pool location variance, as well as the data reception angle variance in this work; hence, we can investigate in future work how the efficacy of our approach would vary in a large scale print, in which the melt pool location variation on the build plate and data reception angle difference are non-negligible. 
\section*{Acknowledgement}
\label{sec:acknow}
This research is supported by the Eaton Cooperation Award Number 001145471 and the United States Army Research Laboratory under Cooperative Agreement Number W911NF-20-2-0175. We thank Dr.~Zhongshu Ren and Prof.~Tao Sun for useful discussions and paper proofreading. We thank Xingyang Li, Evan Diewald, Tong Lin, Xinjian Li, Dr.~Wentai Zhang, Dr.~Niloofar Sanaei, and Dr.~Chi-Ta Yang for useful discussions. We thank Dr.~Sandra DeVincent Wolf and Liza Allison for their support and guidance on the LPBF experiments. Haolin Liu thanks Silin Liu and Kuanren Qian for their help with paper proofreading and figure generation. We thank Xueting Pu for help with dataset preparation. 
\bibliographystyle{elsarticle-num-names}
\bibliography{references}

\begin{thebibliography}{68}
\expandafter\ifx\csname natexlab\endcsname\relax\def\natexlab#1{#1}\fi
\providecommand{\url}[1]{\texttt{#1}}
\providecommand{\href}[2]{#2}
\providecommand{\path}[1]{#1}
\providecommand{\DOIprefix}{doi:}
\providecommand{\ArXivprefix}{arXiv:}
\providecommand{\URLprefix}{URL: }
\providecommand{\Pubmedprefix}{pmid:}
\providecommand{\doi}[1]{\href{http://dx.doi.org/#1}{\path{#1}}}
\providecommand{\Pubmed}[1]{\href{pmid:#1}{\path{#1}}}
\providecommand{\bibinfo}[2]{#2}
\ifx\xfnm\relax \def\xfnm[#1]{\unskip,\space#1}\fi
\bibitem[{Narra et~al.(2023)Narra, Rollett, Ngo, Scannapieco, Shahabi, Reddy,
  Pauza, Taylor, Gobert, Diewald, Dugast, To, Wicker, Beuth, and
  Lewandowski}]{Narra:2023aa}
\bibinfo{author}{S.~P. Narra}, \bibinfo{author}{A.~D. Rollett},
  \bibinfo{author}{A.~Ngo}, \bibinfo{author}{D.~Scannapieco},
  \bibinfo{author}{M.~Shahabi}, \bibinfo{author}{T.~Reddy},
  \bibinfo{author}{J.~Pauza}, \bibinfo{author}{H.~Taylor},
  \bibinfo{author}{C.~Gobert}, \bibinfo{author}{E.~Diewald},
  \bibinfo{author}{F.~X. Dugast}, \bibinfo{author}{A.~To},
  \bibinfo{author}{R.~Wicker}, \bibinfo{author}{J.~Beuth},
  \bibinfo{author}{J.~J. Lewandowski},
\newblock \bibinfo{title}{Process qualification of laser powder bed fusion
  based on processing-defect structure-fatigue properties in ti-6al-4v},
\newblock \bibinfo{journal}{Journal of Materials Processing Technology}
  \bibinfo{volume}{311} (\bibinfo{year}{2023}) \bibinfo{pages}{117775}.
  \DOIprefix\doi{https://doi.org/10.1016/j.jmatprotec.2022.117775}.
\bibitem[{Forien et~al.(2020)Forien, Calta, DePond, Guss, Roehling, and
  Matthews}]{forien2020detecting}
\bibinfo{author}{J.-B. Forien}, \bibinfo{author}{N.~P. Calta},
  \bibinfo{author}{P.~J. DePond}, \bibinfo{author}{G.~M. Guss},
  \bibinfo{author}{T.~T. Roehling}, \bibinfo{author}{M.~J. Matthews},
\newblock \bibinfo{title}{Detecting keyhole pore defects and monitoring process
  signatures during laser powder bed fusion: A correlation between in situ
  pyrometry and ex situ x-ray radiography},
\newblock \bibinfo{journal}{Additive Manufacturing} \bibinfo{volume}{35}
  (\bibinfo{year}{2020}) \bibinfo{pages}{101336}.
\bibitem[{Repossini et~al.(2017)Repossini, Laguzza, Grasso, and
  Colosimo}]{repossini2017use}
\bibinfo{author}{G.~Repossini}, \bibinfo{author}{V.~Laguzza},
  \bibinfo{author}{M.~Grasso}, \bibinfo{author}{B.~M. Colosimo},
\newblock \bibinfo{title}{On the use of spatter signature for in-situ
  monitoring of laser powder bed fusion},
\newblock \bibinfo{journal}{Additive Manufacturing} \bibinfo{volume}{16}
  (\bibinfo{year}{2017}) \bibinfo{pages}{35--48}.
\bibitem[{Lough et~al.(2020)Lough, Escano, Qu, Smith, Landers, Bristow, Chen,
  and Kinzel}]{lough2020situ}
\bibinfo{author}{C.~S. Lough}, \bibinfo{author}{L.~I. Escano},
  \bibinfo{author}{M.~Qu}, \bibinfo{author}{C.~C. Smith},
  \bibinfo{author}{R.~G. Landers}, \bibinfo{author}{D.~A. Bristow},
  \bibinfo{author}{L.~Chen}, \bibinfo{author}{E.~C. Kinzel},
\newblock \bibinfo{title}{In-situ optical emission spectroscopy of selective
  laser melting},
\newblock \bibinfo{journal}{Journal of Manufacturing Processes}
  \bibinfo{volume}{53} (\bibinfo{year}{2020}) \bibinfo{pages}{336--341}.
\bibitem[{Zhang et~al.(2021)Zhang, Yan, Hong, Fuh, Wang, Lin, and
  Ye}]{zhang2021data}
\bibinfo{author}{Y.~Zhang}, \bibinfo{author}{W.~Yan}, \bibinfo{author}{G.~S.
  Hong}, \bibinfo{author}{J.~F.~H. Fuh}, \bibinfo{author}{D.~Wang},
  \bibinfo{author}{X.~Lin}, \bibinfo{author}{D.~Ye},
\newblock \bibinfo{title}{Data fusion analysis in the powder-bed fusion am
  process monitoring by dempster-shafer evidence theory},
\newblock \bibinfo{journal}{Rapid Prototyping Journal} \bibinfo{volume}{28}
  (\bibinfo{year}{2021}) \bibinfo{pages}{841--854}.
\bibitem[{Guo et~al.(2019)Guo, Zhao, Qu, Xiong, Escano, Hojjatzadeh, Parab,
  Fezzaa, Everhart, Sun et~al.}]{guo2019situ}
\bibinfo{author}{Q.~Guo}, \bibinfo{author}{C.~Zhao}, \bibinfo{author}{M.~Qu},
  \bibinfo{author}{L.~Xiong}, \bibinfo{author}{L.~I. Escano},
  \bibinfo{author}{S.~M.~H. Hojjatzadeh}, \bibinfo{author}{N.~D. Parab},
  \bibinfo{author}{K.~Fezzaa}, \bibinfo{author}{W.~Everhart},
  \bibinfo{author}{T.~Sun}, et~al.,
\newblock \bibinfo{title}{In-situ characterization and quantification of melt
  pool variation under constant input energy density in laser powder bed fusion
  additive manufacturing process},
\newblock \bibinfo{journal}{Additive Manufacturing} \bibinfo{volume}{28}
  (\bibinfo{year}{2019}) \bibinfo{pages}{600--609}.
\bibitem[{Hojjatzadeh et~al.(2021)Hojjatzadeh, Guo, Parab, Qu, Escano, Fezzaa,
  Everhart, Sun, and Chen}]{hojjatzadeh2021situ}
\bibinfo{author}{S.~M.~H. Hojjatzadeh}, \bibinfo{author}{Q.~Guo},
  \bibinfo{author}{N.~D. Parab}, \bibinfo{author}{M.~Qu},
  \bibinfo{author}{L.~I. Escano}, \bibinfo{author}{K.~Fezzaa},
  \bibinfo{author}{W.~Everhart}, \bibinfo{author}{T.~Sun},
  \bibinfo{author}{L.~Chen},
\newblock \bibinfo{title}{In-situ characterization of pore formation dynamics
  in pulsed wave laser powder bed fusion},
\newblock \bibinfo{journal}{Materials} \bibinfo{volume}{14}
  (\bibinfo{year}{2021}) \bibinfo{pages}{2936}.
\bibitem[{Javadi et~al.(2020)Javadi, Mohseni, MacLeod, Lines, Vasilev, Mineo,
  Foster, Pierce, and Gachagan}]{javadi2020continuous}
\bibinfo{author}{Y.~Javadi}, \bibinfo{author}{E.~Mohseni},
  \bibinfo{author}{C.~N. MacLeod}, \bibinfo{author}{D.~Lines},
  \bibinfo{author}{M.~Vasilev}, \bibinfo{author}{C.~Mineo},
  \bibinfo{author}{E.~Foster}, \bibinfo{author}{S.~G. Pierce},
  \bibinfo{author}{A.~Gachagan},
\newblock \bibinfo{title}{Continuous monitoring of an
  intentionally-manufactured crack using an automated welding and in-process
  inspection system},
\newblock \bibinfo{journal}{Materials \& Design} \bibinfo{volume}{191}
  (\bibinfo{year}{2020}) \bibinfo{pages}{108655}.
\bibitem[{Grasso et~al.(2018)Grasso, Demir, Previtali, and
  Colosimo}]{grasso2018situ}
\bibinfo{author}{M.~Grasso}, \bibinfo{author}{A.~G. Demir},
  \bibinfo{author}{B.~Previtali}, \bibinfo{author}{B.~M. Colosimo},
\newblock \bibinfo{title}{In situ monitoring of selective laser melting of zinc
  powder via infrared imaging of the process plume},
\newblock \bibinfo{journal}{Robotics and Computer-Integrated Manufacturing}
  \bibinfo{volume}{49} (\bibinfo{year}{2018}) \bibinfo{pages}{229--239}.
\bibitem[{Tenbrock et~al.(2021)Tenbrock, Kelliger, Praetzsch, Ronge, Jauer, and
  Schleifenbaum}]{tenbrock2021effect}
\bibinfo{author}{C.~Tenbrock}, \bibinfo{author}{T.~Kelliger},
  \bibinfo{author}{N.~Praetzsch}, \bibinfo{author}{M.~Ronge},
  \bibinfo{author}{L.~Jauer}, \bibinfo{author}{J.~H. Schleifenbaum},
\newblock \bibinfo{title}{Effect of laser-plume interaction on part quality in
  multi-scanner laser powder bed fusion},
\newblock \bibinfo{journal}{Additive Manufacturing} \bibinfo{volume}{38}
  (\bibinfo{year}{2021}) \bibinfo{pages}{101810}.
\bibitem[{Downey et~al.(2016)Downey, O'Sullivan, Nejmen, Bombinski, O’Leary,
  Raghavendra, and Jemielniak}]{downey2016real}
\bibinfo{author}{J.~Downey}, \bibinfo{author}{D.~O'Sullivan},
  \bibinfo{author}{M.~Nejmen}, \bibinfo{author}{S.~Bombinski},
  \bibinfo{author}{P.~O’Leary}, \bibinfo{author}{R.~Raghavendra},
  \bibinfo{author}{K.~Jemielniak},
\newblock \bibinfo{title}{Real time monitoring of the cnc process in a
  production environment-the data collection \& analysis phase},
\newblock \bibinfo{journal}{Procedia Cirp} \bibinfo{volume}{41}
  (\bibinfo{year}{2016}) \bibinfo{pages}{920--926}.
\bibitem[{Plaza et~al.(2019)Plaza, L{\'o}pez, and
  Gonz{\'a}lez}]{plaza2019efficiency}
\bibinfo{author}{E.~G. Plaza}, \bibinfo{author}{P.~N. L{\'o}pez},
  \bibinfo{author}{E.~B. Gonz{\'a}lez},
\newblock \bibinfo{title}{Efficiency of vibration signal feature extraction for
  surface finish monitoring in cnc machining},
\newblock \bibinfo{journal}{Journal of Manufacturing Processes}
  \bibinfo{volume}{44} (\bibinfo{year}{2019}) \bibinfo{pages}{145--157}.
\bibitem[{Grasso et~al.(2021)Grasso, Remani, Dickins, Colosimo, and
  Leach}]{grasso2021situ}
\bibinfo{author}{M.~Grasso}, \bibinfo{author}{A.~Remani},
  \bibinfo{author}{A.~Dickins}, \bibinfo{author}{B.~M. Colosimo},
  \bibinfo{author}{R.~K. Leach},
\newblock \bibinfo{title}{In-situ measurement and monitoring methods for metal
  powder bed fusion: an updated review},
\newblock \bibinfo{journal}{Measurement Science and Technology}
  \bibinfo{volume}{32} (\bibinfo{year}{2021}) \bibinfo{pages}{112001}.
\bibitem[{Everton et~al.(2016)Everton, Hirsch, Stravroulakis, Leach, and
  Clare}]{everton2016review}
\bibinfo{author}{S.~K. Everton}, \bibinfo{author}{M.~Hirsch},
  \bibinfo{author}{P.~Stravroulakis}, \bibinfo{author}{R.~K. Leach},
  \bibinfo{author}{A.~T. Clare},
\newblock \bibinfo{title}{Review of in-situ process monitoring and in-situ
  metrology for metal additive manufacturing},
\newblock \bibinfo{journal}{Materials \& Design} \bibinfo{volume}{95}
  (\bibinfo{year}{2016}) \bibinfo{pages}{431--445}.
\bibitem[{McCann et~al.(2021)McCann, Obeidi, Hughes, McCarthy, Egan,
  Vijayaraghavan, Joshi, Garzon, Dowling, McNally et~al.}]{mccann2021situ}
\bibinfo{author}{R.~McCann}, \bibinfo{author}{M.~A. Obeidi},
  \bibinfo{author}{C.~Hughes}, \bibinfo{author}{{\'E}.~McCarthy},
  \bibinfo{author}{D.~S. Egan}, \bibinfo{author}{R.~K. Vijayaraghavan},
  \bibinfo{author}{A.~M. Joshi}, \bibinfo{author}{V.~A. Garzon},
  \bibinfo{author}{D.~P. Dowling}, \bibinfo{author}{P.~J. McNally}, et~al.,
\newblock \bibinfo{title}{In-situ sensing, process monitoring and machine
  control in laser powder bed fusion: A review},
\newblock \bibinfo{journal}{Additive Manufacturing} \bibinfo{volume}{45}
  (\bibinfo{year}{2021}) \bibinfo{pages}{102058}.
\bibitem[{Khairallah et~al.(2021)Khairallah, Sun, and
  Simonds}]{khairallah2021onset}
\bibinfo{author}{S.~A. Khairallah}, \bibinfo{author}{T.~Sun},
  \bibinfo{author}{B.~J. Simonds},
\newblock \bibinfo{title}{Onset of periodic oscillations as a precursor of a
  transition to pore-generating turbulence in laser melting},
\newblock \bibinfo{journal}{Additive Manufacturing Letters} \bibinfo{volume}{1}
  (\bibinfo{year}{2021}) \bibinfo{pages}{100002}.
\bibitem[{Bachmann et~al.(2022)Bachmann, Meng, Artinov, and
  Rethmeier}]{bachmann2022elucidation}
\bibinfo{author}{M.~Bachmann}, \bibinfo{author}{X.~Meng},
  \bibinfo{author}{A.~Artinov}, \bibinfo{author}{M.~Rethmeier},
\newblock \bibinfo{title}{Elucidation of the bulging effect by an improved
  ray-tracing algorithm in deep penetration wire feed laser beam welding and
  its influence on the mixing behavior},
\newblock \bibinfo{journal}{Advanced Engineering Materials}
  \bibinfo{volume}{24} (\bibinfo{year}{2022}) \bibinfo{pages}{2101299}.
\bibitem[{Shahabi et~al.(2022)Shahabi, Reddy, Rollett, and
  Narra}]{shahabi2022statistical}
\bibinfo{author}{M.~Shahabi}, \bibinfo{author}{T.~Reddy},
  \bibinfo{author}{A.~D. Rollett}, \bibinfo{author}{S.~P. Narra},
\newblock \bibinfo{title}{A statistical approach to determine data requirements
  for part porosity characterization in laser powder bed fusion additive
  manufacturing},
\newblock \bibinfo{journal}{Materials Characterization} \bibinfo{volume}{190}
  (\bibinfo{year}{2022}) \bibinfo{pages}{112027}.
\bibitem[{Yang et~al.(2020)Yang, Lu, Yeung, and Krishnamurty}]{yang2020scan}
\bibinfo{author}{Z.~Yang}, \bibinfo{author}{Y.~Lu}, \bibinfo{author}{H.~Yeung},
  \bibinfo{author}{S.~Krishnamurty},
\newblock \bibinfo{title}{From scan strategy to melt pool prediction: A
  neighboring-effect modeling method},
\newblock \bibinfo{journal}{Journal of Computing and Information Science in
  Engineering} \bibinfo{volume}{20} (\bibinfo{year}{2020})
  \bibinfo{pages}{051001}.
\bibitem[{Jones et~al.(2021)Jones, Yang, Yeung, Witherell, and
  Lu}]{jones2021hybrid}
\bibinfo{author}{K.~Jones}, \bibinfo{author}{Z.~Yang},
  \bibinfo{author}{H.~Yeung}, \bibinfo{author}{P.~Witherell},
  \bibinfo{author}{Y.~Lu},
\newblock \bibinfo{title}{Hybrid modeling of melt pool geometry in additive
  manufacturing using neural networks},
\newblock in: \bibinfo{booktitle}{International Design Engineering Technical
  Conferences and Computers and Information in Engineering Conference}, volume
  \bibinfo{volume}{85376}, \bibinfo{organization}{American Society of
  Mechanical Engineers}, \bibinfo{year}{2021}, p. \bibinfo{pages}{V002T02A031}.
\bibitem[{Dong et~al.(2022)Dong, Wong, Lestandi, Mikula, Vastola, Jhon, Dao,
  Kizhakkinan, Ford, and Rosen}]{dong2022part}
\bibinfo{author}{G.~Dong}, \bibinfo{author}{J.~C. Wong},
  \bibinfo{author}{L.~Lestandi}, \bibinfo{author}{J.~Mikula},
  \bibinfo{author}{G.~Vastola}, \bibinfo{author}{M.~H. Jhon},
  \bibinfo{author}{M.~H. Dao}, \bibinfo{author}{U.~Kizhakkinan},
  \bibinfo{author}{C.~S. Ford}, \bibinfo{author}{D.~W. Rosen},
\newblock \bibinfo{title}{A part-scale, feature-based surrogate model for
  residual stresses in the laser powder bed fusion process},
\newblock \bibinfo{journal}{Journal of Materials Processing Technology}
  \bibinfo{volume}{304} (\bibinfo{year}{2022}) \bibinfo{pages}{117541}.
\bibitem[{Meier et~al.(2021)Meier, Fuchs, Much, Nitzler, Penny, Praegla,
  Proell, Sun, Weissbach, Schreter et~al.}]{meier2021physics}
\bibinfo{author}{C.~Meier}, \bibinfo{author}{S.~L. Fuchs},
  \bibinfo{author}{N.~Much}, \bibinfo{author}{J.~Nitzler},
  \bibinfo{author}{R.~W. Penny}, \bibinfo{author}{P.~M. Praegla},
  \bibinfo{author}{S.~D. Proell}, \bibinfo{author}{Y.~Sun},
  \bibinfo{author}{R.~Weissbach}, \bibinfo{author}{M.~Schreter}, et~al.,
\newblock \bibinfo{title}{Physics-based modeling and predictive simulation of
  powder bed fusion additive manufacturing across length scales},
\newblock \bibinfo{journal}{GAMM-Mitteilungen} \bibinfo{volume}{44}
  (\bibinfo{year}{2021}) \bibinfo{pages}{e202100014}.
\bibitem[{Li et~al.(2021)Li, Pan, Zhang, Chen, Cui, Yan, and
  Liou}]{li2021deformations}
\bibinfo{author}{L.~Li}, \bibinfo{author}{T.~Pan}, \bibinfo{author}{X.~Zhang},
  \bibinfo{author}{Y.~Chen}, \bibinfo{author}{W.~Cui},
  \bibinfo{author}{L.~Yan}, \bibinfo{author}{F.~Liou},
\newblock \bibinfo{title}{Deformations and stresses prediction of cantilever
  structures fabricated by selective laser melting process},
\newblock \bibinfo{journal}{Rapid Prototyping Journal} \bibinfo{volume}{27}
  (\bibinfo{year}{2021}) \bibinfo{pages}{453--464}.
\bibitem[{Yeung et~al.(2020)Yeung, Yang, and Yan}]{yeung2020meltpool}
\bibinfo{author}{H.~Yeung}, \bibinfo{author}{Z.~Yang},
  \bibinfo{author}{L.~Yan},
\newblock \bibinfo{title}{A meltpool prediction based scan strategy for powder
  bed fusion additive manufacturing},
\newblock \bibinfo{journal}{Additive Manufacturing} \bibinfo{volume}{35}
  (\bibinfo{year}{2020}) \bibinfo{pages}{101383}.
\bibitem[{Yeung and Lane(2020)}]{yeung2020residual}
\bibinfo{author}{H.~Yeung}, \bibinfo{author}{B.~Lane},
\newblock \bibinfo{title}{A residual heat compensation based scan strategy for
  powder bed fusion additive manufacturing},
\newblock \bibinfo{journal}{Manufacturing letters} \bibinfo{volume}{25}
  (\bibinfo{year}{2020}) \bibinfo{pages}{56--59}.
\bibitem[{Liao et~al.(2022)Liao, Webster, Huang, Council, Ehmann, and
  Cao}]{liao2022simulation}
\bibinfo{author}{S.~Liao}, \bibinfo{author}{S.~Webster},
  \bibinfo{author}{D.~Huang}, \bibinfo{author}{R.~Council},
  \bibinfo{author}{K.~Ehmann}, \bibinfo{author}{J.~Cao},
\newblock \bibinfo{title}{Simulation-guided variable laser power design for
  melt pool depth control in directed energy deposition},
\newblock \bibinfo{journal}{Additive Manufacturing} \bibinfo{volume}{56}
  (\bibinfo{year}{2022}) \bibinfo{pages}{102912}.
\bibitem[{Ren et~al.(2023)Ren, Gao, Clark, Fezzaa, Shevchenko, Choi, Everhart,
  Rollett, Chen, and Sun}]{Ren:2023aa}
\bibinfo{author}{Z.~Ren}, \bibinfo{author}{L.~Gao}, \bibinfo{author}{S.~J.
  Clark}, \bibinfo{author}{K.~Fezzaa}, \bibinfo{author}{P.~Shevchenko},
  \bibinfo{author}{A.~Choi}, \bibinfo{author}{W.~Everhart},
  \bibinfo{author}{A.~D. Rollett}, \bibinfo{author}{L.~Chen},
  \bibinfo{author}{T.~Sun},
\newblock \bibinfo{title}{Machine learning--aided real-time detection of
  keyhole pore generation in laser powder bed fusion},
\newblock \bibinfo{journal}{Science} \bibinfo{volume}{379}
  (\bibinfo{year}{2023}) \bibinfo{pages}{89--94}. \URLprefix
  \url{https://doi.org/10.1126/science.add4667}.
  \DOIprefix\doi{10.1126/science.add4667}.
\bibitem[{Drouet and Nadeau(1982)}]{drouet1982acoustic}
\bibinfo{author}{M.~G. Drouet}, \bibinfo{author}{F.~Nadeau},
\newblock \bibinfo{title}{Acoustic measurement of the arc voltage applicable to
  arc welding and arc furnaces},
\newblock \bibinfo{journal}{Journal of Physics E: Scientific Instruments}
  \bibinfo{volume}{15} (\bibinfo{year}{1982}) \bibinfo{pages}{268}.
\bibitem[{Prezelj et~al.(2008)Prezelj, Polajnar et~al.}]{prezelj2008use}
\bibinfo{author}{J.~Prezelj}, \bibinfo{author}{I.~Polajnar}, et~al.,
\newblock \bibinfo{title}{Use of audible sound for on-line monitoring of gas
  metal arc welding process.},
\newblock \bibinfo{journal}{Metalurgija} \bibinfo{volume}{47}
  (\bibinfo{year}{2008}).
\bibitem[{Wang et~al.(2009)Wang, Chen, Chen, and Chen}]{wang2009analysis}
\bibinfo{author}{J.~Wang}, \bibinfo{author}{B.~Chen},
  \bibinfo{author}{H.~Chen}, \bibinfo{author}{S.~Chen},
\newblock \bibinfo{title}{Analysis of arc sound characteristics for gas
  tungsten argon welding},
\newblock \bibinfo{journal}{Sensor review} \bibinfo{volume}{29}
  (\bibinfo{year}{2009}) \bibinfo{pages}{240--249}.
\bibitem[{Cayo and Alfaro(2009)}]{cayo2009non}
\bibinfo{author}{E.~H. Cayo}, \bibinfo{author}{S.~C.~A. Alfaro},
\newblock \bibinfo{title}{A non-intrusive gma welding process quality
  monitoring system using acoustic sensing},
\newblock \bibinfo{journal}{Sensors} \bibinfo{volume}{9} (\bibinfo{year}{2009})
  \bibinfo{pages}{7150--7166}.
\bibitem[{Asif et~al.(2022)Asif, Zhang, Derrible, Indacochea, Ozevin, and
  Ziebart}]{asif2022machine}
\bibinfo{author}{K.~Asif}, \bibinfo{author}{L.~Zhang},
  \bibinfo{author}{S.~Derrible}, \bibinfo{author}{J.~E. Indacochea},
  \bibinfo{author}{D.~Ozevin}, \bibinfo{author}{B.~Ziebart},
\newblock \bibinfo{title}{Machine learning model to predict welding quality
  using air-coupled acoustic emission and weld inputs},
\newblock \bibinfo{journal}{Journal of Intelligent Manufacturing}
  (\bibinfo{year}{2022}) \bibinfo{pages}{1--15}.
\bibitem[{Zhang et~al.(2019)Zhang, Basantes-Defaz, Ozevin, and
  Indacochea}]{zhang2019real}
\bibinfo{author}{L.~Zhang}, \bibinfo{author}{A.~C. Basantes-Defaz},
  \bibinfo{author}{D.~Ozevin}, \bibinfo{author}{E.~Indacochea},
\newblock \bibinfo{title}{Real-time monitoring of welding process using
  air-coupled ultrasonics and acoustic emission},
\newblock \bibinfo{journal}{The International Journal of Advanced Manufacturing
  Technology} \bibinfo{volume}{101} (\bibinfo{year}{2019})
  \bibinfo{pages}{1623--1634}.
\bibitem[{Tempelman et~al.(2022)Tempelman, Wachtor, Flynn, Depond, Forien,
  Guss, Calta, and Matthews}]{tempelman2022detection}
\bibinfo{author}{J.~R. Tempelman}, \bibinfo{author}{A.~J. Wachtor},
  \bibinfo{author}{E.~B. Flynn}, \bibinfo{author}{P.~J. Depond},
  \bibinfo{author}{J.-B. Forien}, \bibinfo{author}{G.~M. Guss},
  \bibinfo{author}{N.~P. Calta}, \bibinfo{author}{M.~J. Matthews},
\newblock \bibinfo{title}{Detection of keyhole pore formations in laser
  powder-bed fusion using acoustic process monitoring measurements},
\newblock \bibinfo{journal}{Additive Manufacturing} \bibinfo{volume}{55}
  (\bibinfo{year}{2022}) \bibinfo{pages}{102735}.
\bibitem[{Drissi-Daoudi et~al.(2022)Drissi-Daoudi, Pandiyan, Log{\'e},
  Shevchik, Masinelli, Ghasemi-Tabasi, Parrilli, and
  Wasmer}]{drissi2022differentiation}
\bibinfo{author}{R.~Drissi-Daoudi}, \bibinfo{author}{V.~Pandiyan},
  \bibinfo{author}{R.~Log{\'e}}, \bibinfo{author}{S.~Shevchik},
  \bibinfo{author}{G.~Masinelli}, \bibinfo{author}{H.~Ghasemi-Tabasi},
  \bibinfo{author}{A.~Parrilli}, \bibinfo{author}{K.~Wasmer},
\newblock \bibinfo{title}{Differentiation of materials and laser powder bed
  fusion processing regimes from airborne acoustic emission combined with
  machine learning},
\newblock \bibinfo{journal}{Virtual and Physical Prototyping}
  \bibinfo{volume}{17} (\bibinfo{year}{2022}) \bibinfo{pages}{181--204}.
\bibitem[{Pandiyan et~al.(2020)Pandiyan, Drissi-Daoudi, Shevchik, Masinelli,
  Log{\'e}, and Wasmer}]{pandiyan2020analysis}
\bibinfo{author}{V.~Pandiyan}, \bibinfo{author}{R.~Drissi-Daoudi},
  \bibinfo{author}{S.~Shevchik}, \bibinfo{author}{G.~Masinelli},
  \bibinfo{author}{R.~Log{\'e}}, \bibinfo{author}{K.~Wasmer},
\newblock \bibinfo{title}{Analysis of time, frequency and time-frequency domain
  features from acoustic emissions during laser powder-bed fusion process},
\newblock \bibinfo{journal}{Procedia CIRP} \bibinfo{volume}{94}
  (\bibinfo{year}{2020}) \bibinfo{pages}{392--397}.
\bibitem[{Pandiyan et~al.(2021)Pandiyan, Drissi-Daoudi, Shevchik, Masinelli,
  Le-Quang, Log{\'e}, and Wasmer}]{pandiyan2021semi}
\bibinfo{author}{V.~Pandiyan}, \bibinfo{author}{R.~Drissi-Daoudi},
  \bibinfo{author}{S.~Shevchik}, \bibinfo{author}{G.~Masinelli},
  \bibinfo{author}{T.~Le-Quang}, \bibinfo{author}{R.~Log{\'e}},
  \bibinfo{author}{K.~Wasmer},
\newblock \bibinfo{title}{Semi-supervised monitoring of laser powder bed fusion
  process based on acoustic emissions},
\newblock \bibinfo{journal}{Virtual and Physical Prototyping}
  \bibinfo{volume}{16} (\bibinfo{year}{2021}) \bibinfo{pages}{481--497}.
\bibitem[{Taheri et~al.(2019)Taheri, Koester, Bigelow, Faierson, and
  Bond}]{taheri2019situ}
\bibinfo{author}{H.~Taheri}, \bibinfo{author}{L.~W. Koester},
  \bibinfo{author}{T.~A. Bigelow}, \bibinfo{author}{E.~J. Faierson},
  \bibinfo{author}{L.~J. Bond},
\newblock \bibinfo{title}{In situ additive manufacturing process monitoring
  with an acoustic technique: Clustering performance evaluation using k-means
  algorithm},
\newblock \bibinfo{journal}{Journal of Manufacturing Science and Engineering}
  \bibinfo{volume}{141} (\bibinfo{year}{2019}) \bibinfo{pages}{041011}.
\bibitem[{Wasmer et~al.(2018)Wasmer, Kenel, Leinenbach, and
  Shevchik}]{wasmer2018situ}
\bibinfo{author}{K.~Wasmer}, \bibinfo{author}{C.~Kenel},
  \bibinfo{author}{C.~Leinenbach}, \bibinfo{author}{S.~Shevchik},
\newblock \bibinfo{title}{In situ and real-time monitoring of powder-bed am by
  combining acoustic emission and artificial intelligence},
\newblock in: \bibinfo{booktitle}{Industrializing Additive
  Manufacturing-Proceedings of Additive Manufacturing in Products and
  Applications-AMPA2017 1}, \bibinfo{organization}{Springer},
  \bibinfo{year}{2018}, pp. \bibinfo{pages}{200--209}.
\bibitem[{Wu et~al.(2016)Wu, Wang, and Yu}]{wu2016situ}
\bibinfo{author}{H.~Wu}, \bibinfo{author}{Y.~Wang}, \bibinfo{author}{Z.~Yu},
\newblock \bibinfo{title}{In situ monitoring of fdm machine condition via
  acoustic emission},
\newblock \bibinfo{journal}{The International Journal of Advanced Manufacturing
  Technology} \bibinfo{volume}{84} (\bibinfo{year}{2016})
  \bibinfo{pages}{1483--1495}.
\bibitem[{Hossain and Taheri(2020)}]{hossain2020situ}
\bibinfo{author}{M.~S. Hossain}, \bibinfo{author}{H.~Taheri},
\newblock \bibinfo{title}{In situ process monitoring for additive manufacturing
  through acoustic techniques},
\newblock \bibinfo{journal}{Journal of Materials Engineering and Performance}
  \bibinfo{volume}{29} (\bibinfo{year}{2020}) \bibinfo{pages}{6249--6262}.
\bibitem[{Gillespie et~al.(2021)Gillespie, Yeoh, Zhao, Parab, Sun, Rollett,
  Lan, and Kube}]{gillespie2021situ}
\bibinfo{author}{J.~Gillespie}, \bibinfo{author}{W.~Y. Yeoh},
  \bibinfo{author}{C.~Zhao}, \bibinfo{author}{N.~D. Parab},
  \bibinfo{author}{T.~Sun}, \bibinfo{author}{A.~D. Rollett},
  \bibinfo{author}{B.~Lan}, \bibinfo{author}{C.~M. Kube},
\newblock \bibinfo{title}{In situ characterization of laser-generated melt
  pools using synchronized ultrasound and high-speed x-ray imaging},
\newblock \bibinfo{journal}{The Journal of the Acoustical Society of America}
  \bibinfo{volume}{150} (\bibinfo{year}{2021}) \bibinfo{pages}{2409--2420}.
\bibitem[{Tan et~al.(2020)Tan, Fang, Li, Liu, Zhu, and Yang}]{tan2020neural}
\bibinfo{author}{Z.~Tan}, \bibinfo{author}{Q.~Fang}, \bibinfo{author}{H.~Li},
  \bibinfo{author}{S.~Liu}, \bibinfo{author}{W.~Zhu},
  \bibinfo{author}{D.~Yang},
\newblock \bibinfo{title}{Neural network based image segmentation for spatter
  extraction during laser-based powder bed fusion processing},
\newblock \bibinfo{journal}{Optics \& Laser Technology} \bibinfo{volume}{130}
  (\bibinfo{year}{2020}) \bibinfo{pages}{106347}.
\bibitem[{Fang et~al.(2022)Fang, Cheng, Glerum, Bennett, Cao, and
  Wagner}]{fang2022data}
\bibinfo{author}{L.~Fang}, \bibinfo{author}{L.~Cheng}, \bibinfo{author}{J.~A.
  Glerum}, \bibinfo{author}{J.~Bennett}, \bibinfo{author}{J.~Cao},
  \bibinfo{author}{G.~J. Wagner},
\newblock \bibinfo{title}{Data-driven analysis of process, structure, and
  properties of additively manufactured inconel 718 thin walls},
\newblock \bibinfo{journal}{npj Computational Materials} \bibinfo{volume}{8}
  (\bibinfo{year}{2022}) \bibinfo{pages}{126}.
\bibitem[{Demir et~al.(2021)Demir, Zhang, Ben-Artzy, Hosemann, and
  Gu}]{demir2021laser}
\bibinfo{author}{K.~Demir}, \bibinfo{author}{Z.~Zhang},
  \bibinfo{author}{A.~Ben-Artzy}, \bibinfo{author}{P.~Hosemann},
  \bibinfo{author}{G.~X. Gu},
\newblock \bibinfo{title}{Laser scan strategy descriptor for defect prognosis
  in metal additive manufacturing using neural networks},
\newblock \bibinfo{journal}{Journal of Manufacturing Processes}
  \bibinfo{volume}{67} (\bibinfo{year}{2021}) \bibinfo{pages}{628--634}.
\bibitem[{Qin et~al.(2022)Qin, Hu, Liu, Witherell, Wang, Rosen, Simpson, Lu,
  and Tang}]{qin2022research}
\bibinfo{author}{J.~Qin}, \bibinfo{author}{F.~Hu}, \bibinfo{author}{Y.~Liu},
  \bibinfo{author}{P.~Witherell}, \bibinfo{author}{C.~C. Wang},
  \bibinfo{author}{D.~W. Rosen}, \bibinfo{author}{T.~W. Simpson},
  \bibinfo{author}{Y.~Lu}, \bibinfo{author}{Q.~Tang},
\newblock \bibinfo{title}{Research and application of machine learning for
  additive manufacturing},
\newblock \bibinfo{journal}{Additive Manufacturing} \bibinfo{volume}{52}
  (\bibinfo{year}{2022}) \bibinfo{pages}{102691}.
\bibitem[{Doh et~al.(2022)Doh, Raju, Raghavan, Rosen, and
  Kim}]{doh2022bayesian}
\bibinfo{author}{J.~Doh}, \bibinfo{author}{N.~Raju},
  \bibinfo{author}{N.~Raghavan}, \bibinfo{author}{D.~W. Rosen},
  \bibinfo{author}{S.~Kim},
\newblock \bibinfo{title}{Bayesian inference-based decision of fatigue life
  model for metal additive manufacturing considering effects of build
  orientation and post-processing},
\newblock \bibinfo{journal}{International Journal of Fatigue}
  \bibinfo{volume}{155} (\bibinfo{year}{2022}) \bibinfo{pages}{106535}.
\bibitem[{Rohe et~al.(2021)Rohe, Stoll, Hildebrand, Reimann, and
  Bergmann}]{rohe2021detecting}
\bibinfo{author}{M.~Rohe}, \bibinfo{author}{B.~N. Stoll},
  \bibinfo{author}{J.~Hildebrand}, \bibinfo{author}{J.~Reimann},
  \bibinfo{author}{J.~P. Bergmann},
\newblock \bibinfo{title}{Detecting process anomalies in the gmaw process by
  acoustic sensing with a convolutional neural network (cnn) for
  classification},
\newblock \bibinfo{journal}{Journal of Manufacturing and Materials Processing}
  \bibinfo{volume}{5} (\bibinfo{year}{2021}) \bibinfo{pages}{135}.
\bibitem[{Acevedo et~al.(2020)Acevedo, Sedlak, Kolman, and
  Fredel}]{acevedo2020residual}
\bibinfo{author}{R.~Acevedo}, \bibinfo{author}{P.~Sedlak},
  \bibinfo{author}{R.~Kolman}, \bibinfo{author}{M.~Fredel},
\newblock \bibinfo{title}{Residual stress analysis of additive manufacturing of
  metallic parts using ultrasonic waves: State of the art review},
\newblock \bibinfo{journal}{Journal of Materials Research and Technology}
  \bibinfo{volume}{9} (\bibinfo{year}{2020}) \bibinfo{pages}{9457--9477}.
\bibitem[{Smoqi et~al.(2022)Smoqi, Gaikwad, Bevans, Kobir, Craig, Abul-Haj,
  Peralta, and Rao}]{smoqi2022monitoring}
\bibinfo{author}{Z.~Smoqi}, \bibinfo{author}{A.~Gaikwad},
  \bibinfo{author}{B.~Bevans}, \bibinfo{author}{M.~H. Kobir},
  \bibinfo{author}{J.~Craig}, \bibinfo{author}{A.~Abul-Haj},
  \bibinfo{author}{A.~Peralta}, \bibinfo{author}{P.~Rao},
\newblock \bibinfo{title}{Monitoring and prediction of porosity in laser powder
  bed fusion using physics-informed meltpool signatures and machine learning},
\newblock \bibinfo{journal}{Journal of Materials Processing Technology}
  \bibinfo{volume}{304} (\bibinfo{year}{2022}) \bibinfo{pages}{117550}.
\bibitem[{Liu et~al.(2021)Liu, Liu, and Zhang}]{liu2021physics}
\bibinfo{author}{R.~Liu}, \bibinfo{author}{S.~Liu}, \bibinfo{author}{X.~Zhang},
\newblock \bibinfo{title}{A physics-informed machine learning model for
  porosity analysis in laser powder bed fusion additive manufacturing},
\newblock \bibinfo{journal}{The International Journal of Advanced Manufacturing
  Technology} \bibinfo{volume}{113} (\bibinfo{year}{2021})
  \bibinfo{pages}{1943--1958}.
\bibitem[{Fang et~al.(2021)Fang, Tan, Li, Shen, Liu, Song, Zhou, Yang, and
  Wen}]{fang2021situ}
\bibinfo{author}{Q.~Fang}, \bibinfo{author}{Z.~Tan}, \bibinfo{author}{H.~Li},
  \bibinfo{author}{S.~Shen}, \bibinfo{author}{S.~Liu},
  \bibinfo{author}{C.~Song}, \bibinfo{author}{X.~Zhou},
  \bibinfo{author}{Y.~Yang}, \bibinfo{author}{S.~Wen},
\newblock \bibinfo{title}{In-situ capture of melt pool signature in selective
  laser melting using u-net-based convolutional neural network},
\newblock \bibinfo{journal}{Journal of Manufacturing Processes}
  \bibinfo{volume}{68} (\bibinfo{year}{2021}) \bibinfo{pages}{347--355}.
\bibitem[{Pandiyan et~al.(2022)Pandiyan, Drissi-Daoudi, Shevchik, Masinelli,
  Le-Quang, Loge, and Wasmer}]{pandiyan2022deep}
\bibinfo{author}{V.~Pandiyan}, \bibinfo{author}{R.~Drissi-Daoudi},
  \bibinfo{author}{S.~Shevchik}, \bibinfo{author}{G.~Masinelli},
  \bibinfo{author}{T.~Le-Quang}, \bibinfo{author}{R.~Loge},
  \bibinfo{author}{K.~Wasmer},
\newblock \bibinfo{title}{Deep transfer learning of additive manufacturing
  mechanisms across materials in metal-based laser powder bed fusion process},
\newblock \bibinfo{journal}{Journal of Materials Processing Technology}
  \bibinfo{volume}{303} (\bibinfo{year}{2022}) \bibinfo{pages}{117531}.
\bibitem[{Shevchik et~al.(2018)Shevchik, Kenel, Leinenbach, and
  Wasmer}]{shevchik2018acoustic}
\bibinfo{author}{S.~A. Shevchik}, \bibinfo{author}{C.~Kenel},
  \bibinfo{author}{C.~Leinenbach}, \bibinfo{author}{K.~Wasmer},
\newblock \bibinfo{title}{Acoustic emission for in situ quality monitoring in
  additive manufacturing using spectral convolutional neural networks},
\newblock \bibinfo{journal}{Additive Manufacturing} \bibinfo{volume}{21}
  (\bibinfo{year}{2018}) \bibinfo{pages}{598--604}.
\bibitem[{Gordon et~al.(2020)Gordon, Narra, Cunningham, Liu, Chen, Suter,
  Beuth, and Rollett}]{GORDON2020101552}
\bibinfo{author}{J.~V. Gordon}, \bibinfo{author}{S.~P. Narra},
  \bibinfo{author}{R.~W. Cunningham}, \bibinfo{author}{H.~Liu},
  \bibinfo{author}{H.~Chen}, \bibinfo{author}{R.~M. Suter},
  \bibinfo{author}{J.~L. Beuth}, \bibinfo{author}{A.~D. Rollett},
\newblock \bibinfo{title}{Defect structure process maps for laser powder bed
  fusion additive manufacturing},
\newblock \bibinfo{journal}{Additive Manufacturing} \bibinfo{volume}{36}
  (\bibinfo{year}{2020}) \bibinfo{pages}{101552}.
  \DOIprefix\doi{https://doi.org/10.1016/j.addma.2020.101552}.
\bibitem[{Arts and van~den Broek(2022)}]{arts2022fast}
\bibinfo{author}{L.~P. Arts}, \bibinfo{author}{E.~L. van~den Broek},
\newblock \bibinfo{title}{The fast continuous wavelet transformation (fcwt) for
  real-time, high-quality, noise-resistant time--frequency analysis},
\newblock \bibinfo{journal}{Nature Computational Science} \bibinfo{volume}{2}
  (\bibinfo{year}{2022}) \bibinfo{pages}{47--58}.
\bibitem[{Schubert et~al.(2017)Schubert, Sander, Ester, Kriegel, and
  Xu}]{schubert2017dbscan}
\bibinfo{author}{E.~Schubert}, \bibinfo{author}{J.~Sander},
  \bibinfo{author}{M.~Ester}, \bibinfo{author}{H.~P. Kriegel},
  \bibinfo{author}{X.~Xu},
\newblock \bibinfo{title}{Dbscan revisited, revisited: why and how you should
  (still) use dbscan},
\newblock \bibinfo{journal}{ACM Transactions on Database Systems (TODS)}
  \bibinfo{volume}{42} (\bibinfo{year}{2017}) \bibinfo{pages}{1--21}.
\bibitem[{Maybeck(1990)}]{maybeck1990kalman}
\bibinfo{author}{P.~S. Maybeck},
\newblock \bibinfo{title}{The kalman filter: An introduction to concepts},
\newblock in: \bibinfo{booktitle}{Autonomous robot vehicles},
  \bibinfo{publisher}{Springer}, \bibinfo{year}{1990}, pp.
  \bibinfo{pages}{194--204}.
\bibitem[{Kolouri et~al.(2018)Kolouri, Rohde, and Hoffmann}]{kolouri2018sliced}
\bibinfo{author}{S.~Kolouri}, \bibinfo{author}{G.~K. Rohde},
  \bibinfo{author}{H.~Hoffmann},
\newblock \bibinfo{title}{Sliced wasserstein distance for learning gaussian
  mixture models},
\newblock in: \bibinfo{booktitle}{Proceedings of the IEEE Conference on
  Computer Vision and Pattern Recognition}, \bibinfo{year}{2018}, pp.
  \bibinfo{pages}{3427--3436}.
\bibitem[{Ronneberger et~al.(2015)Ronneberger, Fischer, and
  Brox}]{ronneberger2015u}
\bibinfo{author}{O.~Ronneberger}, \bibinfo{author}{P.~Fischer},
  \bibinfo{author}{T.~Brox},
\newblock \bibinfo{title}{U-net: Convolutional networks for biomedical image
  segmentation},
\newblock in: \bibinfo{booktitle}{Medical Image Computing and Computer-Assisted
  Intervention--MICCAI 2015: 18th International Conference, Munich, Germany,
  October 5-9, 2015, Proceedings, Part III 18},
  \bibinfo{organization}{Springer}, \bibinfo{year}{2015}, pp.
  \bibinfo{pages}{234--241}.
\bibitem[{Isola et~al.(2017)Isola, Zhu, Zhou, and Efros}]{isola2017image}
\bibinfo{author}{P.~Isola}, \bibinfo{author}{J.-Y. Zhu},
  \bibinfo{author}{T.~Zhou}, \bibinfo{author}{A.~A. Efros},
\newblock \bibinfo{title}{Image-to-image translation with conditional
  adversarial networks},
\newblock in: \bibinfo{booktitle}{Proceedings of the IEEE conference on
  computer vision and pattern recognition}, \bibinfo{year}{2017}, pp.
  \bibinfo{pages}{1125--1134}.
\bibitem[{Tang et~al.(2017)Tang, Pistorius, and Beuth}]{tang2017prediction}
\bibinfo{author}{M.~Tang}, \bibinfo{author}{P.~C. Pistorius},
  \bibinfo{author}{J.~L. Beuth},
\newblock \bibinfo{title}{Prediction of lack-of-fusion porosity for powder bed
  fusion},
\newblock \bibinfo{journal}{Additive Manufacturing} \bibinfo{volume}{14}
  (\bibinfo{year}{2017}) \bibinfo{pages}{39--48}.
\bibitem[{Scime and Beuth(2019)}]{scime2019melt}
\bibinfo{author}{L.~Scime}, \bibinfo{author}{J.~Beuth},
\newblock \bibinfo{title}{Melt pool geometry and morphology variability for the
  inconel 718 alloy in a laser powder bed fusion additive manufacturing
  process},
\newblock \bibinfo{journal}{Additive Manufacturing} \bibinfo{volume}{29}
  (\bibinfo{year}{2019}) \bibinfo{pages}{100830}.
\bibitem[{Virtanen et~al.(2020)Virtanen, Gommers, Oliphant, Haberland, Reddy,
  Cournapeau, Burovski, Peterson, Weckesser, Bright, {van der Walt}, Brett,
  Wilson, Millman, Mayorov, Nelson, Jones, Kern, Larson, Carey, Polat, Feng,
  Moore, {VanderPlas}, Laxalde, Perktold, Cimrman, Henriksen, Quintero, Harris,
  Archibald, Ribeiro, Pedregosa, {van Mulbregt}, and {SciPy 1.0
  Contributors}}]{2020SciPy-NMeth}
\bibinfo{author}{P.~Virtanen}, \bibinfo{author}{R.~Gommers},
  \bibinfo{author}{T.~E. Oliphant}, \bibinfo{author}{M.~Haberland},
  \bibinfo{author}{T.~Reddy}, \bibinfo{author}{D.~Cournapeau},
  \bibinfo{author}{E.~Burovski}, \bibinfo{author}{P.~Peterson},
  \bibinfo{author}{W.~Weckesser}, \bibinfo{author}{J.~Bright},
  \bibinfo{author}{S.~J. {van der Walt}}, \bibinfo{author}{M.~Brett},
  \bibinfo{author}{J.~Wilson}, \bibinfo{author}{K.~J. Millman},
  \bibinfo{author}{N.~Mayorov}, \bibinfo{author}{A.~R.~J. Nelson},
  \bibinfo{author}{E.~Jones}, \bibinfo{author}{R.~Kern},
  \bibinfo{author}{E.~Larson}, \bibinfo{author}{C.~J. Carey},
  \bibinfo{author}{{\.I}.~Polat}, \bibinfo{author}{Y.~Feng},
  \bibinfo{author}{E.~W. Moore}, \bibinfo{author}{J.~{VanderPlas}},
  \bibinfo{author}{D.~Laxalde}, \bibinfo{author}{J.~Perktold},
  \bibinfo{author}{R.~Cimrman}, \bibinfo{author}{I.~Henriksen},
  \bibinfo{author}{E.~A. Quintero}, \bibinfo{author}{C.~R. Harris},
  \bibinfo{author}{A.~M. Archibald}, \bibinfo{author}{A.~H. Ribeiro},
  \bibinfo{author}{F.~Pedregosa}, \bibinfo{author}{P.~{van Mulbregt}},
  \bibinfo{author}{{SciPy 1.0 Contributors}},
\newblock \bibinfo{title}{{{SciPy} 1.0: Fundamental Algorithms for Scientific
  Computing in Python}},
\newblock \bibinfo{journal}{Nature Methods} \bibinfo{volume}{17}
  (\bibinfo{year}{2020}) \bibinfo{pages}{261--272}.
  \DOIprefix\doi{10.1038/s41592-019-0686-2}.
\bibitem[{Huang et~al.(2022)Huang, Fleming, Clark, Marussi, Fezzaa,
  Thiyagalingam, Leung, and Lee}]{huang2022keyhole}
\bibinfo{author}{Y.~Huang}, \bibinfo{author}{T.~G. Fleming},
  \bibinfo{author}{S.~J. Clark}, \bibinfo{author}{S.~Marussi},
  \bibinfo{author}{K.~Fezzaa}, \bibinfo{author}{J.~Thiyagalingam},
  \bibinfo{author}{C.~L.~A. Leung}, \bibinfo{author}{P.~D. Lee},
\newblock \bibinfo{title}{Keyhole fluctuation and pore formation mechanisms
  during laser powder bed fusion additive manufacturing},
\newblock \bibinfo{journal}{Nature Communications} \bibinfo{volume}{13}
  (\bibinfo{year}{2022}) \bibinfo{pages}{1170}.
\bibitem[{Caprio et~al.(2020)Caprio, Demir, and
  Previtali}]{caprio2020observing}
\bibinfo{author}{L.~Caprio}, \bibinfo{author}{A.~G. Demir},
  \bibinfo{author}{B.~Previtali},
\newblock \bibinfo{title}{Observing molten pool surface oscillations during
  keyhole processing in laser powder bed fusion as a novel method to estimate
  the penetration depth},
\newblock \bibinfo{journal}{Additive Manufacturing} \bibinfo{volume}{36}
  (\bibinfo{year}{2020}) \bibinfo{pages}{101470}.
\bibitem[{Pyeon et~al.(2021)Pyeon, Aroh, Jiang, Verma, Gould, Ramlatchan,
  Fezzaa, Parab, Zhao, Sun et~al.}]{pyeon2021time}
\bibinfo{author}{J.~Pyeon}, \bibinfo{author}{J.~Aroh},
  \bibinfo{author}{R.~Jiang}, \bibinfo{author}{A.~K. Verma},
  \bibinfo{author}{B.~Gould}, \bibinfo{author}{A.~Ramlatchan},
  \bibinfo{author}{K.~Fezzaa}, \bibinfo{author}{N.~Parab},
  \bibinfo{author}{C.~Zhao}, \bibinfo{author}{T.~Sun}, et~al.,
\newblock \bibinfo{title}{Time-resolved geometric feature tracking elucidates
  laser-induced keyhole dynamics},
\newblock \bibinfo{journal}{Integrating Materials and Manufacturing Innovation}
  \bibinfo{volume}{10} (\bibinfo{year}{2021}) \bibinfo{pages}{677--688}.
\bibitem[{Bitharas et~al.(2022)Bitharas, Parab, Zhao, Sun, Rollett, and
  Moore}]{bitharas2022interplay}
\bibinfo{author}{I.~Bitharas}, \bibinfo{author}{N.~Parab},
  \bibinfo{author}{C.~Zhao}, \bibinfo{author}{T.~Sun},
  \bibinfo{author}{A.~Rollett}, \bibinfo{author}{A.~Moore},
\newblock \bibinfo{title}{The interplay between vapour, liquid, and solid
  phases in laser powder bed fusion},
\newblock \bibinfo{journal}{Nature Communications} \bibinfo{volume}{13}
  (\bibinfo{year}{2022}) \bibinfo{pages}{2959}.

\end{thebibliography}
\clearpage

\appendix

\section{Algorithm of density-based spatial clustering of applications with noise (DBSCAN) image segmentation}
\label{appendix:dbscan}
The usage of DBSCAN method in our work is described in Sec.~\ref{subsubsec:segmentation} and Fig.~\ref{fig:straighten}. In this section, we introduce additional details on the DBSCAN clustering algorithm implementation. The pseudo-code of the entire DBSCAN algorithm is described as Algorithm~\ref{alg:dbscan}. 

\begin{algorithm}[!t]
    \SetKwInOut{Input}{Input}
    \SetKwInOut{Output}{Output}
    
    \Input{$U(p)$, $Q$, $B$, $\sigma$, $m(x,y)$, $\epsilon$, $\mathrm{minPts}$. }
    
    \Output{Pixel clusters $S_c(p) (c\in [0,C])$}

    \Begin{
        Set $\sigma \gets 0.8$, $\epsilon \gets 2$, $\mathrm{minPts} \gets 25$, $C\gets 0$, $Q \gets \emptyset$;

        Set $m(x,y) \coloneqq \| x-y\|_2^2$;

        Set subset $v=\{p|\sigma\leq U(p)\leq 1\}\subseteq U$;
        
        Set $\mathrm{label}(p\in v) \gets \mathrm{undefined}$; 
 
        \ForEach{$p_a\in v$}{
            \If{$\mathrm{label}(p_b)\neq \mathrm{undefined}$}{
                    Continue;
            }
                
            Set $B \gets \emptyset$;
            
            \ForEach{$p_b\in v, p_b\neq p_a$}{
                \If{$m(p_b,p_a)\leq\epsilon$}{
                    $B\coloneqq B\cup \{p_b\}$;
                }
            }
            
            \If{$|B|<\mathrm{minPts}$}{
                $\mathrm{label}(p_a)\gets\mathrm{Noise}$;
                
                Continue;
            }
            
            $C\coloneqq C+1$;
            
            $\mathrm{label}(p)\gets C$;
            
            $Q\coloneqq B$;
            
            \ForEach{$p_q\in Q$}{
                Set $B\gets\emptyset$;

                \eIf{$\mathrm{label}(p_q)\neq\mathrm{undefined}$ and $\mathrm{label}(p_q)\neq\mathrm{Noise}$}{
                    Continue;
                }{
                    $\mathrm{label}(p_q)\gets C$;
                }
                
                \ForEach{$p_r\in v, p_r\neq p_q$}{
                    \If{$m(p_r,p_q)\leq\epsilon$}{
                        $B\coloneqq B\cup \{p_r\}$;
                    }
                }
                
                \If{$|B|\geq \mathrm{minPts}$}{
                    $Q\coloneqq Q\cup B$;
                }
            }
        }

        \ForEach{$c=0:C$}{
            Set $S_c(p)\gets\emptyset$;
            
            \ForEach{$p\in v$}{
                \If{$\mathrm{label}(p)==c$}{
                    $S_c(p)\coloneqq S_c(p)\cup \{p\}$;
                }
            }
        }
    }
    \caption{DBSCAN clustering}
    \label{alg:dbscan}
\end{algorithm}

\section{Kalman filter melt pool tracking}
\label{appendix:kalman}
In this section, we introduce how we track the position of the melt pool in a consecutive sequence of high-speed melt pool images. 

After implementing DBSCAN clustering for the first high-speed image frame of a scanline, we locate the melt pool by selecting the largest pixel cluster in the image. Then, we calculate the center point of the melt pool pixel cluster and set it as the initial melt pool center location. Since we explicitly know the scanning velocity and orientation of the scanline, we can build a linear ordinary Kalman filter model and calculate the melt pool location estimate together with its covariance matrix. This helps us robustly track the melt pool location in the image frame and avoid detecting large spatters as melt pools. Algorithm~\ref{alg:kalman} shows the details of our Kalman filter implementation in $k\textsuperscript{th}$ scanline.

\begin{algorithm}[!t]
    \SetKwInOut{Input}{Input}
    \SetKwInOut{Output}{Output}
    
    \Input{Melt pool position \& velocity vector $c_M\left(x_{k-1}, y_{k-1}, u_{k-1}, v_{k-1}\right)$, covariance matrix $P_{k-1}$, prediction model $F$, prediction noise $J$, measurement (observation) $z_k$, measurement (sensor) model $H$, measurement noise $R$, initial melt pool centroid $c_0$, total time steps $K$. }
    
    \Output{$c_M\left(x_{k}, y_{k}, u_{k}, v_{k}\right)$, $P_k$. }

    \Begin{
        Set $F = \begin{bmatrix}
          1 & 0 & \frac{1}{f_I} & 0 \\
          0 & 1 & 0 & \frac{1}{f_I} \\
          0 & 0 & 1 & 0 \\
          0 & 0 & 0 & 1 
          \end{bmatrix}$;

        Set $J = R = 0.01\cdot I_{\mathrm{4}\times\mathrm{4}}$;

        Set $H = I_{\mathrm{4}\times\mathrm{4}}$;
        
        \ForEach{$k=1:K$}{
            \If{$k==1$}{
                $P_{k-1} = 0.01\cdot I_{\mathrm{4}\times\mathrm{4}}$;

                $c_M\left(x_{k-1}, y_{k-1}, u_{k-1}, v_{k-1}\right) \gets c_0$;  
            }

            $\hat{c_M}\left(x_k, y_k, u_k, v_k\right) = F\cdot c_M\left(x_{k-1}, y_{k-1}, u_{k-1}, v_{k-1}\right)$; \Comment{Estimate new melt pool position and speed. }

            $\hat{P_k}\left(\hat{c_M}\right) = F\cdot P_{k-1}F_k^{\top} + J$; \Comment{Update the covariance of the prediction. }

            $K = P_k\left(\hat{c_M}\right)\cdot H^{\top}\left(H\hat{P_k}H^{\top} + R\right)^{-1}$; \Comment{Calculate the Kalman gain of this time step. }

            $c_M\left(x_k, y_k. u_k. v_k\right) = \hat{c_M}\left(x_k, y_k, u_k, v_k\right) + K\cdot\left(\overrightarrow{z_k}-H\hat{c_M}\left(x_k, y_k, u_k, v_k\right)\right)$; \Comment{Update $\hat{c_M}$ with the Kalman gain and measurements. }
    
            $P_k\left(c_M\right) = \hat{P_k}\left(\hat{c_M}\right) - K\cdot H\cdot P_k\left(\hat{c_M}\right)$; \Comment{Update the covariance of melt pool position \& speed. }
        }
    }
    \caption{Kalman filter melt pool tracking}
    \label{alg:kalman}
\end{algorithm}

In Algorithm~\ref{alg:kalman}, $x_k$, $y_k$, $u_k$ and $v_k$ denote $x$ coordinate, $y$ coordinate, velocity along with $x$-axis and velocity along with $y$-axis, respectively. The linear prediction model and the measurement model are assumed consistent throughout the optimization process. Algorithm~\ref{alg:kalman} ensures the accuracy of the melt pool position tracking that can increase the approach's robustness on high-speed image data processing. 

\section{\textit{Q} calculation between melt pool high-speed images}
\label{appendix:wasserstein}

In Sec.~\ref{subsec:evaluation_metric}, we propose a new physical quantity-based distance metric to assess the similarity between the ground-truth and prediction melt pool images. In this section, we showcase image comparison examples of high, medium and low $Q$ values to demonstrate that $Q(u,v)$ is sufficient for differentiating similar and dissimilar image couples. 

\begin{figure}[!htb]
\center{\includegraphics[width=\linewidth]
{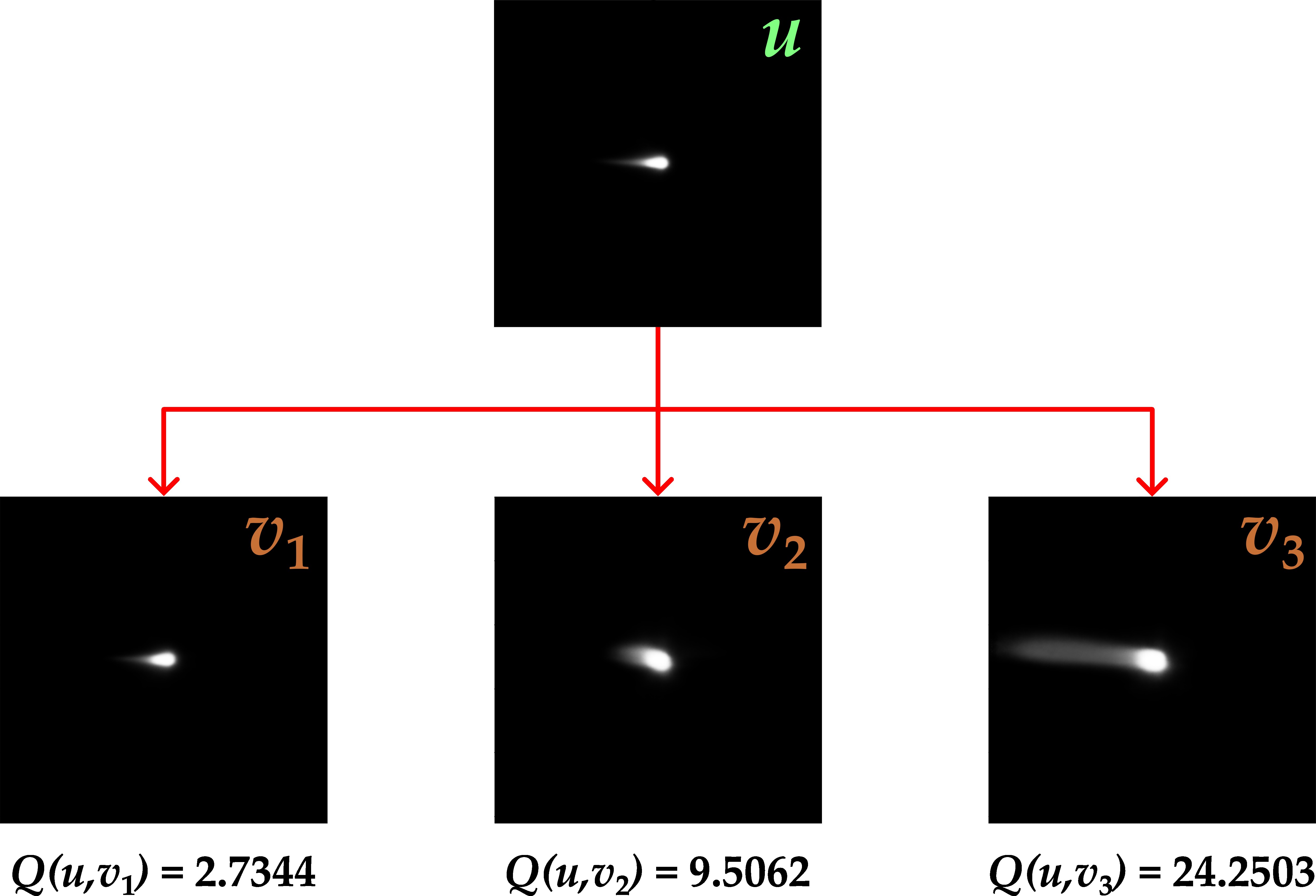}
\caption{Three examples of $Q$ scores demonstrating low (left), medium (middle), and high (right) similarity, respectively. Results reflect that our proposed $Q$ formulation is suitable for assessing the melt pool similarity as well as melt pool image prediction accuracy. }
\label{fig:q_examples}
}
\end{figure}

Fig.~\ref{fig:q_examples} shows that the lowest $Q$ score goes to the most similar image couple while the highest $Q$ score belongs to the most dissimilar image couple. Furthermore, we can set a bar/threshold value as $Q = 10$ such that $Q$ below the threshold indicates high melt pool similarity with respect to melt pool width, length, area, and thermal distribution. 

\section{Proof of $Q(u,v)$ being a metric in melt pool image space}
\label{appendix:metric}
In this section, we provide a mathematical proof of our proposed $Q(u,v)$ formulation being a valid distance metric in the melt pool image space. In particular, we pay attention to the axiom of triangle inequality, which guarantees that a distance measure is unambiguous and generalizable in a metric space. 

Our proposed $Q(u,v)$ similarity metric in Sec.~\ref{subsec:evaluation_metric} is as follows:

\begin{equation}
\label{eqn:q_appendix}
\begin{aligned}[b]
    \resizebox{1.0\linewidth}{!}{$Q(u,v) = \frac{1}{4}\times \left(\sqrt{\|w_0\left(u,v\right)\|_2^2 + \|w_1\left(u,v\right)\|_2^2} + \sqrt{\frac{|A_u-A_v|}{\pi}} + |L_u-L_v| + |W_u-W_v| \right)$}
\end{aligned}
\end{equation}

\noindent where $u$ and $v$ are two arbitrary melt pool images from the melt pool image space $M$, $L$ is the melt pool length, $W$ is the melt pool width, $A$ is the melt pool area, and $w_0$ and $w_1$ denote the sliced Wasserstein distance~\cite{kolouri2018sliced} along with the melt pool length and width axes, respectively. We intend to prove the following proposition:
\begin{proposition}
\label{prop:q}
    $Q(u,v)$ is a metric on the image space $M$ $(u,v\in M)$, which means it satisfies the following three axioms: 
    \begin{enumerate}
        \item (Semi-positivity) $Q(u,v)\geq 0, u,v\in M$;
    
        \item (Symmetry) $Q(u,v) = Q(v,u), u,v\in M$;
    
        \item (Triangle inequality) $Q(u,z)\leq Q(u,v)+Q(v,z), u,v,z\in M$. 
    \end{enumerate}
\end{proposition}

\begin{proof}
From Eqn.~\ref{eqn:q_appendix}, we can observe that the axioms of semi-positivity and symmetry obviously hold. Also, it is apparent that $Q(u,v)=0$ if and only if the length, width, area and thermal distribution of the two melt pools $u$ and $v$ are exactly identical. 

To prove that $Q(u,v)$ satisfies the triangle inequality axiom, we assume there is a metric space $(M, d)$ where $M$ is the aforementioned image space and $d$ is a metric on $M$. We define $d(u,v): M\times M\rightarrow\mathbb{R}$, $u,v\in M$. Apparently, $L_u-L_v$, $W_u-W_v$, $A_u-A_v$, $w_0$ and $w_1$ are all special cases of $d$ and thereby valid metrics on the space $M$. 

We introduce the following lemmas to assist in the proof of our proposition:

\begin{lemma}
\label{lemma:a}
    $\sqrt{d(u,v)}$ is a metric on $M$. 
\end{lemma}

\begin{proof}
    Let $c(u,v)=\sqrt{d(u,v)}$. It is obvious to show that $c(u,v)$ satisfies the semi-positivity and symmetry axioms. To prove that the triangle inequality holds, we find an arbitrary new point $z\in M$, $z\neq u,v$. Then we have:

    \begin{equation}
    \begin{aligned}
        c(u,z) + c(z,v) &= \sqrt{d(u,z)} + \sqrt{d(z,v)} \\
        &= \sqrt{d(u,z)+d(z,v)+2\sqrt{d(u,z)\cdot d(z,v)}} \\
        &\geq \sqrt{d(u,z)+d(z,v)} \\
        &\geq \sqrt{d(u,v)}=c(u,v)
    \end{aligned}
    \end{equation}

    \noindent where $d\geq 0$ comes from the semi-positivity of $d$, and $d(u,z)+d(z,v)\geq d(u,v)$ is the triangle inequality for $d$ being an unambiguous metric on $M$. Therefore, $c(u,v)=\sqrt{d(u,v)}$ satisfies the triangle inequality and is a metric on $M$.
\end{proof}

\begin{lemma}
\label{lemma:b}
    $\left|d(u,v)\right|$ is a metric on $M$.  
\end{lemma}

\begin{proof}
    Similar to the proof of Lemma~\ref{lemma:a}, let $c(u,v)=\left|d(u,v)\right|$. It is again obvious to show that $c(u,v)$ satisfies the axioms of semi-positivity and symmetry. We find an arbitrary new point $z\in M$, $z\neq u,v$ to prove the triangle inequality. Then we have:

    \begin{equation}
    \begin{aligned}
        c(u,z) + c(z,v) &= \left|d(u,z)\right| + \left|d(z,v)\right| \\
        &\geq \left|d(u,z)+d(z,v)\right| && \text{(Triangle inequality of}~d\text{)}\\
        &\geq \left|d(u,v)\right| = c(u,v)
    \end{aligned}
    \end{equation}

    \noindent Therefore, $c(u,v)=\left|d(u,v)\right|$ satisfies the triangle inequality and is a metric on $M$.
\end{proof}

\begin{lemma}
\label{lemma:c}
    (Metric linearity) If there is another metric $f$ on the space $M$ and $f\neq d$, then $\alpha\cdot d(u,v)+\beta\cdot f(u,v)$ is a metric on $M$ ($\alpha, \beta\in\mathbb{R}$).  
\end{lemma}

\begin{proof}
    Let $c(u,v)=\alpha\cdot d(u,v)+\beta\cdot f(u,v)$. It is obvious to show that $c(u,v)$ satisfies the semi-positivity and symmetry axioms. To prove the triangle inequality, we find an arbitrary new point $z\in M$, $z\neq u,v$. Then we have:

    \begin{equation}
    \begin{aligned}
        c(u,z) + c(z,v) &= \alpha d(u,z) + \beta f(u,z) + \alpha d(z,v) + \beta f(z,v) \\
        &= \alpha\left[d(u,z) + d(z,v)\right] + \beta\left[f(u,z) + f(z,v)\right] \\
        &\geq \alpha\cdot d(u,v) + \beta\cdot f(u,v) = c(u,v)
    \end{aligned}
    \end{equation}

    \noindent where $d(u,z) + d(z,v)\geq d(u,v)$ and $f(u,z) + f(z,v)\geq f(u,v)$ come from the triangle inequality property of $d$ and $f$. Therefore, $c(u,v)=\alpha\cdot d(u,v)+\beta\cdot f(u,v)$ satisfies the triangle inequality and is a metric on $M$.
\end{proof}

\begin{lemma}
\label{lemma:d}
    If there is another metric $f$ on the space $M$ and $f\neq d$, then $\sqrt{\left[d(u,v)\right]^2+\left[f(u,v)\right]^2}$ is a metric on $M$.  
\end{lemma}

\begin{proof}
    Let $c(u,v)=\sqrt{\left[d(u,v)\right]^2+\left[f(u,v)\right]^2}$. It is obvious to show that $c(u,v)$ satisfies the semi-positivity and symmetry axioms. To prove that the triangle inequality holds, we find an arbitrary new point $z\in M$, $z\neq u,v$. Then we have:

    \begin{equation}
    \begin{aligned}
        c(u,v) &= \sqrt{\left[d(u,v)\right]^2+\left[f(u,v)\right]^2} \\
        &\leq \sqrt{\left[d(u,z)+d(z,v)\right]^2 + \left[f(u,z)+f(z,v)\right]^2} 
    \end{aligned}
    \end{equation}

    \noindent where $d(u,v)\leq d(u,z)+d(z,v)$, $f(u,v)\leq f(u,z)+f(z,v)$ come from the triangle inequality of $d$ and $f$. Then, according to Minkowski inequality for $L^p$ normal vector space, we further have: 

    \begin{equation}
    \begin{aligned}
        c(u,v) &\leq\sqrt{\left[d(u,z)+d(z,v)\right]^2 + \left[f(u,z)+f(z,v)\right]^2} \\
        &\leq \sqrt{\left[d(u,z)\right]^2+\left[f(u,z)\right]^2} + \sqrt{\left[d(z,v)\right]^2+\left[f(z,v)\right]^2} \\
        &= c(u,z) + c(z,v)
    \end{aligned}
    \end{equation}
    
    \noindent Therefore, $c(u,v)=\sqrt{\left[d(u,v)\right]^2+\left[f(u,v)\right]^2}$ satisfies the triangle inequality and is a metric on $M$.
\end{proof}

Now that all the necessary lemmas have been introduced, we can finish the proof of Proposition~\ref{prop:q}: 

According to Lemma~\ref{lemma:b}, we can easily know that $\left|L_u-L_v\right|$ and $\left|W_u-W_v\right|$ are metrics on the space $M$. 

According to Lemma~\ref{lemma:a}, Lemma~\ref{lemma:b} and Lemma~\ref{lemma:c}, we can easily know that $\sqrt{\frac{\left|A_u-A_v\right|}{\pi}}$ is a metric on the space $M$. 

According to Lemma~\ref{lemma:d}, we can easily know that $\sqrt{\|w_0\left(u,v\right)\|_2^2 + \|w_1\left(u,v\right)\|_2^2}$ is a metric on the space $M$. 

Finally, based on Lemma~\ref{lemma:c}, we can know that $Q(u,v)$ is a metric on the image space $M$, thereby proving Proposition~\ref{prop:q}. 
\end{proof}

Furthermore, considering the physical interpretation of each term in the formulation of $Q(u,v)$, it allows users to include however many and whatever melt pool physical quantities they want as long as the newly introduced quantities are well-weighted and satisfy the metric requirements. It is also worth noting that our proposed $Q$ metric actually represents a characteristic distance between melt pool images $u$ and $v$ that encodes the dissimilarity information based on our selected critical melt pool quantities.  We believe this framework of $Q(u,v)$ can be a useful evaluation metric to assess the melt pool prediction performance of any kind of image-reconstruction-based melt pool characterization method. 

\section{Baseline model details}
\label{appendix:baseline}
\subsection{U-Net}
The U-Net model is a fully convolutional model that takes as input a $64\times64$ image with two grayscale channels (acoustic/photodiode scalograms) and outputs a $64\times64$ melt pool prediction. Input images are passed through two $3\times3$ convolutions with stride and zero-padding of 1, each followed by a ReLU activation and batch normalization. Then $2\times2$ max pooling with stride 2 reduces the image sizes to $32\times32$. The convolutions and pooling steps are performed four times, with each level's convolutions using 16, 32, 64, 128, and 256 channels respectively. Next, the steps are reversed using max un-pooling to bring the image resolution back to $64\times 64$. Skip-connections concatenate feature maps from the outputs of the encoder side of the network to the inputs of the decoder side to preserve fine details. The mean squared error is used as a loss function, and we trained the U-Net for 50 epochs (batch size 100) using the Adam optimizer with a learning rate of 0.01.

\subsection{Pix2Pix}
Pix2Pix is a conditional GAN that translates image to image. The generator of Pix2Pix takes in an input image consisting of a $256\times256$ color image and outputs an image with the exact resolution. The convolution layers use a filter size of 4 with strides size of 2 followed by batch normalization and LeakyReLU activation function. Skip connections are established between the encoder and decoder. Pix2Pix uses a PatchGAN with an output size of $30\times 30\times1$ for the discriminator. The discriminator receives 2 image inputs that are concatenated together. The input images to the discriminator consist of the input image and the target image which it should classify as real. Or the input image and the generated image which it should classify as fake. We trained the Pix2Pix for 7 epochs (batch size 1) using the Adam optimizer with a learning rate of 0.0002.

\section{Theoretical baseline model of melt pool width prediction based on Rosenthal Eqn. }
\label{appendix:LOF_baseline}
According to Tang \textit{et al.}~\cite{tang2017prediction}, melt pool width ($W$) can be estimated by numerically solving the following equation: 

\begin{equation}
\label{eqn:rosenthal_width}
    \epsilon P = \pi k\left(T_m-T_\infty\right)W + \frac{e\pi\rho C\left(T_m-T_\infty\right)VW^2}{8}
\end{equation}

\noindent where $P$ and $V$ are the laser power and laser scanning velocity, respectively. $\epsilon$, $\rho$, $C$, $k$, and $T_m$ denote laser absorptivity, density, specific heat capacity, thermal conductivity, and melting point of Ti-64, respectively. $T_\infty$ denotes the ambient/preheating temperature of the build. The material properties of Ti-64 and related process parameters (other than $P$ and $V$) in Eqn.~\ref{eqn:rosenthal_width} are listed in Tab.~\ref{tab:Ti64_property}. 

\begin{table}[!htb]
    \centering{
    \caption{Ti-6Al-4V metal material properties and related process parameters}
    \vspace{-2mm}
    \resizebox{\linewidth}{!}{
    \begin{tabular}{c|c|c|c|c|c|c}
        \toprule
        \textbf{Prop.} & $\epsilon$ & $\rho~\left[\texttt{kg/}\texttt{m}^3\right]$ & $C~\left[\texttt{J/(kg}\cdot\texttt{K)}\right]$ & $k~\left[\texttt{W/(m}\cdot\texttt{K)}\right]$ & $T_m~\left[\texttt{K}\right]$ & $T_\infty~\left[\texttt{K}\right]$ \\
        \midrule
        \textbf{Val.} & $0.48$ & $4430$ & $526$ & $6.7$ & $1913$ & $308.15$ \\
        \bottomrule
    \end{tabular}}
    \label{tab:Ti64_property}}
\end{table}

Empirically, for LOF-affiliated printing parameters, the obtained $W$ from Eqn.~\ref{eqn:rosenthal_width} is roughly $30\%$ less than the actual melt pool width for most types of alloys (including Ti-64)~\cite{tang2017prediction}. Therefore, we eventually used $W^{*}$ as the theoretical melt pool width, which is calculated as: 

\begin{equation}
    W^{*}=\frac{W}{0.7}
\end{equation}

\section{Additional analysis on constructed latent space}
\label{appendix:additional_latent}

\begin{figure*}[!htb]
\center{\includegraphics[width=\linewidth]
{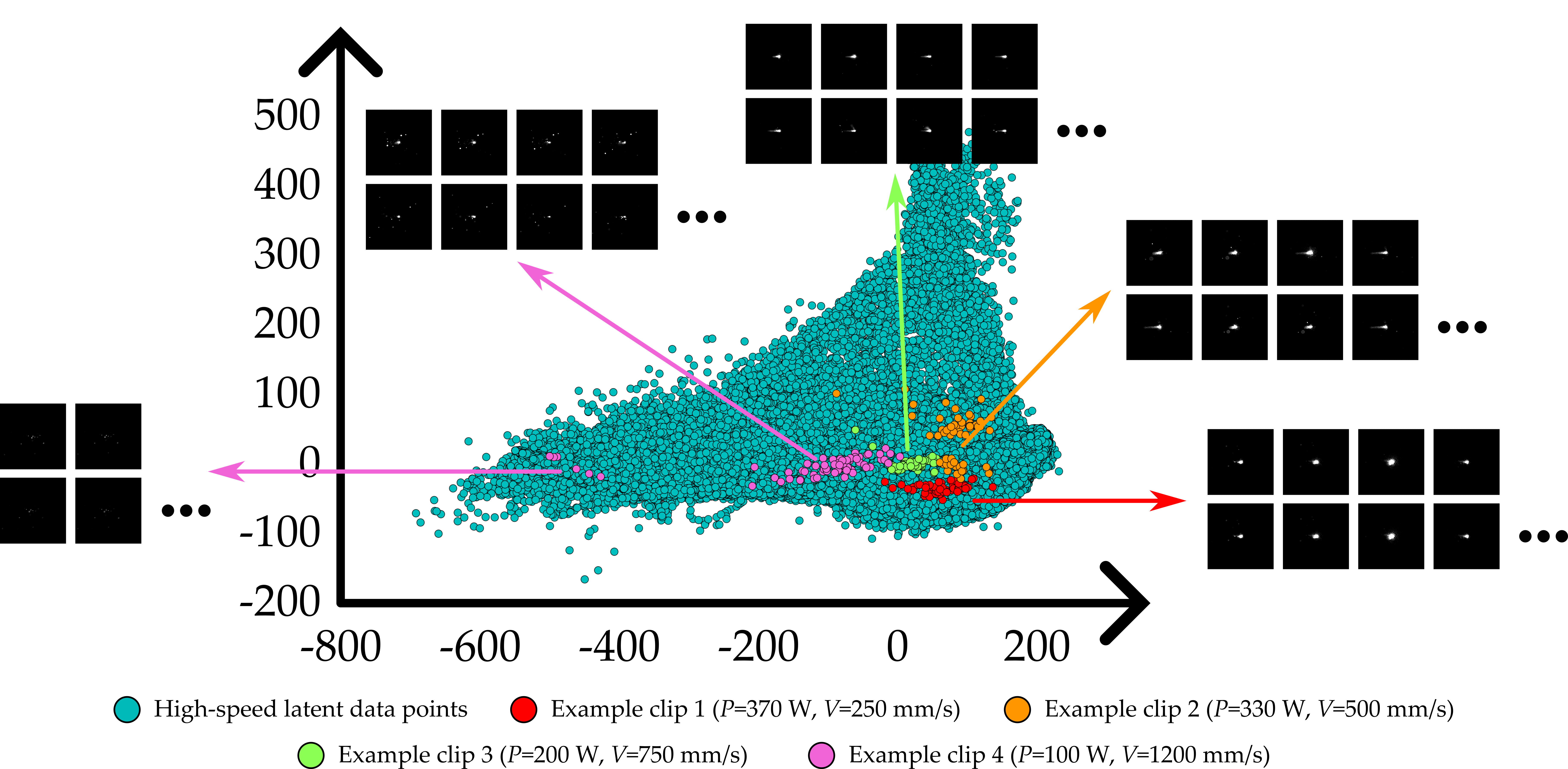}
\caption{A t-SNE plot of the latent space distribution constructed by the trained autoencoder. Highlighted data points showcase examples of specific clips with different printing conditions, demonstrating that similar high-speed melt pool images are close to each other in the constructed latent space. }
\label{fig:ae_latent_space}
}
\end{figure*}

We report latent space construction results in this section. Fig.~\ref{fig:ae_latent_space} shows the t-SNE plot of the embedding distribution in the $4\times 1$ latent space. The result indicates that latent vectors of high-speed images belonging to the same acoustic and photodiode scalograms are also spatially close to each other. This ensures a high accuracy of image averaging, which is critical for the moving average implementation as described in Sec.~\ref{subsubsec:wavelet}. 

\begin{figure*}[!htb]
\center{\includegraphics[width=\linewidth]
{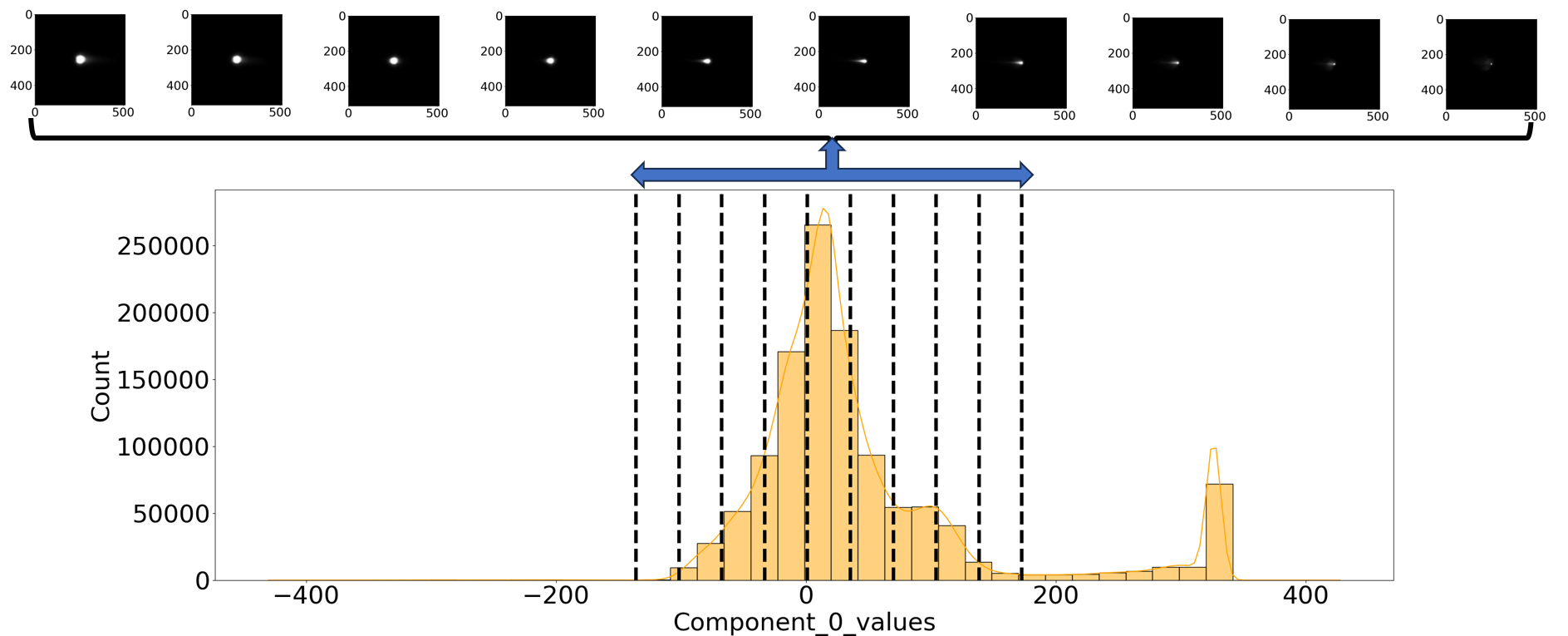}
\caption{Results of disentanglement analysis of the $1^{\mathrm{st}}$ latent vector component (Component 0). Histograms demonstrate the value distribution of Component 0. The value of Component 0 clearly captures the physical quantities, namely the melt pool width and the melt pool area, and shows a linear correlation with them. }
\label{fig:latent_disentangle_0}
}
\end{figure*}

\begin{figure*}[!htb]
\center{\includegraphics[width=\linewidth]
{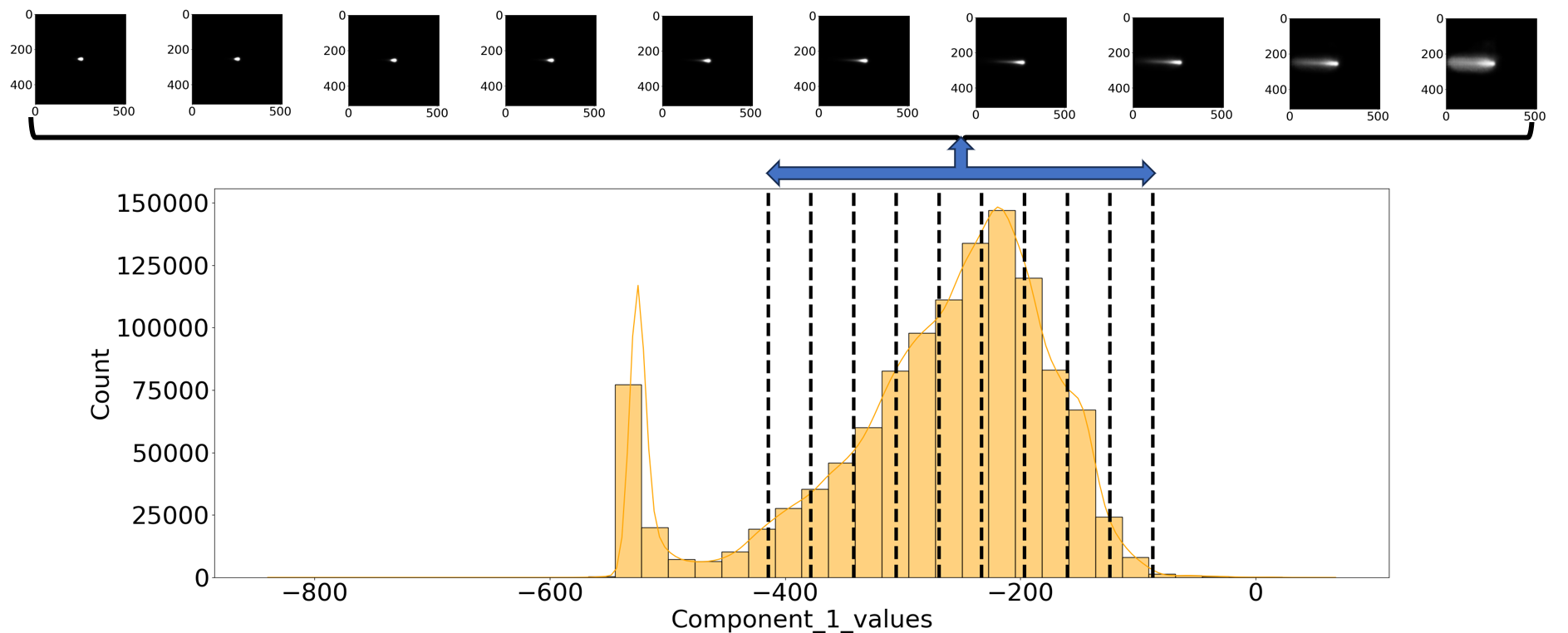}
\caption{Results of disentanglement analysis of the $2^{\mathrm{nd}}$ latent vector component (Component 1). Histograms demonstrate the value distribution of Component 1. The value of Component 1 clearly captures the physical quantity, namely the melt pool length, and shows a linear correlation with it. }
\label{fig:latent_disentangle_1}
}
\end{figure*}

Next, to investigate the properties of our constructed latent space, we perform a latent space disentanglement analysis that introduces component-wise variations to the latent vectors and sees how the trained decoder interprets them. For the convenience of demonstration, we investigated a $2\times 1$ latent space and disentangle each component separately. 

Figure~\ref{fig:latent_disentangle_0} demonstrates the results of the disentanglement analysis of the first component of the latent vector, where we fixed the value of the second component of the latent vector at its median. As we vary its value within a range comparable to its distribution, shown as the histogram in Fig.~\ref{fig:latent_disentangle_0}, we selected 10 data points indicated as the vertical black dash lines and visualized their corresponding reconstruction results via our trained decoder. From the reconstructed melt-pool images, we clearly see that the first component dominates the width and the area of the melt pool. 

Similarly, Fig.~\ref{fig:latent_disentangle_1} demonstrates the results of the disentanglement analysis of the second component of the latent vector, where we fixed the value of the first component of the latent vector at its median. As we vary its value within a range comparable to its distribution, shown as the histogram in Fig.~\ref{fig:latent_disentangle_1}, we selected 10 data points indicated as the vertical black dash lines and visualized their corresponding reconstruction results via our trained decoder. From the reconstructed melt-pool images, we clearly see that the second component dominates the length of the melt pool. 

Our disentanglement analysis demonstrates that our proposed approach possesses the potential to capture a physics-aware data-driven correlation between acoustic emission, thermal emission, and melt pool geometric characteristics.

\end{document}